\documentclass[10pt,journal]{IEEEtran}
\usepackage{array}
\usepackage{amsmath}
\usepackage{amsthm}
\usepackage{amsmath}
\usepackage{amssymb,bbm}
\usepackage[vlined, ruled, linesnumbered]{algorithm2e}
\usepackage[export]{adjustbox}
\usepackage{balance}
\usepackage{bm}
\usepackage{booktabs, tabularx}
\usepackage[skip=4pt,font={bf}, margin=4pt]{caption}  \usepackage{comment}
\usepackage{enumitem}
\usepackage{float}
\usepackage{graphicx}
\graphicspath{ {./images/} }

\usepackage{hyperref}
\usepackage{cleveref} \usepackage{multirow}
\usepackage{subfigure}
\usepackage{siunitx}
\usepackage{tikz} 
\usepackage{thmtools, thm-restate}
\usepackage{xcolor}
\usepackage{accents}
\newcommand{\ubar}[1]{\underaccent{\bar}{#1}}
\usepackage{bibunits}
\usepackage{pdfpages}

\newcommand{\CD}{\mathcal{D}}
\newcommand{\CI}{\mathcal{I}}

\newcommand{\CN}{\mathcal{N}}

\newcommand{\CF}{\mathcal{F}}

\newcommand{\CC}{\mathcal{C}}
\newcommand{\CS}{\mathcal{S}}
\newcommand{\CT}{\mathcal{T}}
\newcommand{\CA}{\mathcal{A}}

\newcommand{\CP}{\mathcal{P}}

 \ifx\techreport\undefined
\newcommand{\onlytech}[1]{\ignorespaces}
  \newcommand{\onlypaper}[1]{#1}
\else
\newcommand{\onlytech}[1]{#1}
  \newcommand{\onlypaper}[1]{\ignorespaces}
\makeatother
\fi

\makeatletter \@ifclassloaded{acmart}
{}{
\usepackage{substitutefont, newtxtext} }
\makeatother

\newcolumntype{Y}{>{\centering\arraybackslash}X}
\def\tablename{Table}
\renewcommand{\thetable}{\arabic{table}}  
\definecolor{red}{HTML}{E51400}  \definecolor{blue}{HTML}{0100EF} \definecolor{green}{HTML}{008A00} \definecolor{purple}{HTML}{AA00FF} \definecolor{dark-red}{rgb}{0.4, 0.15, 0.15}
\definecolor{dark-blue}{rgb}{0.15, 0.15, 0.4}
\definecolor{medium-red}{rgb}{0.3, 0, 0}
\definecolor{medium-blue}{rgb}{0, 0, 0.3}
\definecolor{light-red}{rgb}{0.7, 0, 0}
\definecolor{light-blue}{rgb}{0, 0, 0.7}

\hypersetup{
  colorlinks,
  linkcolor={medium-red},
  citecolor={medium-blue},
  urlcolor={medium-blue}
}

\SetCommentSty{mycommfont}
\DontPrintSemicolon
\SetKw{Break}{break}
\SetKwProg{myproc}{Procedure}{}{}


\pgfdeclarelayer{background}
\pgfdeclarelayer{foreground}
\pgfsetlayers{background,main,foreground}

 \newtheorem{corollary}{{\bf Corollary}}
\newtheorem{assumption}{{\bf Assumption}}
\newtheorem{definition}{{\bf Definition}}

\newtheorem{lem}{{\bf Lemma}}
\newtheorem{theorem}{{\bf Theorem}}
\newtheorem{proposition}{{\bf Proposition}}

\makeatother

\newcommand{\COLOR}{\color{black}}
\newcommand{\MARK}{\color{black}}
\newcommand{\ICNP}{\color{black}}
\newcommand{\SHEEP}{\color{black}}

\newcommand{\TON}{\color{black}}
\newcommand{\TONminor}{\color{black}}

\title{Quantifying Deployability \& Evolvability of Future Internet Architectures via Economic Models}

\makeatletter 
\author{
  \IEEEauthorblockN{
    Li Ye\IEEEauthorrefmark{2},
    Hong Xie\IEEEauthorrefmark{1}\thanks{$^\ast$Hong Xie is the corresponding author},
    John C.S. Lui\IEEEauthorrefmark{2},
    Kenneth L. Calvert\IEEEauthorrefmark{3}
  }\\
\IEEEauthorblockA{
    \IEEEauthorrefmark{2}The Chinese University of Hong Kong, 
    \IEEEauthorrefmark{1}Chongqing University,
    \IEEEauthorrefmark{3}University of Kentucky 
  } 
  \vspace{-0.2in}
}

\makeatother \hypersetup{draft}
\begin{document}

\setlength{\textfloatsep}{0.2\textfloatsep}
  \setlength{\dbltextfloatsep}{0.2\dbltextfloatsep}
  \setlength{\floatsep}{0.2\floatsep}
  \setlength{\dblfloatsep}{0.2\dblfloatsep}
  \setlength{\belowdisplayskip}{0.2\baselineskip}
  \setlength{\abovedisplayskip}{0.2\baselineskip}
  \captionsetup[subfloat]{captionskip=2pt}  

\maketitle

\begin{abstract}
Emerging new applications demand the current Internet to
provide new functionalities.
Although many future Internet architectures {\ICNP and protocols} have been proposed
to fulfill such needs, ISPs have been reluctant to deploy many of these architectures.
We believe technical issues are not the main reasons as
many of these new proposals are technically sound.
In this paper, we take an economic perspective
and seek to answer:
{\it Why do most new Internet architectures fail to be deployed?}
{\it How can the deployability of a new architecture be enhanced?}
We develop a game-theoretic model to 
characterize the {\MARK outcome of an architecture's deployment} through the
\textit{equilibrium} of ISPs' decisions.
This model enables us to: 
(1) analyze several key factors of the deployability of a new
  architecture such as the number of critical ISPs and the change of routing path; 
(2) explain the deploying outcomes of some previously proposed
architectures/protocols such as
IPv6, DiffServ, CDN, etc., and shed light on the \textit{``Internet flattening
phenomenon''}; 
(3)  predict the deployability of a new
architecture such as NDN, 
and compare its deployability with competing architectures.  
  Our study suggests that the difficulty to deploy a new Internet
  architecture comes from the \textit{``coordination''} of distributed ISPs. 
Finally, we design a mechanism to enhance the deployability 
of new architectures. 
\end{abstract} 

\section{\bf Introduction}
\label{sec:introduction}

Emerging applications create a constant push for the Internet
to be \emph{``evolvable''} to provide new functionalities.
For example, streaming video traffic from Netflix and other services
  {\SHEEP requires}
highly efficient content delivery across the Internet.
{\SHEEP Also,} users of online social network services like Facebook
want their private chats to be securely protected.
The increasing number of mobile phones
and IoT devices require better mobility and security support, and so on.
However, many of these needs are not being
supported by {\SHEEP the IPv4 network}.
To meet these emerging needs,
researchers have been developing new architectures and protocols,
and more importantly, exploring how to make the Internet \emph{``evolvable''}
so as to incorporate new functionalities.
Unfortunately, many of these research efforts ultimately fail to
result in wide scale deployment.

{Different Internet architectures/protocols experience different degrees of
deployment difficulties.} In the 1990s, the protocol IP~version~6 (IPv6)
was introduced to overcome certain shortcomings of IP~version~4.
In particular, IPv6 provides a larger address space and
additional features such as security.
However, after more than 20 years of effort, only a minority of
current Internet traffic is based on IPv6~\cite{ipv6_stat}.
Differentiated service (DiffServ)\cite{diff_serv} was designed to
provide Quality of Service (QoS) guarantees.
Although it is supported by many commercial routers~\cite{diff_serv},
only a few Internet service providers (ISPs) are willing to turn on
the DiffServ function.
In contrast, content delivery network (CDN) {\SHEEP
technology} has enjoyed rapid growth and wide deployment;
at the time of this writing,
over 50\% of the Internet traffic is delivered by CDNs~\cite{cdn_stat}.
In the past decade, a number of novel Internet architectures
were proposed to address challenges in the
the current IP network, through substantial changes to the network protocols.
For example, Named-Data Networking (NDN)~\cite{ndn} natively
facilitates content distribution, while the eXpressive Internet
Architecture (XIA)~\cite{xia} enables
incremental deployment of future protocols and features intrinsic
security.
MobilityFirst\cite{mobilityfirst} provides first-class support for
mobile devices.
Although NDN, XIA and MobilityFirst all have functional prototype systems,
as of this moment they do not have the wide-scale deployment.

All of the above {\ICNP architecture/protocol} proposals claimed to improve the
current Internet if they are successfully deployed.
However, only {\SHEEP the CDN technology} is smoothly deployed in the Internet,
while most of the others are not.
This motivates us to explore:
\textit{Why do many new Internet architectures fail to be deployed?
How can the deployability of a new architecture/protocol be enhanced?}
The failure of many proposed architectures{\ICNP /protocols}
to reach wide deployment is probably not due to
technical issues. In fact, many of these proposed architectures
feature technically superior designs, compared to the present {\SHEEP
  IPv4 network}. 
Instead, \emph{economic factors} are crucial in determining the
deployability of a new Internet architecture/protocol.

\vspace{-0.1in}
\begin{figure}[htb] 
  \centering
  \includegraphics[width=0.5\textwidth]{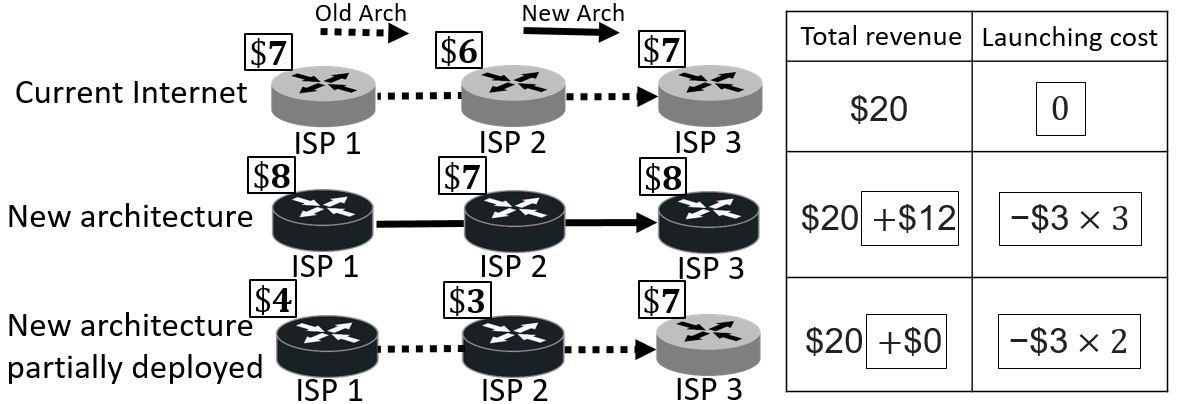}
  \caption{Economic difficulty to deploy a new Internet architecture}
  \label{fig:Intro:example} 
\end{figure}
\vspace{-0.05in}

To illustrate, let us consider the following example.
Fig.~\ref{fig:Intro:example} depicts a simple network with three ISPs, and
a traffic flow from ISP 1 to ISP 3.
Suppose that under the current Internet architecture,
the whole network can gain a total revenue of \$20.
Assume that a new Internet architecture, if it is fully deployed, will
increase the total revenue of the whole network to \$32
(i.e., improve the revenue by \$12).
Each ISP has a launching cost of \$3 to deploy this new architecture.
Suppose the revenue improvement is evenly distributed among ISPs,
i.e., each ISP gains \$12/3=\$4, or each ISP will earn \$4 more by
investing \$3 to upgrade the architecture, which yields a net gain of \$4{-}\$3{=}\$1.
However, the deployment requires a ``full-path'' participation,
i.e. ISP~1, ISP~2 and ISP~3 all need to deploy the new architecture in
order to achieve the improved functionality.
Fig.~\ref{fig:Intro:example} shows that
when only ISP~1 and ISP~2 deploy (ISP~3 does not),
the total revenue will not be improved, because
the new functionality is not enabled.

Although the above example illustrates
the potential profit gain for each ISP, unfortunately, the
new architecture will not necessarily be deployed in the network.
The reason is no ISP can be certain of gaining the benefit by
deploying the new architecture unilaterally.
In fact, if an ISP deploys, it will gain \$1 only if the other two ISPs
also deploy; otherwise the ISP will lose \$3.
The main difficulty is that each ISP is uncertain about 
other ISPs' deployment decisions.
Given this uncertainty,
ISPs tend to be conservative, and this
leads to the failure of deployment.
The above example highlights that
a new Internet architecture can be difficult to deploy
even if it can bring higher profits.

This paper studies the economic issues
for the deployment of new Internet architecture/protocol, and we aim to answer:
\begin{itemize}[topsep=0pt]
\item
\textit{Why have many Internet architectures/protocols
(e.g., IPv6, NDN, XIA) failed
to achieve large scale acceptance and wide deployment by ISPs?}

\item
\textit{What are the key factors that influence the deployability of a new 
architecture/protocol?   
How to enhance the deployability of a new architecture/protocol?
}
\end{itemize}
In addition, we analyze the economic impact of some engineering
mechanisms, e.g., tunneling,
that have been proposed to support incremental deployment of new
Internet architectures.
We also study the \textit{``Internet flattening phenomenon''}, where
content providers are bypassing transit ISPs and instead placing their servers
in data centers close to the end users.

Our contributions are:
  \begin{itemize}[topsep=0pt]
  \item We present a game-theoretic model of
   ISPs' economic interactions regarding the deployment
   decisions of a new Internet architecture (Sec.~\ref{sec:system_model}).  
Our model captures realistic factors, {\TONminor e.g., the routing paths can
  change during the deployment}. We later extend the model to
compare the deployability of competing
   architectures and study which will eventually get deployed
   (Sec.~\ref{sec:extension}).
 
\item We analyze ISPs' deployment decisions via the notion of
  ``equilibrium'' when ISPs have uncertainties on the benefits
  of the new architecture and are
    risk-neutral.

  \item We identify factors that make an architecture difficult to deploy.
Our analysis suggests that the requirement for many ISPs to coordinate is 
an important hurdle for deployability.
    It explains why architectures like IPv6 or DiffServ are
    difficult to deploy, while deploying CDN and NAT is easy.
Furthermore, the change of routing path and the competition between the old and
the new architectures are \emph{not\/} major factors hindering deployment.

  \item 
  We study several alternatives to enhance the deployability of a new
  architecture. We quantify how incremental deployment mechanisms such
  as IPv6 tunneling improve 
    the deployability of an architecture
    (Sec.~\ref{sec:increment_deployment}).
Our model confirms that by relying on data centers, content providers
can more easily deploy new architectures in the flattened Internet. 
Lastly, we design a coordination mechanism to improve the deployability 
of new architectures (Sec.~\ref{sec:mechanism}).
  \end{itemize}

{\TONminor
The organization of this paper is as follows.  
In Section~\ref{sec:system_model}, we model the ISP-network and the
cost/benefits 
for ISPs to deploy a new architecture, and formulate a strategic game. 
In Section~\ref{sec:equilibrium}, we reason about
ISPs' behaviors on deployment. In Section~\ref{sec:analysis}, we analyze the
impact of different factors on the deployability of an architecture.
Section~\ref{sec:extension} presents an extension of the model to
consider partial deployment and 
competitions of multiple architectures. In Section~\ref{sec:mechanism}, we propose a centralized economic
mechanism to help the deployment.
Section~\ref{sec:experiments} presents numerical experiments.
Section~\ref{sec:relatedwork} describes related works, and Section~\ref{sec:conclusion}
concludes.
}
\vspace{-0.15in}

\section{\bf System Model}
\label{sec:system_model}

We will present models for the old and the new architectures.
Then we formulate a game of architecture deployment.

\subsection{{\bf The Baseline Model of The Old Architecture}}

\begin{figure}
\centering
  \hspace{-0.05in}\includegraphics[width=0.52\textwidth]{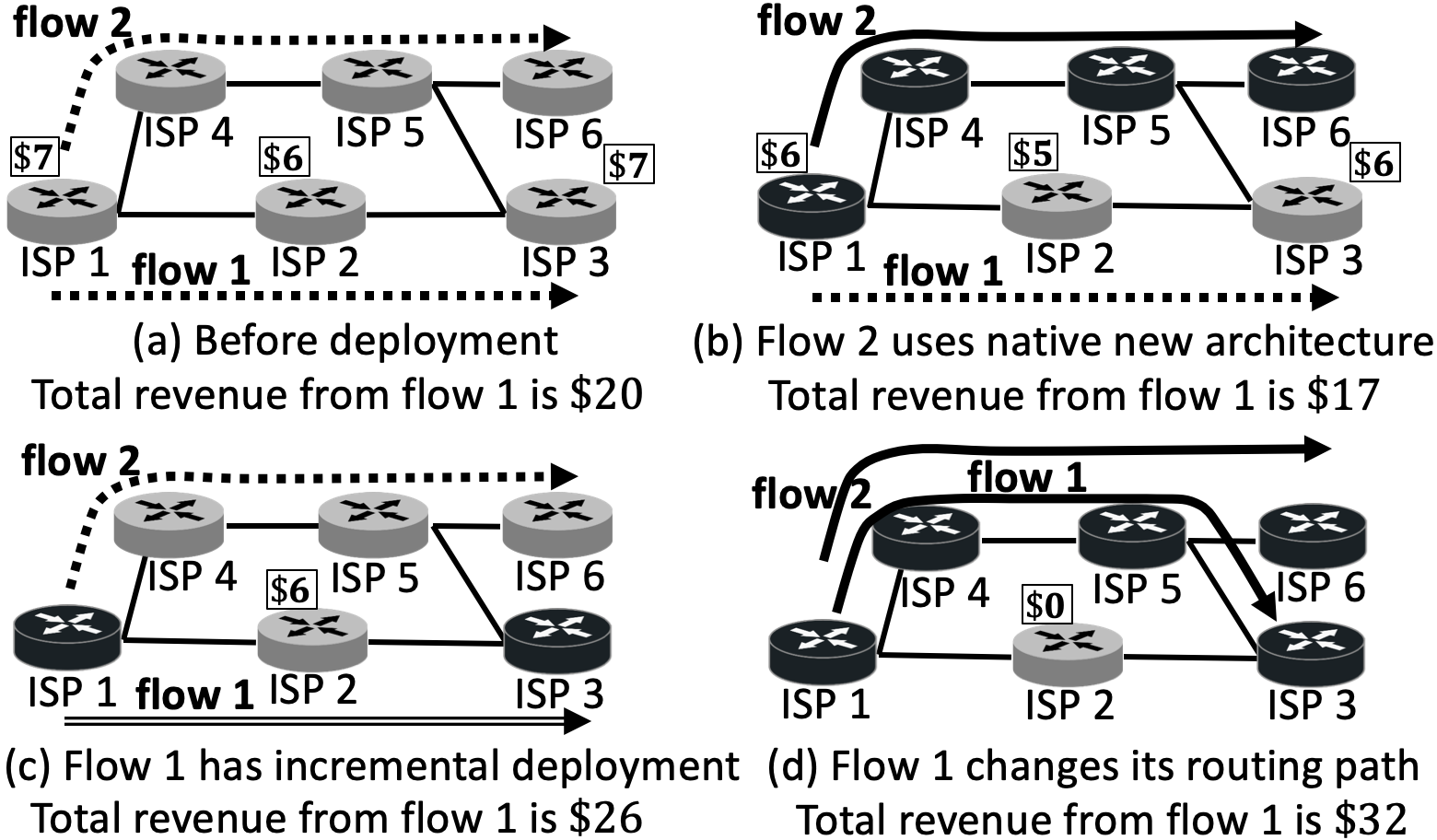}
\caption{Example to deploy IPv6 (different deployment status)}
  \label{fig:example}
\end{figure}

Consider $I {\in} \mathbb{N}_+$ ISPs denoted by $\mathcal{I} {\triangleq} \{1,\ldots, I\} $ 
and an undirected graph $(\mathcal{I},\mathcal{E})$,  
where $\mathcal{E}  {\subseteq}
\{(i,j) | i, j {\in} \mathcal{I}, i {<} j\}$
indicates the connectivity among ISPs 
and we impose $i {<} j$ to eliminate
the redundancy, since edges are undirected.
Fig.~\ref{fig:example}(a) illustrates a set $\mathcal{I} = \{1,2,3,4,5,6\}$
of six ISPs with
$\mathcal{E} {=} \{(1,2), (2,3), (1,4), (3,5), (4,5),(5,6)\}$.

Consider the baseline case where all ISPs use the old architecture.
We define a flow to be aggregation of all packets sent along a
particular routing path (sequence of edges).
Thus, we partition the traffic over the network
$(\mathcal{I}, \mathcal{E})$ into a finite set $\CF$ of flows.
Let $w_f \in (0,1]$ be the fraction of all traffic due to flow $f \in \CF$,
so that $\sum_{f \in \CF} w_f =1$.
{\TONminor For simplicity, we assume $w_f$ is static for the old architecture.}
Let $\vec{p}_f$=(\textit{source},$\ldots$, \textit{destination})
denote the routing path of $f$.
Fig.~\ref{fig:example}(a) shows two flows $\CF {=} \{1, 2\}$
under the old architecture.
Flow $1$ goes through ISPs 1, 2 and 3,
accordingly $\vec{p}_{1} {=} (1{,}2{,}3)$.
Let $r_{i,f} \in \mathbb{R}_{\geq 0}$ denote ISP $i$'s revenue share
from flow $f$,
where $r_{i,f} = 0$ whenever ISP $i$ is not on the routing path of $f$,
i.e., $i \notin \vec{p}_f$.
Then ISP $i$'s revenue share from all the flows is
$\sum_{f \in \mathcal{F}} r_{i,f}$ and
the total revenue generated from $f$ is $\sum_{i \in \mathcal{I}} r_{i,f}$.
Fig.~\ref{fig:example}(a) shows that
$r_{1,1}{=}\$7$, $r_{2,1}{=}\$6$ and $r_{3,1}{=}\$7$.

\subsection{\bf Model for The New Architecture}
\label{sec:model_new}

Given a set of $\mathcal{S} \subseteq \mathcal{I}$ ISPs that deploy
a new architecture, 
we model how this deployment influences 
the routing of flows and the revenues of ISPs.

\noindent
{\bf $\bullet$ The deployment of the new architecture. }
Let $c_i \in \mathbb{R}_+$ denote ISP $i$'s launching cost
to deploy a new architecture.
This cost captures the expense of procuring new hardware,
upgrading software, and payment for engineers to manage the new
infrastructure, etc.
We model two types of deployment: 
(1) \textit{native new architecture}; 
(2) \textit{incrementally deployed new architecture}.  
First, however, we must define critical ISPs.

\begin{definition}
[\bf Critical ISP]  
Define $\CC(\vec{p})$ to be the set of ISPs such that 
the native new architecture can be enabled on routing path $\vec{p}$
if and only if all ISPs in $\CC(\vec{p})$ deploy the new architecture.   
We call an ISP $i \in \CC(\vec{p})$ a critical ISP of $\vec{p}$.  
\end{definition}

\noindent
Note that the critical ISP set $\CC(\vec{p})$ depends on the new
protocol or architecture.
For example, consider the routing path $\vec{p}_{1}$ in Fig.~\ref{fig:example}.
To enable native IPv6 on $\vec{p}_1$, the critical ISPs are
$\CC(\vec{p}_1) = \{1,2,3\}$.
On the other hand, to enable native TCP on $\vec{p}_1$,
{\TONminor none of the ISPs are critical, i.e., $\CC(\vec{p}_1) = \emptyset$.}
The reason is that IPv6 requires ``all'' ISPs along a path to deploy,
while TCP {\TONminor requires only the end devices (not the ISPs) to deploy it.}
For the ease of presentation,
we consider one protocol/architecture at any time, unless otherwise
stated explicitly.
We next define the ``full path participation''.

\begin{definition}
[\bf Full path participation]
A new architecture requires full-path participation,
if for any routing path $\vec{p}$, $\mathcal{C} (\vec{p})$ contains all
ISPs in $\vec{p}$.
\end{definition}

\noindent
Thus, IPv6 needs full path participation while
TCP does not. 

Some architectures can be incrementally deployed,
where a routing path $\vec{p}$ can (partially) enable 
the functionality of the new architecture even when only
a proper subset of critical ISPs in $\mathcal{C} (\vec{p})$ participate.  
A real-world example is {\it ``IPv6 tunneling''}, where encapsulation
is used to carry IPv6 packets through islands of IPv4-only ISPs
between ISPs that have deployed IPv6.
Incremental deployment may introduce 
additional overhead to the network protocols and
thus degrade the performance of the new architecture.
With incremental deployment, performance of the new architecture
may improve as more critical ISPs along a routing path deploy it.

\noindent
{\bf $\bullet$ Change of routing path. }
Let $\CP_f$ denote the ordered set of all alternative routing paths of flow $f$.
The routing protocol determines $\CP_f$ (independent of the new architecture),
ordering the paths in $\CP_f$ according to their priority.  
Note that $p_f \in \CP_f$ and $p_f$ has the highest priority in $\CP_f$.   
In practice, a Border Gateway Protocol (BGP) router typically records multiple
paths to the same destination\cite{cisco_bgp}, and then chooses one with the
highest priority according to ISPs' \emph{routing policy\/}
configuration and the path length.
In our model, if multiple routing paths in $\CP_f$ can use the
{native new architecture}, flow $f$ switches to the one among them
with the highest priority.
If none of the routing paths in $\CP_f$ can use the new architecture, the
routing path $\vec{p}_f$ is not changed.
Here,
we assume a flow changes its routing path {\TONminor only when the
native architecture is available (or otherwise there is no incentive
to change the routing path).
E.g. in Fig.~\ref{fig:example}(d), flow $1$ can use the new architecture
via the route $(1,4,5,3)$, thus flow $1$ switches to this route. }

Given the set of deployed ISPs $\mathcal{S}$ and flow $f$'s alternative routing paths
$\mathcal{P}_f$,
we have a unique routing path for each flow $f$.
Let $\vec{P}_f (\CS)$ denote the routing path of $f$ when ISPs in $\CS$
deploy the new architecture.
It satisfies $\vec{P}_f (\emptyset) = \vec{p}_f$.
Let $\mathcal{F}^{old} (\mathcal{S})$
and $\mathcal{F}^{new} (\mathcal{S})$ denote the set of flows
who use the old and new architecture respectively.
Here,
$\mathcal{F}^{new} (\mathcal{S}) \cap \mathcal{F}^{old} (\mathcal{S})
= \emptyset$,
$\mathcal{F}^{new} (\mathcal{S}) \cup \mathcal{F}^{old} (\mathcal{S})
= \mathcal{F}$, and
$\mathcal{F}^{old} (\emptyset) = \mathcal{F}$.
The set of flows $\mathcal{F}^{new}(\mathcal{S})$ includes flows that use the native new architecture or
incremental deployment mechanisms.
We use $b_{i,f} (\mathcal{S}) \in \{0,1\}$ to indicate whether ISP
$i$ is bypassed in flow $f$, formally 
\[
b_{i,f} (\mathcal{S})
= \left\{
\begin{aligned}
&1, && \text{if $i \in \vec{p}_f$ and $i \notin \vec{P}_f (\mathcal{S})$},
\\
& 0, && \text{otherwise}.
\end{aligned}
\right.
\]

{\TONminor
  We use the example in Fig.~\ref{fig:example} to illustrate our model.
  In Fig.~\ref{fig:example}(a),
  the routing path of flow $1$ is $\vec{P}_1(\emptyset){=}(1,2,3)$.
  In Fig.~\ref{fig:example}(b), only flow $1$ uses the old architecture and
  $\CF^{old}(\{1,4,5,6\}){=}\{1\}$. Meanwhile, only flow $2$ uses the new architecture and
  $\CF^{new}(\{1,4,5,6\}){=}\{2\}$. The possible routing paths for flow $1$ are
  $\CP_1=\{(1,2,3), (1,4,5,3)\}$.  In Fig.~\ref{fig:example}(d),
  flow $1$ has routing path $\vec{P}_1(\{1,3,4,5,6\})=(1,4,5,3)$. In flow $1$, ISP 2 is bypassed, so $b_{2,1}(\{1,3,4,5,6\})=1$.
}

\noindent
{\bf $\bullet$ Change in revenue. }
We first model the revenue change at the flow-level.
Let $\Delta_f (\mathcal{S}) \in \mathbb{R}$ denote the revenue change
on flow $f$ when ISPs in $\mathcal{S}$ deploy the new architecture, i.e.,
the revenue of flow $f$ changes from $\sum_{i \in \mathcal{I}} r_{i,f}$ to
$\sum_{i \in \mathcal{I}} r_{i,f} + \Delta_f (\mathcal{S})$.
We assume the revenue change satisfies
$\Delta_f (\mathcal{S}) \geq 0$ for all $f \in \mathcal{F}^{new} (\mathcal{S})$
and $\Delta_f (\mathcal{S}) \leq 0$ for all $f \in \mathcal{F}^{old} (\mathcal{S})$.
This captures the assumption that the new architecture is
superior to the old architecture in terms of the revenue, and the revenue
of flows that use the old architecture may drop due to competition from
the new architecture.
{\TONminor
  Consider the single (call it $f$) in
  Fig.~\ref{fig:Intro:example}. Before deployment,
  $r_{1,f}=r_{3,f}=7$, $r_{2,f}=6$.
  After deployment, $\Delta_f(\{1,2,3\})=\$12$.
}
The following assumption further characterizes the revenue
from a flow $f \in \mathcal{F}^{new} (\mathcal{S})$.

\begin{assumption}
  \label{asump:revenue_gain}

For each $f \in \mathcal{F}^{new} (\mathcal{S})$,
$\Delta_f (\mathcal{S})$ satisfies:
\begin{enumerate}
\item
  $\Delta_f(\CS) \le \Delta_f(\CT)$ if $\CS\subset \CT$,

\item
  $\Delta_f(\CS){=}\Delta_f(\CT)$ if $n_f( \CS){=}n_f(\CT)$ and
  $\vec{P}_f (\CS) {=} \vec{P}_f (\CT)$, 
\end{enumerate}
where $n_f(\CS) {\triangleq}
|\CC(\vec{P}_f(\CS)) \cap \CS|$ denotes the number of critical
ISPs in the routing path of $f$ that deploy the new architecture.
\label{asump:revenueFun}
\end{assumption}
\vspace{-0.2in}
(This quantity will play an important role in later results.)
Assumption \ref{asump:revenueFun} captures that:
1) The revenue from a flow is non-decreasing
as more ISPs deploy the new architecture; and
2) The revenue generated from a flow is determined by the number of
critical ISPs who deploy the new architecture.

Now we model the revenue change at the level of individual ISPs.  
Here we only focus on the ISPs who do not deploy the architecture, 
as they share the revenue according to the existing contracts (or
peering agreement) on the old architecture.  
We will address ISPs that deploy the new architecture in Sec.~\ref{sec:game}. 
Let $\delta_{i,f} (\mathcal{S}) \in \mathbb{R}$
denote the revenue change of ISP $i\not\in \CS$ on flow $f$.
Namely, the revenue share of ISP $i\not\in\CS$ from flow
$f$ changes from $r_{i,f}$ to $r_{i,f} + \delta_{i,f} (\mathcal{S})$
when some other ISPs deploy the new architecture.
The following proposition characterizes
desirable properties of $\delta_{i,f}(\CS)$.

\begin{proposition}
The revenue change $\delta_{i,f}(\CS)$ of $i{\not\in}\CS$ satisfies:
\begin{enumerate}

\item
For each $f \in \mathcal{F}^{old} (\mathcal{S})$:
\[
\textstyle
i \in \vec{p}_f \Rightarrow \delta_{i,f} (\mathcal{S})\leq 0 
\text{ and } 
i\not\in \vec{p}_f \Rightarrow \delta_{i,f}(\mathcal{S})=0. 
\]

\item
For each $f\in\CF^{new}(\CS)$:
\[
\textstyle
b_{i,f} (\mathcal{S}) {=} 1 
{\Rightarrow}  
\delta_{i,f} (\mathcal{S}) {=} - r_{i,f} 
\text{ and } 
b_{i,f}(\CS) {=} 0 
{\Rightarrow} 
\delta_{i,f}(\CS) {=} 0.  
\]  
\end{enumerate}
\label{Assum:RevDis}
\end{proposition}

\vspace{-0.1 in}
\noindent
Proposition \ref{Assum:RevDis} captures that 
ISPs that are in the routing path of a flow that uses the old
architecture may have revenue losses 
(e.g. ISPs of flow 1 in Fig.~\ref{fig:example}(b)), while
those not in the path do not have revenue loss from the flow.  
Furthermore, for a flow that uses the new architecture: 
when an ISP $i\not\in\CS$ is bypassed by a flow,
the ISP's revenue share from that
flow becomes 0 (e.g. ISP 2 in Fig.~\ref{fig:example}(d));
otherwise 
the ISP's revenue is unchanged (e.g. ISP 2 in Fig.~\ref{fig:example}(c)).

\subsection{{\bf The ISP's Decision Model}}
\label{sec:game}
{\color{red}
}

\begin{figure}
  \centering
  \begin{minipage}{0.315\textwidth}
\includegraphics[width=\textwidth]{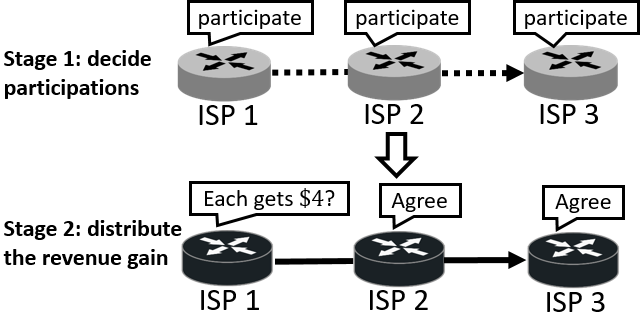}
  \caption{Two stages of ISPs' game}
  \label{fig:two_stage_game}
  \end{minipage}
  \begin{minipage}{0.167\textwidth}
  \centering
  \includegraphics[width=\textwidth]{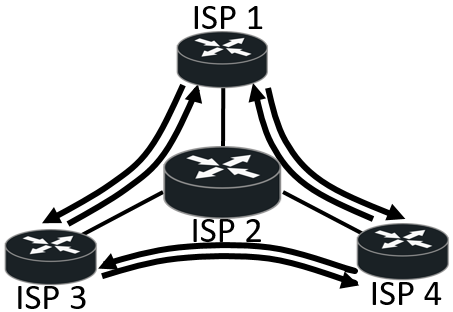}
  \caption{Example of flows, which imply multiple equilibria}
  \label{fig:example_flow}
  \end{minipage}
\end{figure}

We formulate a two-stage game as depicted in Fig.~\ref{fig:two_stage_game}, 
where in the first stage, ISPs decide whether to deploy the new architecture 
and in the second stage, ISPs decide how to distribute the revenue.
We will reason about the game via {\em backward-induction}, i.e. firstly analyze
how ISPs distribute the revenue, and then analyze how ISPs decide whether to deploy.

\noindent{\bf $\bullet$ The second stage: distributing net revenue
  gain.}
We first focus on how the
 net revenue gain among ISPs in
  $\CS$ will be distributed. 
We will describe a
  ``proposal-agreement'' process, for negotiating a distribution
  among the deploying ISPs, and we state an assumption
  (supermodularity) that ensures that a stable
  equilibrium exists, and can be computed \emph{a priori}. 

Let $v(\mathcal{S})$ denote the net revenue gain to ISPs $\CS$.
This is the total revenue increment of all ISPs, minus the revenue
increment of all the non-deploying ISPs:
\begin{align*}
v(\mathcal{S})
=
&
\sum\nolimits_{f \in \mathcal{F} } \Delta_f (\mathcal{S})
-
\sum\nolimits_{i \not\in \mathcal{S}}
\sum\nolimits_{f \in \mathcal{F}} \delta_{i,f} (\mathcal{S}).  
\end{align*}
As a simple consequence, if $v(\mathcal{I}) < 0$, 
it is impossible for all ISPs to deploy, 
as they will suffer revenue loss as a whole; thus we 
assume that $v(\CS){\ge} 0$. Moreover, we assume as follows:

\begin{assumption}[supermodular]
\label{asmp:supermodular}
For all $\CS{\subseteq} \CT$, $i {\in} \CI$ and $i{\not\in} \CT$:
\[
v(\CS \cup \{i\}){-}v(\CS)
\leq
v(\CT \cup \{i\}) {-} v(\CT).
\]
\end{assumption}

\noindent
Assumption \ref{asmp:supermodular} states that
the net revenue gain from one new ISP deploying the architecture cannot
be smaller if a larger set of ISPs have deployed beforehand.
That is, different ISPs are complementary in their effect on net revenue
gain through deployment of the new architecture.
Let
$\phi(\CS, v){=}(\phi_1,{\ldots},\phi_{|\CS|}){\in}\mathbb{R}^{|\CS|}$
denote a distribution mechanism for ISPs in $\CS$,
where $\phi_i(\CS{,}v)$ is the net revenue gain distributed to ISP
$i{\in}\mathcal{S}$ and
$\sum_{i\in\CS} \phi_i(\CS{,}v){=}v(\CS)$.

We consider a ``proposal-agreement'' process described
by Hart and Mas-Colell~\cite{hart1996bargaining} for ISPs 
to reach a fair distribution mechanism.
It works in the following multi-round process.
In each round there is a set of ``active''
players $\CT{\subseteq} \CS$, and a ``proposer'' $i\in \CT$.
In the first round, $\CT{=}\CS$. The proposer is chosen uniformly
at random from $\CT$. The proposer proposes a
distribution mechanism $\phi^\prime$ that satisfies
$\sum_{i\in\CT}\phi_i^\prime(\CT,v)=v(\CS)$. If all members of $\CT$ accept it,
then each ISP in $\CT$ will be distributed the revenue gain defined by
$\phi_i^\prime(\CT,v)$.
If the proposal is rejected by at least one member of $\CT$, then we move
to the next round, where with probability $\rho\ge 0$, the proposer
$i$ drops out, and the set of active players becomes $\CT\backslash \{i\}$.
The dropped-out proposer $i$ gets a final distribution of $0$.
This process models how ISPs bargain to get an acceptable (``fair'')
distribution of revenue after the deployment.
The bargaining power of an ISP comes from the ability to reject
proposals it considers ``unfair'' and propose an alternative, and the risk
of dropping out if the ISP proposes an unfair distribution.

{\TONminor
  Note that other processes, such as bargaining~\cite{gul1989bargaining} or
  bidding~\cite{perez2001bidding} can result in ISPs reaching the
  equilibrium whose existence is established in the next section.
}

\noindent{\bf $\bullet$ The first stage: the architecture deployment game.}
Given the mechanism $\phi$,
how much revenue increment an ISP will share is determined by its own action
(i.e., whether to deploy the architecture or not)
and other ISPs' actions.
Thus, we formulate a strategic-form game to characterize
ISPs' strategic behavior in deciding whether to deploy the new architecture.
We denote by $\tilde{\CC}
{\triangleq}
\bigcup\nolimits_{{f}\in \CF} \CC(\vec{p}_{f})
$
the set of all critical ISPs. If an ISP is not critical to any flow, it
is irrelevant and will not get involved in the deployment.
Hence the players of our interest are all the critical ISPs $\tilde{\CC}$.
Each ISP has two possible actions denoted by $\CA {\triangleq} \{0,1\}$,
where $1$ indicates that an ISP deploys the new architecture and
$0$ indicates not.
Let $a_i {\in} \CA$ be the action of ISP $i {\in} \tilde{\CC}$
and let $\bm{a} {\triangleq} (a_i)_{i\in \tilde{\CC}}$
denote the action profile of all critical ISPs.
Given an action profile $\bm{a}$, the corresponding set of ISPs who
deploy the new architecture is denoted by
$
\CS_{\bm{a}}
{\triangleq}
\{i | a_i {=}1, i {\in} \tilde{\CC}\}.
$
The utility (or profit) $u_i(\bm{a})$ of ISP $i$ is the
distributed revenue increment minus its launching cost, 
plus the change of revenue
  from the old architecture, i.e.
\begin{align}
  \label{eq:utility}
u_i(\bm{a})
\triangleq
\begin{cases}
\phi_i(\CS_{\bm{a}}{,}v) {-} c_i
&
\text{if } a_i{=}1,\\
 \sum_{f\in \CF} \delta_{i,f}(\CS_{\bm{a}})& \text{if }a_i{=}0,
\end{cases}
\end{align}
where $c_i$ is ISP $i$'s cost in deploying the new architecture.
We denote this {\it ``architecture deployment game''} by a tuple
$G {\triangleq} \langle{\tilde{\CC}, \CA, \bm{u}}  \rangle$,
where $\bm{u} {\triangleq} (u_1, {\dots}, u_{|\tilde{\CC}|})$ is a vector of functions.

\section{\bf Analyzing ISPs' Decisions via Equilibrium}
\label{sec:equilibrium}

We first consider ISPs' deployment decisions under a general setting, and 
show that multiple equilibria are possible
in the architecture deployment game.  
We also derive a potential function to characterize the equilibria 
and show that when ISPs face uncertainty, the equilibrium that
maximizes the potential function will be reached.  
Lastly, we study a special case to derive more closed-form results  
revealing more insights.  

\vspace{-0.1in}
\subsection{\bf Equilibrium of the Architecture Deployment Game} 

\noindent 
{\bf $\bullet$ Stable distribution mechanism in the second stage. }
In the following lemma, we apply the technique of Hart and
Mas-Colell~\cite{hart1996bargaining} 
to prove the existence and uniqueness
of a stable distribution mechanism 
for the proposal-agreement process defined in Sec.~\ref{sec:game},
and derive a closed-form expression for the 
net revenue share for each ISP.   

\begin{lem}Under Assumption~\ref{asmp:supermodular}, 
the proposal-agreement process in Sec.~\ref{sec:game} leads to a unique stationary subgame-perfect equilibrium, 
in which ISPs divide the net revenue gain as
\begin{align}
& 
\phi_i(\CS,v) = 
\varphi_i(\CS, \tilde{v}) + \sum\nolimits_{f\in \CF}
\delta_{i,f}(\CS\backslash \{i\}), 
&& 
\forall i \in \mathcal{S},  
\label{eq:NetRevDisGeneral}
\end{align}
where $\varphi_i(\CS, \tilde{v})$ and $\tilde{v}(\CS)$ are defined as 
\begin{align}
& 
\varphi_i(\CS, \tilde{v}) {\triangleq}
\!\sum_{\mathcal{T} \subseteq \CS \setminus \{i\}}\!\!\!\!
\frac{|\mathcal{T}|! (|\CS| {-} |\mathcal{T}|{-}1)!}{|\CS|!}
[\tilde{v}(\mathcal{T} {\cup} \{i\}) {-} \tilde{v}(\mathcal{T})], 
\label{eq:sharplyvaluedef}
\\
& 
\tilde{v}(\CS)\triangleq v(\CS)-\sum\nolimits_{i\in
    \CS}\sum\nolimits_{f\in \CF} \delta_{i,f}(\CS \backslash \{i\}).  
\label{eq:v_prime} 
\end{align}
\label{lem:EquilibriumSecondStage}
\end{lem} 
\vspace{-0.18 in}

{\TONminor
\begin{definition}
The stable distribution mechanism 
is a distribution mechanism in the unique stationary subgame-perfect
equilibrium of the game with the proposal-agreement 
process.  
\end{definition}
}
The physical meaning of this stationary subgame-perfect equilibrium is that 
no ISP in $\mathcal{S}$ can increase its net revenue gain share 
by proposing other distribution mechanisms or 
rejecting the stable distribution mechanism.   
For a mathematically rigorous definition of the 
stationary subgame-perfect equilibrium, 
we refer the reader to \cite{hart1996bargaining}.   
ISPs may have other mechanisms to do non-cooperative bargaining, 
one can also apply other
techniques~\cite{gul1989bargaining,Ma:2010:IEU} 
to show that they will reach the same stable distribution
mechanism as in Lemma~\ref{lem:EquilibriumSecondStage}.   
In Lemma~\ref{lem:EquilibriumSecondStage}, 
the $\varphi_i(\CS, \tilde{v})$ derived in
Eq.~(\ref{eq:sharplyvaluedef}) is a Shapley value 
with a value function $\tilde{v}(\CS)$ defined in Eq.~(\ref{eq:v_prime}) 
instead of the net revenue $v(\mathcal{S})$. This is because
$\tilde{v}(\CS)$ considers ISPs' potential
revenue loss from not participating in the proposal-agreement process.  

\noindent
{\bf $\bullet$ Deployment equilibria in the first stage. }  
Based on the stable distribution mechanism in Lemma~\ref{lem:EquilibriumSecondStage}, 
we analyze ISPs' deployment decisions via the notion of Nash
equilibrium.

\begin{definition}
  \label{def:nash_eq}
An action profile $\bm{a}^* {\in} \CA^{|\tilde{\CC}|}$ is a strict
pure Nash equilibrium of the game $G$,
if for $\forall$ $i {\in} \tilde{\CC}$, and $\forall a_i {\in} \CA$, $a {\neq} \bm{a}^*_i$,
\[
u_i(\bm{a}^*) > u_i(\bm{a}^*_{[i:=a]}),
\]
where the notation $\bm{a}_{[i:=x]}$ denotes the action profile
$\bm{a}$ with $i$'s action $a_i$ replaced by $x$, i.e.
$(a_1,\ldots,a_{i-1},x,a_{i+1},\ldots, a_{|\tilde{\CC}|})$.
\end{definition}
\noindent
In other words, at such equilibrium, no ISP can increase its utility
by unilaterally deviating from its current action.  

The game $G$ may have multiple equilibria.  
To illustrate, consider Fig.~\ref{fig:example_flow}.
Suppose that the launching cost of each ISP is \$3.
There are two equilibria. The first one is all ISPs do not deploy the new
architecture. This is because an ISP's unilateral deviation to deploy the new
architecture will result in a loss of $\$3$. The second one is all ISPs
deploy the new architecture. This is because all ISPs can have positive revenue
gain when the new architecture is successfully deployed in the network.

We apply Topkis's results\cite{topkis2011supermodularity} to show
that the game $G$ has a smallest (or largest) equilibrium with the smallest
(or largest) set of deployed ISPs in our game in Lemma~\ref{lemma:super_eq}.

\begin{lem}\label{lemma:super_eq}
{The set of equilibria of $G$ has a smallest element $\underline{\bm{a}}^*$,
  and a largest element $\overline{\bm{a}}^*$, such that
  $\underline{\bm{a}}^\ast\le \bm{a}^*\le \overline{\bm{a}}^\ast$ for
  any other equilibrium $\bm{a}^*$, where ordering is component-wise, or
 equivalently $\CS_{\underline{\bm{a}}^\ast} {\subseteq} \CS_{\bm{a}^*} {\subseteq}
  \CS_{\overline{\bm{a}}^\ast}$}.
\end{lem}

\begin{corollary}
(1) If $\forall i v(\{i\}) {\leq} c_i$,
 then $(0,{\ldots}, 0)$ is an equilibrium.
(2) If $\forall i$, $\varphi_i(\tilde{\CC}, \tilde{v}) {\geq} c_i$,
 then $(1,\ldots, 1)$ is an equilibrium.
\label{cor:EquiFailSucc}
\end{corollary}
Corollary \ref{cor:EquiFailSucc} states that {\em ``all critical ISPs decide not to deploy''} is an equilibrium, if no
ISP's revenue gain exceeds its launching cost when only that ISP deploys.
{Many architectures such as DiffServ and IPv6 satisfy this condition since
  their functionality can not be used under the deployment of a single ISP.}
Also, 
  full deployment is an equilibrium if every critical ISP's revenue gain
  is greater than its cost of deployment.

\noindent 
{\bf $\bullet$ Which equilibrium will be reached in the first stage?} 
We apply two approaches to show that a unique equilibrium 
will be reached.  
The first one lets ISPs dynamically change their decisions in
response to others' actions.  
The second one lets ISPs reason about other ISPs' actions.
These two approaches reflect ISPs' uncertainties on the benefits from
deployment.
In particular, before the deployment, ISP $i$ perceives a utility
$u_i(\bm{a}){+}\varepsilon_i(\bm{a})$, where $\varepsilon_i(\bm{a})$ 
denotes the error or noise in the perception.   
When ISPs choose not to deploy the architecture (i.e. $a_i{=}0$), they
are certain that there will be no revenue improvement, that is,
$\varepsilon_i(\bm{a})=0$ when $a_i=0$.
To facilitate analysis, we need the following lemma.  

\begin{lem}[\cite{monderer1996potential}]
If Eq. (\ref{eq:sharplyvaluedef}) holds,
then $G$ is a potential game, i.e., $\exists$ a function $\Phi:
\CA^{|\CI|} \mapsto \mathbb{R}$
such that for all $\bm{a}$ and $i$,
\begin{align}
    \label{eq:potential}
    u_i(\bm{a}_{[i:=1]}) - u_i(\bm{a}_{[i:=0]})
    = \Phi(\bm{a}_{[i:=1]}) - \Phi(\bm{a}_{[i:=0]})
\end{align}
\end{lem}

\noindent
The above lemma states that the change of an ISP's utility is equal
to the change of a potential function. 
Therefore an ISP will have a positive profit to deploy an architecture
if and only if her deployment increases the potential function.
Therefore, there exists a one-to-one mapping
between the equilibria of game $G$ and the local maxima of
the potential function $\Phi$.

\noindent 
{\bf  \em (1) Logit response dynamics.}
In this part, the error $\varepsilon_i(\bm{a})\in \mathbb{R}$ follows a logistic distribution with c.d.f.
$\mathbb{P}[\varepsilon_i(\bm{a}){<}x]={1}/({1+e^{-x/\beta_i}})$ when $a_i{=}1$.
When $a_i{=}0$, the error $\varepsilon_i(\bm{a})=0$ with probability 1.
 Here, the parameter $\beta_i{\ge}0$
represents ISP $i$'s degree of uncertainty. 
For example, when $\beta_i{\rightarrow}0$, $\varepsilon_i(\bm{a})$ is always close to $0$ which means
that ISP $i$
has little uncertainty about its utility.  
We assume that an ISP will choose to deploy when its perceived utility 
of deploying is greater than the perceived utility of not deploying.

ISPs' sequential dynamics are summarized in Algorithm~\ref{alg:SA}.
We divide time into slots, i.e.,
$t{\in}\{1,\ldots, T\}$.
  Let $\bm{a}^{(t)}$ denote the action profile at time slot
$t$, and ISPs start with some initial action profile $\bm{a}^{(0)}$.
At time slot $t$, we randomly pick one ISP, let's
say $i\in\tilde{\CC}$, to make a decision based on
 ISPs' actions in the last time
slot $\bm{a}^{(t-1)}$. 
{This setting captures that ISPs
  sequentially make decisions.}
More specifically, ISP $i$ chooses each action $a_i^{(t)}{\in}\{0,1\}$ with
a probability that is logit-weighted by utility as shown in
Line~\ref{line:5} of the Algorithm. The setting that ISPs can switch
their decisions between ``to deploy'' and ``not to deploy'' captures
that ISPs can buy (or rent) and sell (or stop renting) devices for the
new architecture.  
Note that in the {\em Logit Response Dynamics}, we ignore the cost for
an ISP to switch from one architecture to another.

\begin{algorithm}
 \caption{  Logit Response Dynamics}\label{alg:SA}
 {ISPs have some initial action profile $\bm{a}^{(0)}$}\\
 \For{$t=1,\ldots,T$}{
   Pick an ISP $i$ uniform randomly from all ISPs\\
   choose ISP $i$'s action $a\in\{0,1\}$ with probability
   \begin{align}
\mathbb{P}[a_i^{(t)}=a|\bm{a}^{(t-1)}]=
 \frac{e^{u_i(\bm{a}_{[i:=a]}^{(t-1)})/\beta_i}}
 {e^{u_i(\bm{a}_{[i:=0]}^{(t-1)})/\beta_i} +  
    e^{u_i(\bm{a}_{[i:=1]}^{(t-1)})/\beta_i}}.\nonumber 
   \end{align}
   \vspace{-0.05in}
   \label{line:5}
}
\end{algorithm} 

\begin{lem}[\cite{alos2010logit}\cite{alos2017convergence}]
Following Algorithm~\ref{alg:SA}, 
 ISPs' strategy profile converges to a unique
(strict) equilibrium as the time $T {\rightarrow} \infty,
\beta_i{\rightarrow} 0, \forall i{\in}\CN$.
Given $\beta_i {=} \beta, \forall i{\in}\CN$, 
as $T {\rightarrow} \infty$ the limiting distribution  
of ISPs' action profiles is (if the limit exists)
\begin{align}
     \label{eq:stationary}
     \lim_{T\rightarrow \infty}\mathbb{P}(\bm{a}^{(T)}=\bm{a}) 
    \stackrel{\text{if exists}}{=}
     e^{\Phi(\bm{a})/\beta}
     /
     \sum\nolimits_{\bm{a}\in \CA^{|\tilde{\CC}|}} e^{\Phi(\bm{a})/\beta}.
\end{align}
where $\Phi$ is the potential function of game $G$ defined in (\ref{eq:potential}).
\end{lem}

\noindent
In the potential game $G$, 
ISPs are more
likely to stay in the deployment status $\bm{a}$ with a higher potential value $\Phi(\bm{a})$.
When ISPs' 
uncertainties about the benefits of deployment vanishes, i.e.
$\beta_i{\rightarrow} 0$,
ISPs' dynamics lead to a unique equilibrium   
that maximizes the potential function, i.e. $\arg\max_{\bm{a}\in \CA^{|\tilde{\CC}|}}\Phi(\bm{a})$.

{\TONminor
Algorithm~\ref{alg:SA} provides a dynamic model for a static game.
We also introduce the following process for ISPs to reason about their
decisions, which do not require ISPs' dynamic plays.\\
}
\noindent 
{\bf \em (2) Iterative elimination of dominated strategies. }
Another perspective is that an ISP infers other ISPs' behaviors.
Before deployment,
ISP $i$ perceives a revenue change $(1{+}\theta_i) \Delta_f(\CS_{\bm{a}})$
for the flow ${f}$.
Then, we have
  $
  \varepsilon_i(\bm{a}) \triangleq \theta_i \phi_i(v, \CS_{\bm{a}}),
  $
where $\theta_i\in\mathbb{R}$ is a random variable that represents { ISP $i$'s
uncertainty towards the new architecture}, and
different $\theta_i's$ are independent.
A negative (or positive) perception error $\varepsilon_i$ means ISP $i$ is
pessimistic (or optimistic) about the new architecture.
{The perceived revenue change $(1+\theta_i)
  \Delta_f(\CS_{\bm{a}})$
is only known to ISP $i$,
while the distribution of $\theta_i$ is known to all the ISPs.
}
We denote the strategy of an ISP $i$ as a function from the
uncertainty parameter $\theta_i$ to its action: 
$s_i(\theta_i): \mathbb{R}{\mapsto} \{0,1\}$.
Since $\theta_i$ fully describes ISP $i$'s perceived utility, the
strategy functions $s_i(\cdot)$ specify each ISP's action facing
different perceived utilities.

To investigate in ISPs' strategies,
we use the concept called {\it iterative strict dominance} 
\cite{frankel2003complementarity}.
The basic idea is that ISPs will not choose those actions which are known
  to have worse profit {\ICNP in expectation.
For example, an ISP will not deploy IPv6 if it will lose money by
deploying it.  
Due to page limits, we omit the
  detailed procedure of the \textit{iterative strict dominance}~\cite{arXiv_version}.

\vspace{-0.04in}
\begin{lem}[\cite{frankel2003complementarity}]
Under Assumption \ref{asmp:supermodular},
as $\theta_i$'s distribution concentrates around zero for each $i\in\CS$,
ISPs have a unique strategy profile after the iterative elimination of
dominated strategies. Moreover, when the perception error
$\theta_i\rightarrow 0$, ISPs have the 
unique actions $(s^\ast_i (0))_{i\in\tilde{\CC}} {\in}
 \arg\max_{\bm{a}\in \CA}\Phi(\bm{a})$.\label{lemma:equiGlobalGame} 
\end{lem}
\vspace{-0.04in}

\noindent 
This lemma states that under the iterative elimination of dominated
strategies, the equilibrium that maximizes the potential
function will be reached as the perception error diminishes.
{\TONminor
\begin{definition}
The ``robust equilibrium'' in our potential game $G$ is the equilibrium that maximizes the potential function.  
\end{definition}
}
\vspace{-0.05in}
Some economic experiments were carried
out~\cite{heinemann2004experiments} that coincide with the prediction of ``robust equilibrium''.
\vspace{-0.08in}

\subsection{\bf A Case Study} 
\vspace{-0.02in}
\label{sec:case_study}
We study a special case, 
under which we derive more closed-form results  
revealing more insights.   
In this particular case, the routing path is not allowed to change:
\begin{align}
&
\CP_f = \{\vec{p}_f\}, 
&& 
\forall f{\in}\CF,    
\label{eq:CondRoutPath}
\end{align}
and no revenue loss is caused by the
competition between the old architecture and new architecture, i.e.,  
\begin{align}
& 
\Delta_{f}(\CS) 
=0, 
&& 
\forall f{\in}\CF^{old} ( \CS).     
\label{eq:RevLoss}
\end{align}
Note that Eq.~(\ref{eq:NetRevDisGeneral}) is NP-hard to solve, 
because the Shapley value is NP-hard to compute in
general~\cite{deng1994complexity}.
Using Eq.~(\ref{eq:CondRoutPath}) and (\ref{eq:RevLoss}), 
we derive a simpler expression of Eq.~(\ref{eq:NetRevDisGeneral}) 
with computational complexity significantly reduced.  

\begin{theorem}
  \label{thm:Shapley_value} 
Suppose Eq. (\ref{eq:CondRoutPath}), Eq. (\ref{eq:RevLoss}) 
and Assumption~\ref{asump:revenue_gain} hold.    
Then, Eq. (\ref{eq:NetRevDisGeneral}) can be simplified to 
\begin{align}
    \label{eq:Shapley}
&
\phi_i(\CS,v) {=} \sum\nolimits_{{f} \in \CF}
  \mathbbm{1}_{\{i \in \CC(\vec{p}_f)\}}
  \frac{\Delta_f(\CS)}
  {n_f(\CS)},~ \forall i {\in} \CS.
  \end{align}
\end{theorem}
where $\mathbbm{1}_{A}\in \{0,1\}$ is an indicator function for the event $A$.

\noindent
Due to page limit, proofs are in the supplement.  
Recall that $n_f(\CS)$ is the number of critical ISPs of flow $f$ that
deploy the new architecture.  
Eq.~(\ref{eq:Shapley}) in Theorem~\ref{thm:Shapley_value} states that
the revenue gain of a flow ${f}$
is evenly distributed to all the $n_f(\CS)$ critical ISPs that deploy the new
architecture, and the Shapley value of an ISP is the sum of such distributions from
different flows.

To illustrate, let us consider Fig.~\ref{fig:example_flow},
which depicts a network consisting of
4 ISPs and 6 flows where each flow passes through three ISPs.
Suppose the new architecture requires a full-path participation and all four ISPs $\CS{=}\{1,2,3,4\}$ upgrade to the new architecture.
Each flow has a revenue gain of \$3,
i.e. ${\Delta_f(\CS) {=} \$3},
\forall {f} {\in} \CF$.
According to (\ref{eq:Shapley}), each ISP shares \$3/3=\$1 in each flow.
In total, ISP 1, 3 and 4 gain \$4 because they participate in 4 flows.
One can see that ISP 2 shares a higher Shapley value of \$6, because all 6
flows must go through it. In other words, ISP 2 has a higher contribution to the
revenue gain of the new architecture.
Based on Theorem \ref{thm:Shapley_value} , 
we next derive a closed-form potential function for the 
game $G$.

\begin{theorem}
\label{thm:potential_game}
Given the same conditions in Theorem \ref{thm:Shapley_value}, the potential
function that satisfies Eq.~(\ref{eq:potential}) can be 
\begin{align}
  \label{eq:potential_func}
\Phi(\bm{a})
\triangleq
\sum\nolimits_{{f} \in \CF}
\sum\nolimits_{m=1}^{n_f(\CS_{\bm{a}})}
\frac{{ \tilde{\Delta}_f}(m)}{m}
-
\sum\nolimits_{i \in \tilde{\CC}} a_i c_i,
\end{align}
where $\tilde{\Delta}_f(m)$ is the value of $\Delta_f(\CS)$
when $n_f(\CS) {=} m$. We call $B(\CS_{\bm{a}})
{\triangleq}
\sum\nolimits_{{f} {\in} \CF}
\sum\nolimits_{m=1}^{n_f(\CS_{\bm{a}})}
{\frac{ \tilde{\Delta}_f(m)}{m}}
$ {the total immediate benefits}.
\end{theorem}

This potential function $\Phi(\bm{a})$
has insightful physical meanings.
The term $\tilde{\Delta}_f(m)/m$ is the revenue gain distributed
to the $m^{th}$ deployer in flow ${f}$ immediately after the $m^{th}$ deployer's deployment. The term $\sum_{m=1}^{n_f(\CS_{\bm{a}})}
{\tilde{\Delta}_f(m)}/{m}$ is all such immediate benefits that have
been distributed to the past deployers in flow ${f}$. Summing over all flows, $B(\CS_{\bm{a}})$ is the {\em total immediate benefits}
that ISPs $\CS_{\bm{a}}$ receive immediately
when they deploy. Also, $\sum_{i \in \tilde{\CC}} a_i c_i$ is the {\em total
launching costs} of ISPs $\CS_{\bm{a}}$. Therefore, the potential function
$\Phi(\bm{a})$ is the {\em total immediate benefits} minus the {\em total launching costs}.

\section{\bf Deployability \& Evolvability: Theoretical Analysis}
\label{sec:analysis}

Based on the equilibrium analysis in the last section, 
we first present general conditions on the deployability of a new
architecture.   
We analyze the following four factors 
to reveal their impacts on the deployability: (1) the number of
critical ISPs, (2) incremental deployment, (3) change of routing path, and (4)
revenue loss from old functionalities.

\vspace{-0.1in}
\subsection{\bf General Conditions on Deployability}
{\TONminor
\begin{definition}
An architecture is {\em ``deployable''} (or {\em``successfully deployed''}) if all
ISPs in $\tilde{\CC}$ deploy
in the robust equilibrium.  
\end{definition}
}
We next define a ``profitable'' architecture, whose benefit can
cover the total launching cost of all critical ISPs $\tilde{\CC}$.

\begin{definition}
  \vspace{-0.08in}
 An architecture is profitable if
 \begin{align}
   \label{eq:profitable}
   v(\tilde{\CC}) \geq \sum\nolimits_{i\in \tilde{\CC}} c_i .
 \end{align}
\end{definition}

\noindent
An architecture needs to be ``profitable'' to be successfully deployed.
However, as we will derive in
a more refined condition, some profitable architectures
may not be deployed.

{\COLOR
\begin{corollary}[Necessary Condition for Deployment]
 \label{corollary:necessary_condition}
An architecture is successfully deployed only if the potential function
satisfies $\Phi(\bm{1})\ge \Phi(\bm{0})$. Under conditions
(\ref{eq:CondRoutPath}) and (\ref{eq:RevLoss}), 
this necessary condition has a closed-form expression:
  \begin{align}
    \label{eq:necessary_condition}
B(\tilde{\CC})
\geq
\sum\nolimits_{i \in \tilde{\CC}} c_i
  \end{align}
Condition (\ref{eq:necessary_condition})
 implies (\ref{eq:profitable}), but (\ref{eq:profitable}) does not imply (\ref{eq:necessary_condition}).
\end{corollary}
}

\vspace{-0.05in}
\noindent{\bf Remark.}
{\MARK This corollary comes from the requirement that $\Phi(\bm{1})$
  should be the maximum value of the potential function, for
``all critical ISPs to deploy'' to be a robust equilibrium.}
It shows why a ``profitable'' architecture
may not be successfully deployed.
To illustrate, consider a network of three ISPs
connected in a line topology.
There is only one flow and all three ISPs are critical.
Consider an architecture with a total benefit that is twice the
total launching cost:
$v(\tilde{\CC}){=}2\sum_{i\in \tilde{\CC}}c_i$.
Then $B(\tilde{\CC}) {=} \frac{1}{3}v(\tilde{\CC}){<}\sum_{i\in \CC} c_i$,
which violates condition (\ref{eq:necessary_condition}) for successful
deployment.
Interestingly, even when the
total benefit is twice the total launching cost,
the new architecture still cannot be successfully deployed,
because the {\em total immediate benefits} is less than the total launching cost.

\begin{corollary}[Sufficient Condition for Deployment]
\label{corollary:sufficient}
If condition $\Phi(\bm{1})\ge \Phi(\bm{0})$ holds,
then in the robust equilibrium, a non-empty set of ISPs will deploy the new architecture.
\end{corollary}

This corollary states that condition $\Phi(\bm{1})\ge \Phi(\bm{0})$
is sufficient to guarantee that at least some
of the ISPs (if not all) will deploy the new architecture.
Together with Corollary~\ref{corollary:necessary_condition}, we could see that
the condition (\ref{eq:necessary_condition}) is both necessary and
sufficient to determine whether an architecture is deployable.

\subsection{\bf Impact of The Number of Critical ISPs}
\label{sec:ImpNumCritISP}
We start our analysis with the simple setting of no incremental
deployment mechanisms, i.e.,
\begin{align}
\forall 
\mathcal{S} {\subseteq} \mathcal{I}, 
f {\in} \mathcal{F}: 
\mathcal{C} ( \vec{P}_f (\mathcal{S}) ) 
{\setminus} 
\mathcal{S} 
{\neq} \emptyset 
\Rightarrow
f {\in} \mathcal{F}^{old} (\mathcal{S}).
\label{eq:CondNoIncreDeploy}
\end{align}
We also suppose conditions (\ref{eq:CondRoutPath}), (\ref{eq:RevLoss}) to hold 
(i.e., no change of routing path 
and no revenue loss by competitions between old
and new architectures).     
Now, condition (\ref{eq:necessary_condition}) is equivalent to
\begin{align}
  \label{eq:condition_2}
  {v(\tilde{\CC})}/{ \left( \sum\nolimits_{i\in \tilde{\CC}}c_i \right)}\ge 
\gamma, 
\end{align}
where $\gamma {\triangleq} v(\tilde{\CC}) / B(\tilde{\CC})$ 
denotes the ratio between the total benefits $v(\tilde{\CC})$ 
and the ``total immediate benefits'' $B(\tilde{\CC})$.  
Condition (\ref{eq:condition_2}) states that the ratio between the total benefit
and total launching cost ${v(\tilde{\CC})}/{ \sum_{i\in \tilde{\CC}}c_i }$
(``benefit-cost ratio'' in short) should
be higher than a {\it ``threshold''} $\gamma$, for a new architecture to be deployable.  
As there is no incremental deployment mechanism 
(i.e. $\Delta_f(\CS){=}0$ if $n_f(\CS){<}|\CC(\vec{p}_f)|$), we have
\begin{align}
  \label{eq:critical_nodes}
  \gamma =
  \frac{v(\tilde{\CC})}{B(\tilde{\CC})}
  =
  \frac{ \sum_{{f}\in \CF} {\Delta_f(\CC(\vec{p}_{f}))}}{\sum_{f\in \CF}
  \left( {\Delta_f(\CC(\vec{p}_{f}))} / {|\CC(\vec{p}_{f})|} \right) }.
\end{align}
Eq.~(\ref{eq:critical_nodes}) comes from the facts that
$
v(\tilde{\CC})=\sum_{{f}\in \CF} {\Delta_f(\CC(\vec{p}_{f}))}
$
and
$
B(\tilde{\CC}){=} \sum_{{f}\in \CF}
{\Delta_f(\CC({\vec{p}_f}))} / {|\CC(\vec{p}_{f})|}$.
Eq. (\ref{eq:critical_nodes}) states that $\gamma$ is the harmonic mean of the number
of critical ISPs $|\CC(\vec{p}_{f})|$ of the flows $f{\in}\CF$, weighted by each flow's maximum benefit $\Delta_f(\CC(\vec{p}_f))$.  
Here, $|\CC(\vec{p}_{f})|$ is the
{\it ``degree of coordination''} required by the new architecture for flow ${f}$.
Then, the physical meaning of $\gamma$ is the {\it``average degree of coordination''} 
over the whole network.
If an architecture requires a small number of critical ISPs to deploy for each flow, 
then $\gamma$ is small, 
and its deployability is high.  
Let us see some real-world cases.

\noindent{\bf Case 1: Deployment difficulty of DiffServ. }
To have QoS guarantees offered by DiffServ, all ISPs along the path are
critical.  
{\TON
  Consider {\TONminor the network topology of} a European education
  network G\'EANT~\cite{uhlig2006providing}.
  If {\TONminor the revenue change is proportional to the weight,
    i.e., $\Delta_f(\CC(\vec{p}_f))\propto w_f$ for each flow 
  $f{\in}\CF$}, then the ratio $\gamma{=}3.3$ 
(more details are in Section \ref{sec:experiments}).  
  Hence, Eq. (\ref{eq:condition_2}) states that only if the total benefits of
  DiffServ is higher than 3.3 times of its
  total launching cost, ISPs will deploy DiffServ in the G\'EANT network.  
  This high benefit-cost ratio make DiffServ difficult to be deployed.  
}
It provides an explanation why we see little adoption of DiffServ
even if QoS guarantee is urgently needed in the current Internet.

\noindent{\bf Case 2: The Internet flattening phenomenon. }
We are witnessing a flattening Internet~{\SHEEP\cite{Gill:2008:FIT,
    chiu2015we}}. This happens as large content 
providers such as Google and Facebook place their data centers near
end-users. Hence, the routing paths
become shorter and many intermediate ISPs are bypassed.
For many new architectures that require full-path
participation (e.g. IPv6, DiffServ), the flattening Internet reduces the number
of critical ISPs and makes these architectures more deployable.
In fact, with co-located data centers, many Internet
flows may traverse data centers within a single ISP (or the content provider).
For those intra-data-center flows, one provider owns the entire
topology, so there is only one critical node.
Then $\gamma{=}1$ regardless of the revenue change; thus architectures
like DiffServ can be deployed as soon as its total benefit exceeds its
total launching cost.
This explains why many proposed innovations for data centers are deployed.

\subsection{\bf Impact of Incremental Deployment Mechanism}  
\label{sec:increment_deployment}

Let us relax the settings of Sec. \ref{sec:ImpNumCritISP} to 
allow incremental deployment mechanisms.  
Then, we can quantify the impact of incremental deployment mechanisms on 
$\gamma$ as follows:
\begin{align}
  \label{eq:incremental_deployment}
  \gamma {=}
  \frac{v(\tilde{\CC})}{B(\tilde{\CC})}
  {=}
  \frac{ \sum_{{f}\in \CF} {\Delta_f(\CC(\vec{p}_{f}))}}{\sum_{{f}\in \CF}
  \left(
  \frac{\Delta_f(\CC(\vec{p}_{f}))}{|\CC(\vec{p}_{f})|}
  +\sum_{m=1}^{|\CC(\vec{p}_{f})|-1}\frac{\tilde{\Delta}_f(m)}{m}
  \right) }.
\end{align}
{\TONminor
\makebox[0ex][l]{\raisebox{6ex}{\hspace{1.45in}{$\underbrace{\hspace{0.15\textwidth}}_{\textbf{\small the term for incremental
    deployment}}$
}}}
}
The benefit from incremental deployment
is reflected by $\tilde{\Delta}_f(m)$ in (\ref{eq:incremental_deployment})
where $m{\le}|\CC(\vec{p}_{f})|{-}1$.
In contrast to (\ref{eq:critical_nodes}),
the incremental benefits brought by the mechanisms reduces the
ratio $\gamma$, {\ICNP as the denominator in (\ref{eq:incremental_deployment})
  becomes larger than that in (\ref{eq:critical_nodes})}. According to
(\ref{eq:condition_2}), we know incremental
deployment mechanisms improve the deployability of an architecture, as
illustrated in the following cases.

\noindent{\bf Case 3: Incremental deployment mechanisms of IPv6.}
Different incremental
deployment mechanisms~\cite{mukerjee2013tradeoffs} enable IPv6 in the current
Internet by selecting ingress/egress points to bypass the non-IPv6 areas.
Despite many of these mechanisms, almost
all the IPv6 traffic are using the native IPv6\cite{ipv6_stat}, which
means these 
mechanisms are mostly not used. Based on this fact, we speculate that ISPs
do not have significant revenue gain from these incremental mechanisms, i.e.,
$\tilde{\Delta}_f(m)$ is neglegible when $m{<}|\CC(\vec{p}_f)|$.
Otherwise many ISPs would use these mechanisms to improve their
revenue. Comparing (\ref{eq:critical_nodes}) 
and (\ref{eq:incremental_deployment}), {\TON we see that incremental
  deployment mechanisms for IPv6 do not significantly reduce the
  value of $\gamma$,} and thus do not increase
the deployability of IPv6. In a word, these
mechanisms for IPv6 failed because they did not provide significant
incremental benefits to ISPs.

\noindent{\bf Case 4: XIA.}
XIA~\cite{han2012xia} is a future Internet architecture proposed recently that
aims for an evolvable Internet.
XIA has an intent-fallback system.
If routers cannot operate on the primary ``intent'', ``fallbacks'' will allow communicating parties to specify alternative actions.
However, incremental benefits may be limited due to the characteristics of the
intended functionality. For example, it is almost
impossible to have QoS guaranteed without the participation of every
ISP along the path.
{\TONminor
  This means that $\tilde{\Delta}_f(m){=}0$ when $m{<}|\CC(\vec{p}_f)|$.
}
{\SHEEP Our model predicts that XIA has low deployability,} because it
is a network layer protocol and thus has a large number of critical ISPs.
{\TON
  For example, XIA can be deployed in G\'EANT network only if its total benefit
  is 3.3 times higher than
  its total launching cost, {\TONminor in the settings
    of Case 1}.}
The problem is that the benefit for ISPs from deploying XIA is not clear.

\vspace{-0.1in}
\subsection{\bf Impact of Change of Routing Path}
\label{sec:routing_path}
We relax the setting of Sec. \ref{sec:ImpNumCritISP} to allow a flow
 to change its routing path during the deployment.  
First, we decompose ISPs' net revenue gain 
$\tilde{v}(\CS)$ (defined in Lemma~\ref{lem:EquilibriumSecondStage})
into two components, $v_g$ and $v_l$, as follows:
\[
  \tilde{v}(\CS){=}
  \overbrace{\sum_{f\in \CF} \Delta_f(\CS) }^{v_{g}(\CS)}
  \overbrace{{-} \sum_{i\not\in \CS} \sum_{f\in\CF} \delta_{i,f}(\CS) {-}
  \sum_{i\in\CS}\sum_{f\in\CF}\delta_{i,f}(\CS\backslash \{i\}) }^{v_{l}(\CS)}.
\]
Then, we can characterize 
the impact of change of routing path on the net revenue distribution 
in the following lemma.  

\begin{lem}
The Shapley value $\varphi_i(\CS,\tilde{v})$ defined in Eq. (\ref{eq:sharplyvaluedef}) 
can be expressed as
$\varphi_i(\CS,\tilde{v}){=}\varphi_i(\CS,v_l){+}\varphi_i(\CS,v_g)$.
If (\ref{eq:RevLoss}) and (\ref{eq:CondNoIncreDeploy}) hold,   
then $\varphi_i(\CS,v_l) \geq 0$ and $\varphi_i(\CS,v_g)$ satisfies 
\begin{align*}
  \varphi_i(\CS,v_g){=}
 \sum_{f\in\CF}\sum_{\CP \subseteq \CP_f} 
\frac{(-1)^{|\CP|-1}\Delta_f(\tilde{\CC}) }{|\bigcup_{\vec{p}\in \CP}\CC(\vec{p})|} 
\mathbbm{1}_{ \left\{ \substack{i\in \bigcup_{\vec{p}\in \CP}\CC(\vec{p}),\\ \bigcup_{\vec{p}\in \CP}\CC(\vec{p})\subseteq \CS} \right\}}.
\end{align*} 
\label{lemma:shapley_routing_path}
\end{lem}
\vspace{-0.05in}

\begin{figure}
\centering
  \includegraphics[width=0.49\textwidth]{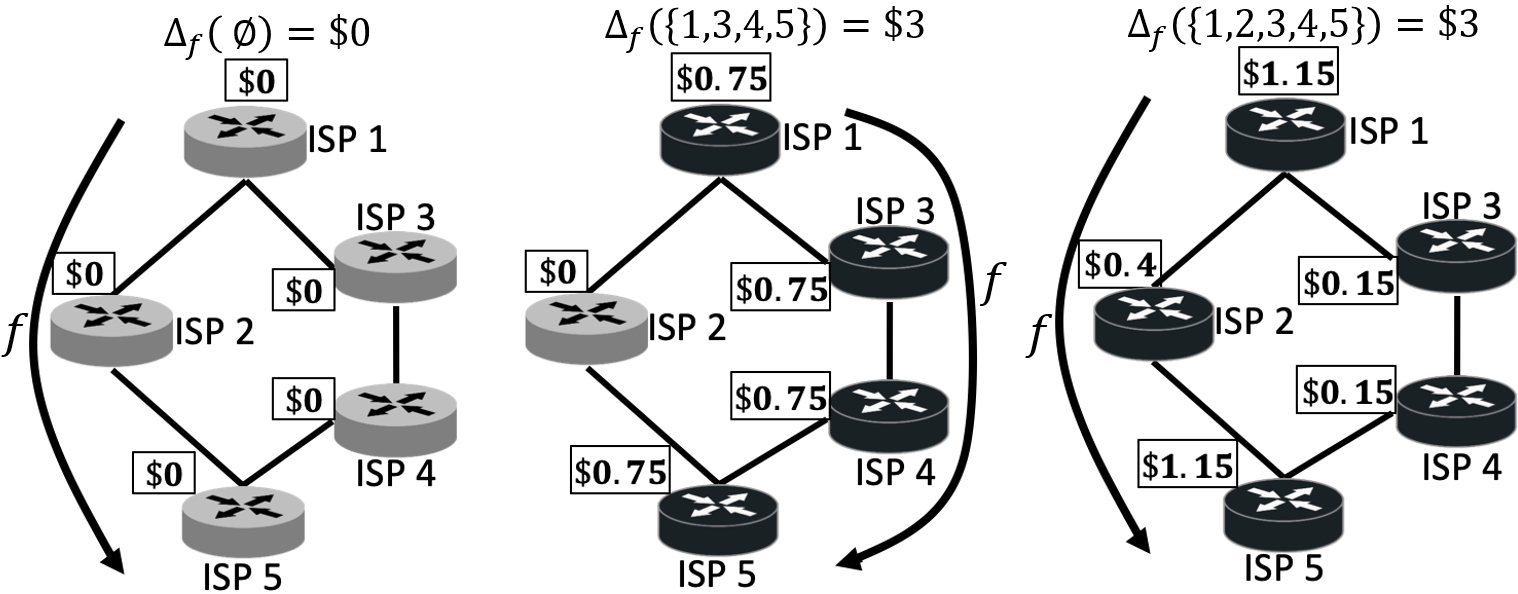}
\caption{The routing path and ISPs' $\varphi_i(\CS,v_g)$ - before deployment (left), partial
  deployment (middle) and after deployment (right)}
  \label{fig:before_after_deployment}
\end{figure}
 
\noindent
We use the Shapley value $\varphi(\CS,v_g)$ in Lemma~\ref{lemma:shapley_routing_path} 
to illustrate the impact of change of routing path.   
Consider Fig.~\ref{fig:before_after_deployment} 
which depicts one flow $f$ and its two alternative routing paths
$\CP_f{=}\{(1,2,5),(1,3,4,5)\}$.  
All the ISPs along a path are critical to the new architecture.  
Before the deployment (left figure of Fig.~\ref{fig:before_after_deployment}), $\Delta_f(\emptyset){=}0$, so $\varphi_i(\CS,{v}_g){=}\$0$ for $1{\le}i{\le}5$.  
When ISPs $\CS{=}\{1,3,4,5\}$
deploy the new architecture (middle figure of Fig.~\ref{fig:before_after_deployment}), the total revenue increment from the
new architecture is $v_g(\CS){=}\Delta_f(\CS){=}\$3$. 
Then, each of these four ISPs has a net revenue share 
$\varphi_i(\CS,{v}_g){=}\$0.75$ for $i{=}1,3,4,5$.  
When all ISPs $\CI{=}\{1,2,3,4,5\}$ deploy the new
architecture (right figure of Fig.~\ref{fig:before_after_deployment}), 
the routing path of $f$ is (1,2,5).  
Then, ISP 1 has a Shapley value 
$\varphi_1(\CI,{v}_g){=}3{\times}(1{/}3{+}1{/}4{-}1{/}5){=}\$1.15$ according to Lemma~\ref{lemma:shapley_routing_path}.  
This is because ISP 1 is a critical ISP for both routing paths $(1,2,5)$ and $(1,3,4,5)$.
Similarly, ISP 3 has a Shapley value $\varphi_3(\CI,v_g){=}3{\times}(1/4{-}1/5){=}\$0.15$.
We observe that ISP 1 has a higher Shapley value than ISP 3. This is because ISP
1 is critical in more routing paths, thus it has a larger bargaining power to
share the revenue gain.

Based on the Shapley value in Lemma~\ref{lemma:shapley_routing_path}, we have
the following Theorem for the impact of change of routing paths.
\vspace{-0.15in}
\begin{theorem} 
If all conditions in Lemma \ref{lemma:shapley_routing_path} hold, 
then $\Phi(\bm{1})\ge \Phi(\bm{0})$ under $|\CP_f| {=} 1$ 
implies $\Phi(\bm{1})\ge \Phi(\bm{0})$ under $|\CP_f|{\ge} 1$.  
\label{theorem:deployability_routing_path}
  \vspace{-0.03in}
\end{theorem}

\noindent 
Theorem \ref{theorem:deployability_routing_path} shows that 
the change of routing paths increases the
deployability of a new architecture.
This is because the change of routing paths will let 
the non-deploying ISPs be bypassed and lose revenue, thus, give them incentives to deploy.

\subsection{\bf Impact of Revenue Loss from Old Architecture}
\label{sec:old_new}

Let us relax the setting of Sec. \ref{sec:ImpNumCritISP} to study 
the impact of revenue loss caused by the competition between the old
and the new architectures.  
Recall that $w_f$ is the proportional traffic volume of flow $f$.
We model the revenue change of a flow $f{\in} \CF^{old}(\CS)$ that uses the old architecture as 
\begin{align}
& 
\textstyle
\Delta_f(\CS){=} - \sigma |\vec{p}_f| w_f 
  \sum_{h \in \CF^{new}(\CS)} w_{h} , 
&&
\forall f{\in} \CF^{old}(\CS). 
\label{eq:RevLossModel}
\end{align}
Eq. (\ref{eq:RevLossModel}) captures that 
the revenue loss of the old architecture is proportional to the total volume of
flows that can use the new architecture, i.e., $\sum_{h{\in}\CF^{new}} w_{h}$.  
Here, the parameter $\sigma{\ge} 0$ captures the scale of revenue losses.
An ISP $i{\not\in} \CS$ that does not deploy the new architecture has the following revenue
loss:
\begin{align}
 \delta_{i,f}(\CS)
  {=}
\begin{cases}
  -\sigma w_f \sum_{h{\in}\CF^{new}(\CS)}w_{h},  & i {\in} \vec{p}_f
  \text{ and }f{\in}\CF^{old}(\CS), \\
  0, & \text{otherwise}.
\end{cases}
  \label{eq:loss_function}
\end{align}
\vspace{-0.03in}

Let $\bar{N}_c {\triangleq} 
( {\sum_{f\in\CF}w_f|\CC(\vec{p}_f)|^2} )
/
( {\sum_{h \in\CF}w_h |\CC(\vec{p}_h)|})$
denote the weighted average number of critical ISPs over all flows in the network, 
where the weight for a flow $f$ is $w_f|\CC(\vec{p}_f)|$.

\vspace{-0.03in}
\begin{theorem}
  \label{theorem:deployability_old_new}
 Suppose we have no change of routing path, i.e. (\ref{eq:CondRoutPath}), and no
 incremental deployment mechanisms, i.e. (\ref{eq:CondNoIncreDeploy}).
Suppose the revenue loss of flows and ISPs satisfies (\ref{eq:RevLossModel}) 
and (\ref{eq:loss_function}). 
If $\sum_{f{\in}\CF}w_f\mathbbm{1}_{\{i{\in}\vec{p}_f\}} \le 
\sum_{f{\in}\CF} {w_f|\CC(\vec{p}_f)|} / (|\CC(\vec{p}_f)| + \bar{N}_c + 1)$ holds  
for all $i \in \tilde{\CC}$,
then $\Phi(\bm{1})\ge \Phi(\bm{0})$ under $\sigma = 0$ 
implies $\Phi(\bm{1})\ge \Phi(\bm{0})$ under $\sigma > 0$.
\end{theorem}

\noindent
Theorem \ref{theorem:deployability_old_new} states 
that when no ISP participates in more than a
threshold volume of flows, 
ISPs' revenue loss from old architecture will make the new
architecture easier to deploy, rather than the other way around.
{\TONminor Recall that $\bar{N}_c$ is the weighted average number of critical nodes in the flows, and $w_f$ is the
  weight of flow $f$. When all the flows have the same number of
  critical nodes $N_c$, the threshold fraction $\sum_{w\in\CF} w_f{|\CC(\vec{p}_f)|} / (
|\CC(\vec{p}_f)|{+}\bar{N}_c{+}1 )=N_c/(2N_c{+}1)< 0.5$.
}

Let us apply Theorem \ref{theorem:deployability_old_new} to examine a real-world case.  
For G\'EANT network, suppose the new architecture requires a ``full-path
participation'', then the threshold
$\sum_{f{\in}\CF} {w_f|\CC(\vec{p}_f)|} / (|\CC(\vec{p}_f)|{+}\bar{N}_c{+}1)$ is
0.407. Moreover, The condition
that {\em``no ISP participates to more than 40.7\% of all flow volumes''}
holds for every ISP in G\'EANT network.
  From Theorem~\ref{theorem:deployability_old_new}, we know that for the G\'EANT network, a
  new architecture is more deployable if we
  consider ISPs' revenue loss from the old architecture.

{\section{\bf Extensions}
\label{sec:extension}
Our results thus far consider binary actions, i.e., 
each ISP deploys the new architecture either in 
all of its networks, or in none of them. Moreover, we focus on one new architecture.
Now, we extend our model to allow more than two actions 
and multiple competing new architectures respectively.

\subsection{\bf Partial Deployment in Sub-networks}
\label{sec:ISP_division}

We first extend our previous binary action model, where $a_i{\in} \{0,1\}$, 
to allow more actions.  
We model that ISP $i$ has a finite set of devices denoted by $\CD_i$, 
where an device could be a switch, router, etc.  
Each ISP can deploy the new architecture in a subset $\mathcal{A}_i \subseteq \CD_i$ of its devices, 
and we use $\CA_i$ to represent ISP $i$'s action.   
Namely, ISP $i$ has $2^{|\CD_i|}$ possible actions.   
Fig.~\ref{fig:example_ton} illustrates this new action model.
Note that no device belongs to multiple ISPs, so
$\CD_i{\cap} \CD_j{=}\emptyset$ for any $i{\ne} j$.   
Let $\tilde{c}_d$ denote the cost to deploy 
the new architecture in device $d{\in} \bigcup_{i\in \CI} \CD_i$.   

To simplify presentation, we assume that changes of routing 
path are not allowed and there is no competition among the old
architecture and new architecture, i.e, 
conditions (\ref{eq:CondRoutPath}) and (\ref{eq:RevLoss}) hold.  
Note that dropping this assumption only involves a more 
complicated notation system.  
Let $\CD_{i,f} \subseteq \CD_i$ denote ISP $i$'s devices that 
support the routing path of flow $f$, i.e., $\vec{p}_f$.  
Without loss of generality, we assume that there are no dummy devices, i.e.,
$\bigcup_{f \in \mathcal{F}} \CD_{i,f} = \CD_i$.  
We say that an ISP $i$ deploys the new architecture for flow $f$ if 
it deploys this architecture in all of $\CD_{i,f}$, 
i.e., $\CD_{i,f} \subseteq \mathcal{A}_i$. 
Then, we define the set of critical ISPs that deploy the new
architecture for flow $f$ as 
$
\mathcal{S}_f (\bm{\mathcal{A}})
\triangleq 
\{i|\CD_{i,f} \subseteq \mathcal{A}_i, i \in \tilde{\mathcal{C}}\}, 
$
where
$
\bm{\mathcal{A}} 
\triangleq 
(\mathcal{A}_i:i\in \tilde{\mathcal{C}})
$ 
denotes the action profile for all critical ISPs.  
To be consistent with Assumption~\ref{asump:revenue_gain}, 
the revenue from a flow is
determined by the number of critical ISPs that deploy the new architecture for
this flow, i.e. $n_f(\mathcal{S}_f (\bm{\mathcal{A}}))$.   
We still use $\tilde{\Delta}_f(m)$ to denote the
revenue increment of flow $f$ when   
$n_f(\mathcal{S}_f (\bm{\mathcal{A}}))=m$.
Given  action profile $\bm{\mathcal{A}}$, 
the revenue gain distributed to ISP $i \in \tilde{\mathcal{C}}$ is
\[
\phi_i(\bm{\mathcal{A}},v) {=} \sum_{{f} \in \CF}
  \mathbbm{1}_{\{i \in \CC(\vec{p}_{f}), \CD_{i,f} \subseteq \mathcal{A}_i \}}
  \frac{\Delta_f(\mathcal{S}_f (\bm{\mathcal{A}}))}
  {n_f(\mathcal{S}_f (\bm{\mathcal{A}}))}.
\]
Then, under action profile $\bm{\mathcal{A}}$, the
utility of ISP $i$ is 
\[
u_i(\bm{\mathcal{A}})
\triangleq
\begin{cases}
0, & \text{if }\mathcal{A}_i=\emptyset, \\
\phi_i(\bm{\mathcal{A}},v) - \sum_{d\in \mathcal{A}_i} \tilde{c}_d,
&
\text{otherwise}.
\end{cases}
\]

\begin{theorem}
  Suppose (\ref{eq:CondRoutPath}) and (\ref{eq:RevLoss}) hold 
and ISPs' total launching cost is a constant, i.e. $\sum_{i\in\tilde{\CC}}
  c_i{=}
  \sum\nolimits_{i \in \tilde{\mathcal{C}}} 
\sum\nolimits_{d\in \mathcal{D}_i} \tilde{c}_d$.
  Then an architecture is {\em
    ``deployable''} under the partial deployment action model, 
    if and only if this architecture is {\em ``deployable''} 
under the binary action model.
\label{thm:partial_deploy}
\end{theorem}

\noindent
Theorem \ref{thm:partial_deploy} states that 
 whether ISPs can partially deploy a new architecture in
 sub-networks does not affect the architecture's deployability.
 The reason is that the {\em ``degree of
   coordination''} depends on the number of decision makers (i.e. the
 number of ISPs) in a
 flow, rather than the number of devices.

} 

\subsection{\bf Competing Architectures}\label{sec:competition}
We extend our model to study multiple competing 
architectures with similar functionalities.
We will show that a more {\it ``deployable''} architecture
will have a competitive advantage.

\noindent{\bf $\bullet$ Deployment price of an architecture. }
ISPs charge customers for using the new functionality (e.g.
CDN~\cite{azure_cdn}, DDos protection~\cite{amazon_shield}).
We consider a usage-based charging scheme.
When all critical ISPs deploy the architecture,
the unit price for the new functionality is $p {\in} \mathbb{R}_+$.
Then, the revenue gain of a flow $f$ is
$
\Delta_f(\tilde{\CC}){=}p{\times} w_f,
$
where we recall that $w_f$ is the
proportional traffic volume of flow $f$.
In comformance with our previous model,
the unit price for flow $f$ when a subset $\CS{\subset} \tilde{\CC}$ of
ISPs deploy is $p \Delta_f(\CS) / \Delta_f(\tilde{\CC})$.
We define the deployment price of an architecture (denoted by $p_d$) as the minimum unit price such that the
condition (\ref{eq:necessary_condition}) is satisfied.
Namely, ISPs will deploy the architecture when the unit price is above $p_d$.

We illustrate the ``deployment price'' by considering the case
in Sec.~\ref{sec:case_study} where $\CP_f=\{\vec{p}_f\}$ for each $f\in\CF$
and $\Delta_f(\CS)=0$ for each $f\in \CF^{old}(\CS)$.
Then, according to Corollary~\ref{corollary:necessary_condition},
$
p_d
{\triangleq}
\gamma \sum_{i\in \tilde{\CC}}c_i  / \sum_{f\in \CF}w_f.
$
Notice that the deployment price of an architecture depends on
``degree of coordination'' $\gamma$ (which depends on the
number of critical ISPs) and the total launching cost $\sum_{i\in
  \tilde{\CC}}c_i$.

\noindent
{\bf $\bullet$ Multiple architectures under competition. }
Fig.~\ref{fig:competition} is an extension of the example in
Fig.~\ref{fig:example}, where two architectures provide the same new
functionality. Remember that architecture B requires all the ISPs 1,2,3 to deploy, and the launching
cost for each ISP is \$3. Architecture A, on the other hand, requires
only ISP 4, the ISP closest to the end user, to deploy and has a
launching cost \$9 for that ISP. Note that the total launching costs
for the two architectures are both \$9.
For simplicity, we again assume no incremental deployment mechanism
and we have one unit usage volume.
{\it What will be a reasonable price for the new functionality?} As we can see,
only if the price of the
new functionality is higher than the ``deployment price'' \$27 will
any ISP deploy B.
Meanwhile, architecture A will be deployed as long as a
price higher than \$9 can be charged. Clearly
a customer will not \$27 if the same functionality is available for
\$9. Then, ISPs will choose
architecture A, and architecture B will not be deployed
because of the lower price reached by the more evolvable and
competitive architecture A.

\begin{figure}
  \centering
  \begin{minipage}{0.16\textwidth}
  \includegraphics[width=\textwidth]{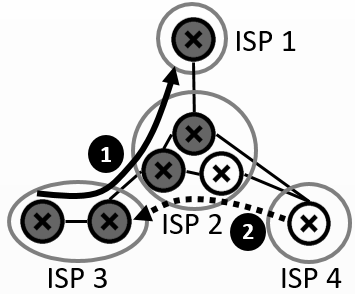}
  \caption{Partial Deployment in routers} \label{fig:example_ton}
  \end{minipage}
\begin{minipage}{0.32\textwidth}
    \hspace{-0.06in}\includegraphics[width=1.05\textwidth,height=0.4\textwidth]{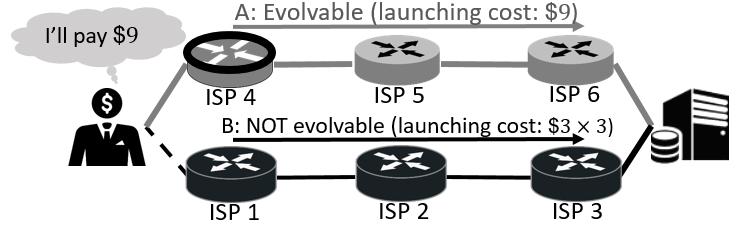}
  \caption{Competing architectures}
  \label{fig:competition}
  \end{minipage}
\end{figure}

Suppose we have $K$ new architectures providing the same
functionality, with deployment prices $p_d^{(1)}{,} {\cdots},p_d^{(K)}$ respectively.
We consider a market where ISPs are highly competitive so the customers
can dictate the price.
Here, we claim without rigorous proof that the
customers will set the price to the lowest deployment price of these
architectures, i.e. $\min_{k}\{p_d^{(k)}\}$.
This is reasonable because the customers will not pay a higher price for a
functionality if they could enjoy the same functionality with a lower price. {\MARK Consequently, other architectures with higher deployment prices will not be
successfully deployed.} 
When competitive architectures have similar deployment price and 
total launching costs, the
architecture with the lowest ``degree of coordination'' $\gamma$ will win.
Let us apply these observations to study the following cases.

\noindent{\bf Case 5: IPv6 vs. NAT.}
IPv6 and NAT (Network Address Translation) both provide similar
functionality of ``addressing hosts''.
In fact, NAT is now deployed in a
great many ISPs, while IPv6 is still not deployed in many countries.
In short, NAT wins and this observation could be explained by our model.
IPv6 is a network-layer protocol which requires full-path
participation. Although
there are incremental deployment mechanisms, as we discussed before, they are
rarely enabled by many ISPs. NAT is an application-layer solution that
can be deployed transparently for most end users who need more addresses.
The average AS path length is around 4~\cite{path_lengths}.
Therefore, IPv6 requires a much higher ``degree of coordination'' (around
four times) compared to NAT.

\noindent{\bf Case 6: NDN vs. CDN.} Content Delivery Network (CDN)
caches data spatially close to
end-users to provide high availability and better performance.
Meanwhile, Named Data
Networking (NDN)~\cite{ndn} is a future Internet architecture that names data instead of their locations. Their
main functionalities are both to provide scalable content delivery.
CDNs operate at the application-layer, so only the
CDN owners need to deploy the CDN infrastructures.
NDN is designed to operate as the network layer~\cite{ndn}; thus
its deployment requires full-path
participation, implying a degree of coordination around four (based on
average AS path length). Suppose NDN has
a comparable total launching cost as CDN.
Then according to our model, NDN would have around four times
higher deployment price than CDN.
In order for NDN to overcome competition from CDN, it 
it would need to provide additional benefits and/or have incremental
deployment mechanisms.
{\ICNP Guided by our analysis, one possible incremental deployment
  mechanism would be to run NDN on top of IP as an overlay~\cite{jiang2014ncdn} to provide incremental benefits. This is possible since NDN is designed as a
  ``universal overlay'', and could make NDN evolvable and competitive with
CDN.  Indeed, Cisco's hybrid-ICN~\cite{hICN} pursues a related
incremental deployment approach.
}

{\SHEEP
  \noindent{\bf Case 7: Multipath TCP vs. Multipath QUIC.}
  Multipath-TCP (MPTCP)\cite{mptcp}
is an extension of TCP, which enables inverse multiplexing of resources, and thus increases TCP throughput.
MPTCP requires middleboxes (e.g., firewalls) in the
  Internet to upgrade so they do not drop its
  packets~\cite{mptcp_middlebox} because they do not recognize them. 
  Therefore, the critical nodes for MPTCP include the
  senders, receivers, and the ISPs with middleboxes. In contrast,
  Multipath-QUIC   (MPQUIC)\cite{de2017multipath} is an extension of
  QUIC\cite{langley2017quic} that achieves the multipath functionality of
  MPTCP. Because QUIC encrypts its packets and 
  headers, MPQUIC avoids interference from middleboxes. Then the
  critical nodes of MPQUIC only include the senders and receivers.
We argue that the total launching cost of MPQUIC is not more than that
of MPTCP. This is because MPTCP and MPQUIC both
require the senders and receivers to
  upgrade their software, but MPTCP additionally requires the
  middleboxes to be upgraded. Also, MPQUIC has lower degree of coordination.
Comparing the total launching cost and the degree of coordination, our
models predict that MPQUIC will be deployed more rapidly than
(or instead of) MPTCP.
}

\section{\bf Mechanism Design to Enhance Deployability}
\label{sec:mechanism}

Based on the observations thus far, 
we first design a coordination mechanism 
to enhance the deployability of a new architecture.  
Then, we improve the practicability of our mechanism 
via the idea of tipping set \cite{heal2006supermodularity}.  
Here, we consider one new architecture, and do not allow partial deployment.

\subsection{\bf Coordination Mechanism Design}

Recall that the difficulty of deployment comes from 
the requirement of coordination among decentralized ISPs.  
To mitigate this difficulty, we design a coordination mechanism.
Before presenting our mechanism, we review some real-world examples to
illustrate the power of coordination.  
From the historical data~\cite{ipv6_stat, ipv6_stat_ISOC} for the
transition from IPv4 to IPv6, we observe that coordinating actions of
are highly correlated with IPv6's deployment. Before the
first World IPv6 Launch Day organized by 
  Internet Society in 2012~\cite{world_ipv6_launch}, less than 1\% of
  users accessed their services over
  IPv6~\cite{ipv6_stat}. In 2018, this number goes to nearly
  25\%~\cite{ISOC_dominant}. 
As another example, the Indian government produced
a roadmap of IPv6's deployment in July 2010~\cite{india_gov} when the
adoption rate was less than 0.5\%. Now, over 30\% of
the traffic in India uses IPv6~\cite{ipv6_stat}. In contrast, the government of
China did not announce a plan to put IPv6 into large-scale use until Nov.
2017~\cite{cn_government}. Now, less than 3\% of traffics in China use
IPv6~\cite{ipv6_stat}.  

Formally, our coordination mechanism contains two steps:  

\begin{enumerate}
\item 
{\bf Quoting:} 
Each ISP $i\in \tilde{\CC}$ submits a quote $q_i {\in} \mathbb{R}_+$
to the coordinator. An ISP's quote is a contract under which
the ISP would deploy the architecture once someone pays more than the
quote. Quoting itself does not cost anything.

\item 
{\bf ISP selection:}  
In this step, the coordinator selects a set of ISPs 
to deploy the architecture, and announces a reward for each of them.  
For each selected ISP, the announced reward is at least as high as
that ISP's quote. Then the selected ISPs deploy the new architecture,
and the coordinator gives ISPs the announced reward.

\end{enumerate}
The coordinator might be an international organization, group of
governments, etc.
{\ICNP In the second step of the mechanism, the coordinator selects the ISPs
via the following optimization: }
\begin{align}
  \label{eq:final_objective}
&  
  \text{maximize}_{\CS \subseteq \tilde{\CC}}
&& |\CS|,\nonumber\\
&  \text{subject to} && \varphi_i(\CS, \tilde{v})\ge q_i, \forall i \in \CS,
\end{align}
where  
$
\tilde{v}(\CS)
$
is defined in Eq.~(\ref{eq:v_prime}).  
Let $\CS^*(\bm{q})$ denote one optimal solution of the above optimization problem, 
where $\bm{q} \triangleq (q_i : i \in \tilde{\mathcal{C}})$.  
Finally, each selected ISP 
$i{\in} \CS^*(\bm{q})$ gets a reward $\varphi_i(\CS^*(\bm{q}), \tilde{v})$, 
which is at least as high as its quote.

ISPs may intentionally quote
lower to increase the chance to be selected, 
or quote higher to ask for more reward.
Our mechanism enforces ISPs to quote exactly their 
launching cost.   

\begin{theorem}  
Suppose there is only one new architecture, 
ISPs have binary actions $\{0,1\}$ 
and Assumption~\ref{asmp:supermodular} holds.  
Quoting $q_i{=}c_i$ is a weakly dominant
strategy for each ISP $i{\in} \tilde{\CC}$. 
\label{thm:truth_telling}
\end{theorem}

We further quantify the efficiency of our mechanism.  

\begin{theorem}  
 \label{thm:max_objective} 
Under the same conditions as Theorem \ref{thm:truth_telling},
if $\CS\subset \CT \implies v(\CS) {<} v(\CT)$
for any $\CS, \CT$, the unique selection $\CS^*(\bm{c})$ yields a
maximal total revenue gain $v(\CS^*(\bm{c}))$ for all ISPs.
\end{theorem} 

\noindent{\TONminor
  Note that Theorem~\ref{thm:truth_telling} and
  \ref{thm:max_objective} still hold when we consider the change of
  routing path or the revenue loss from the old architecture, etc. The
  proof only requires that ISPs distribute the revenue gain via the
  ``stable'' distribution mechanism.


\begin{figure*}
    \centering
  \begin{minipage}{0.245\textwidth}
  \centering
  \includegraphics[width=\textwidth]{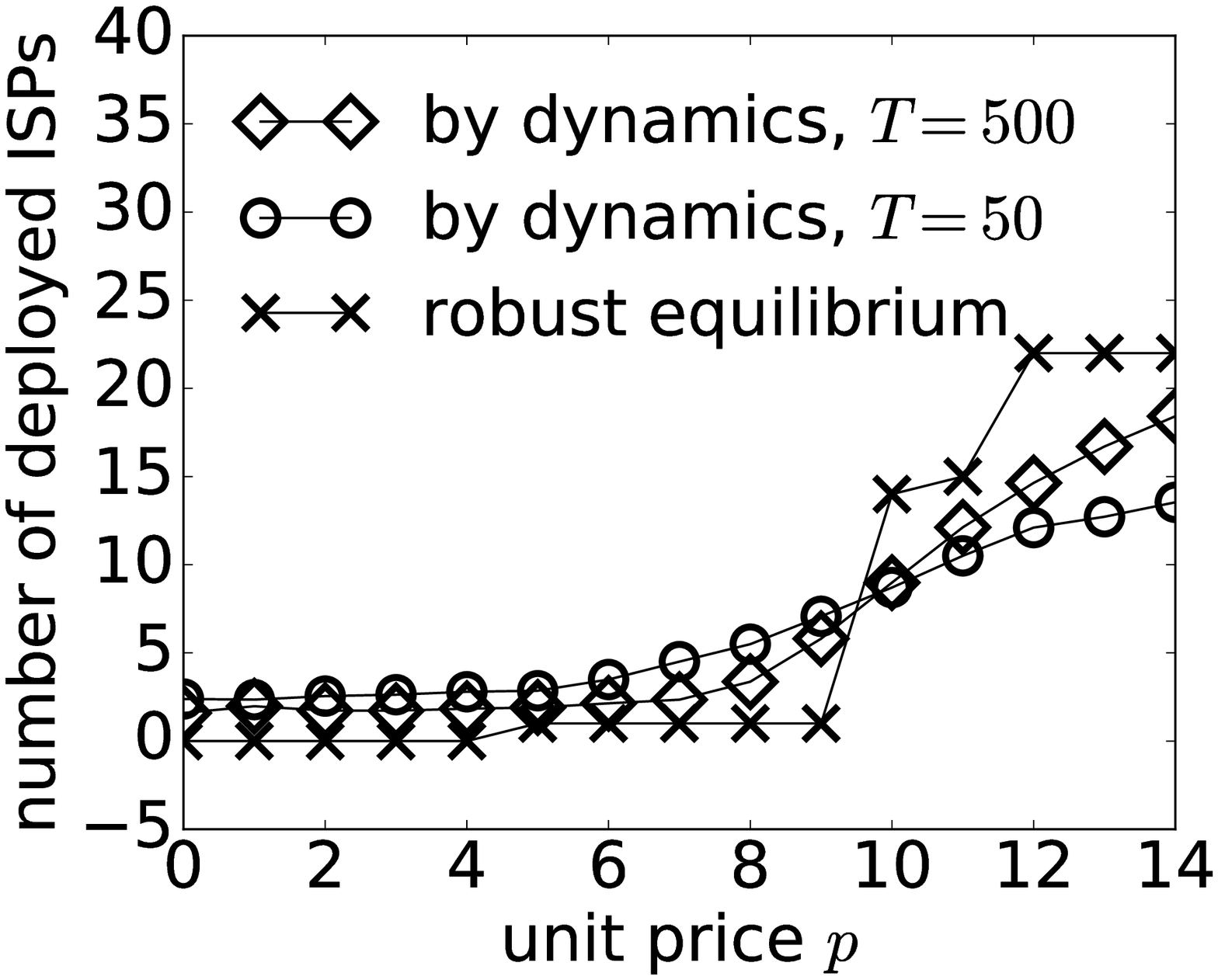}
  \caption{Logit-response dynamics (G\'EANT)}
  \label{fig:logit_response}
  \end{minipage}
\begin{minipage}{0.245\linewidth}
  \centering
  \includegraphics[width=\textwidth]{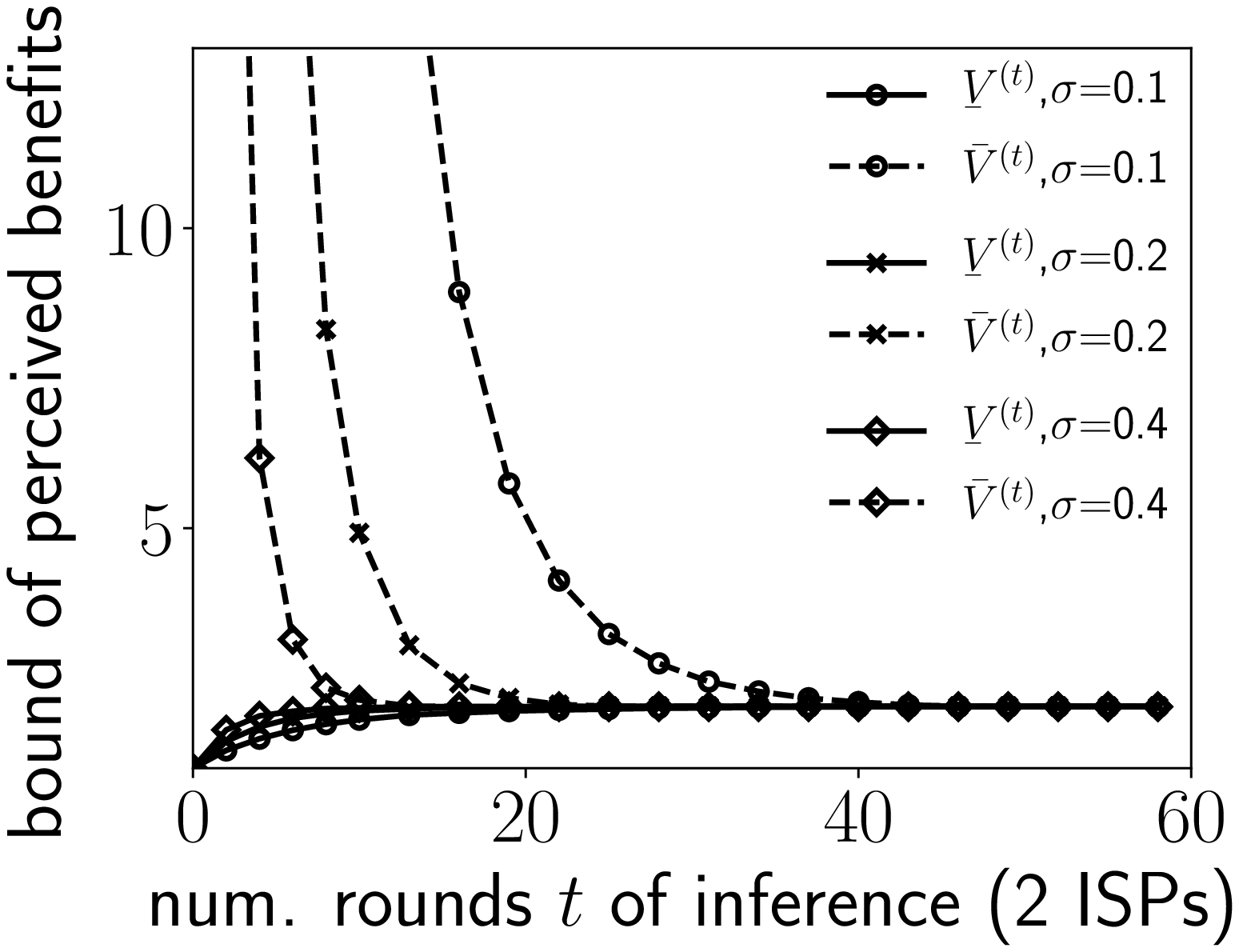}
  \caption{ Two ISPs' dynamics to infer their strategies}
  \label{fig:2ISP_induction}
  \end{minipage}
\begin{minipage}{0.245\linewidth}
  \centering
  \includegraphics[width=\textwidth]{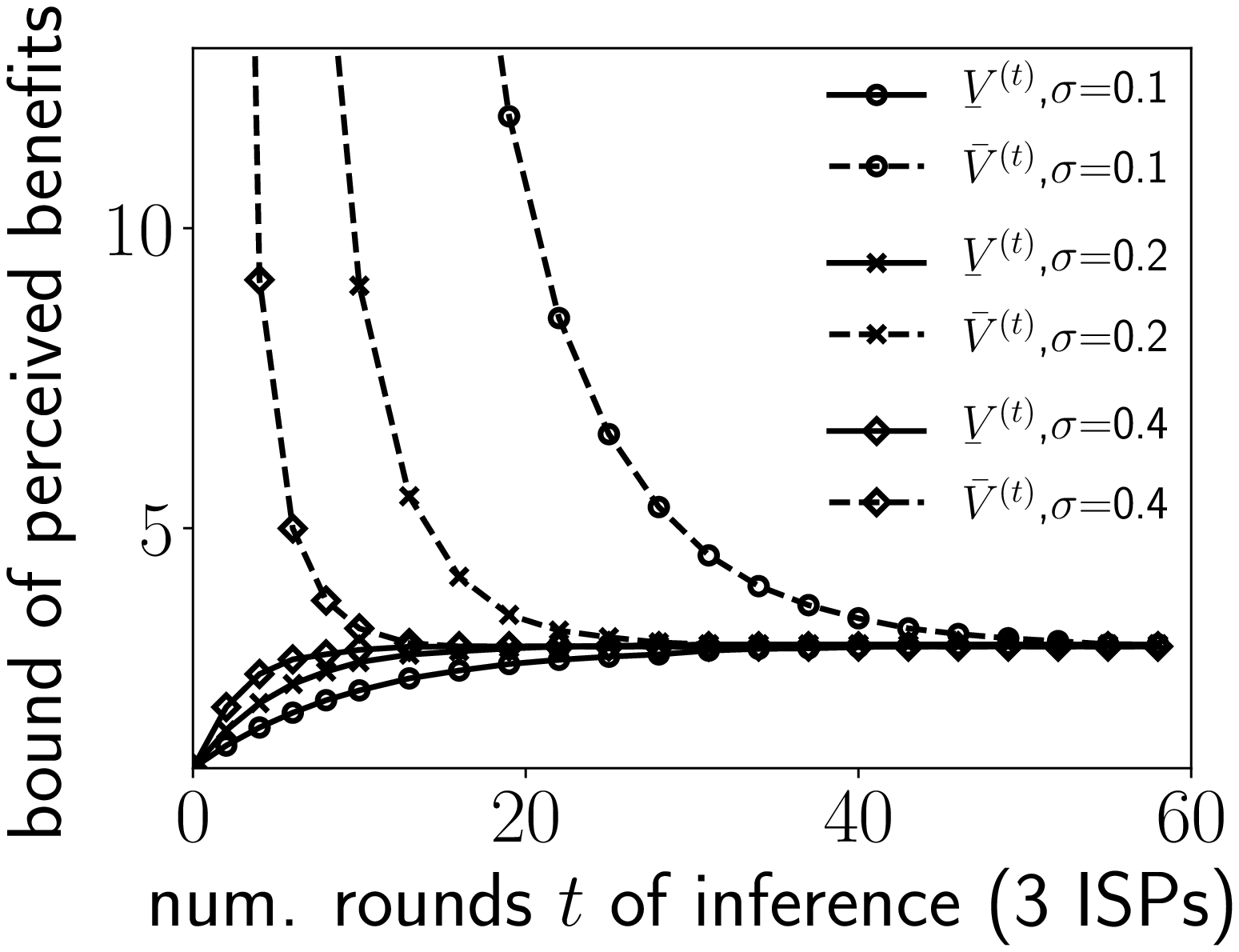}
  \caption{Three ISPs' dynamics to infer their strategies}
  \label{fig:3ISP_induction}
  \end{minipage}
\begin{minipage}{0.245\linewidth}
  \centering
  \includegraphics[width=\textwidth]{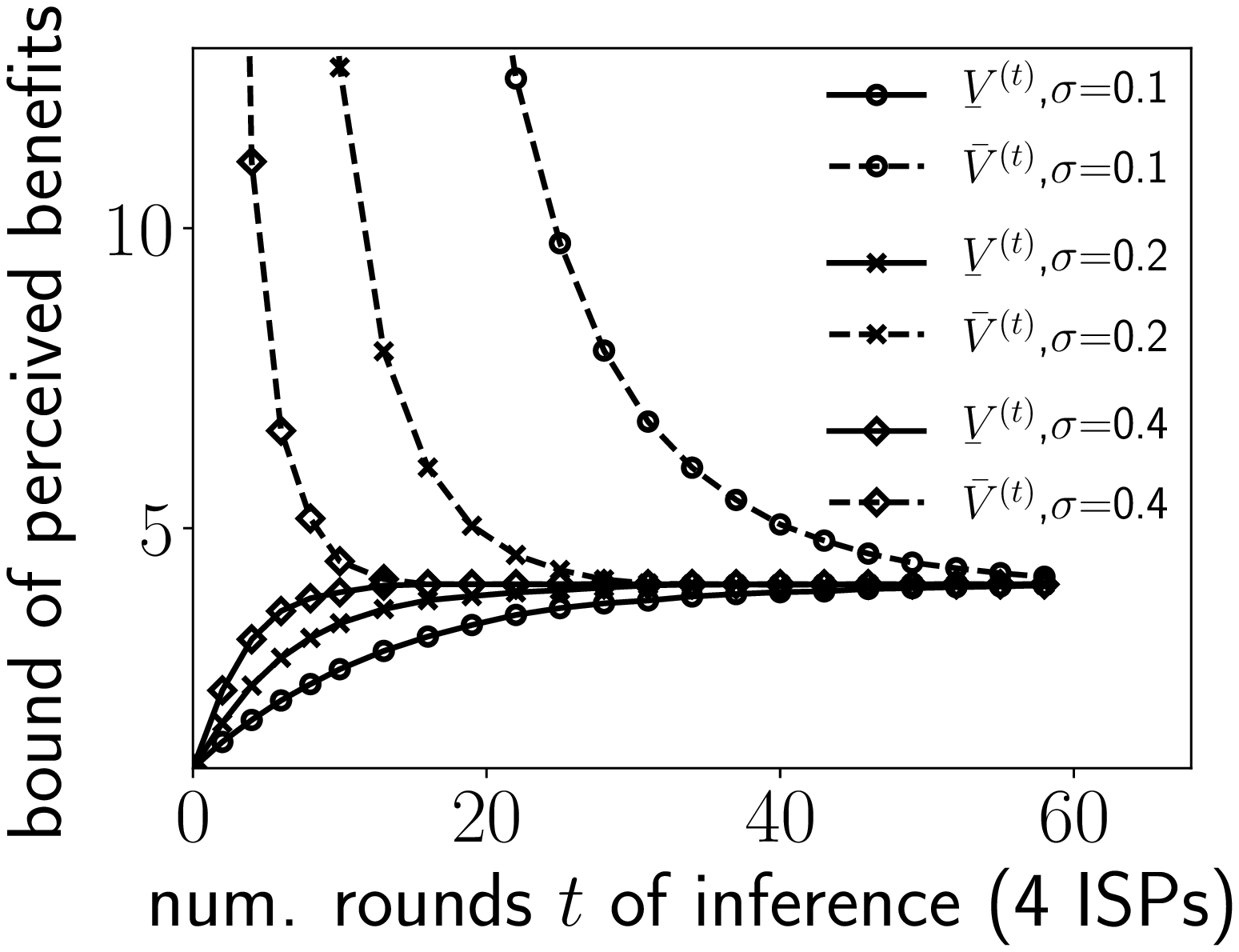}
  \caption{ Four ISPs' dynamics to infer their strategies}
  \label{fig:4ISP_induction}
  \end{minipage}
  \end{figure*}

\begin{figure*}[htb]
  \begin{minipage}{0.235\textwidth}
  \centering
  \includegraphics[width=\textwidth]{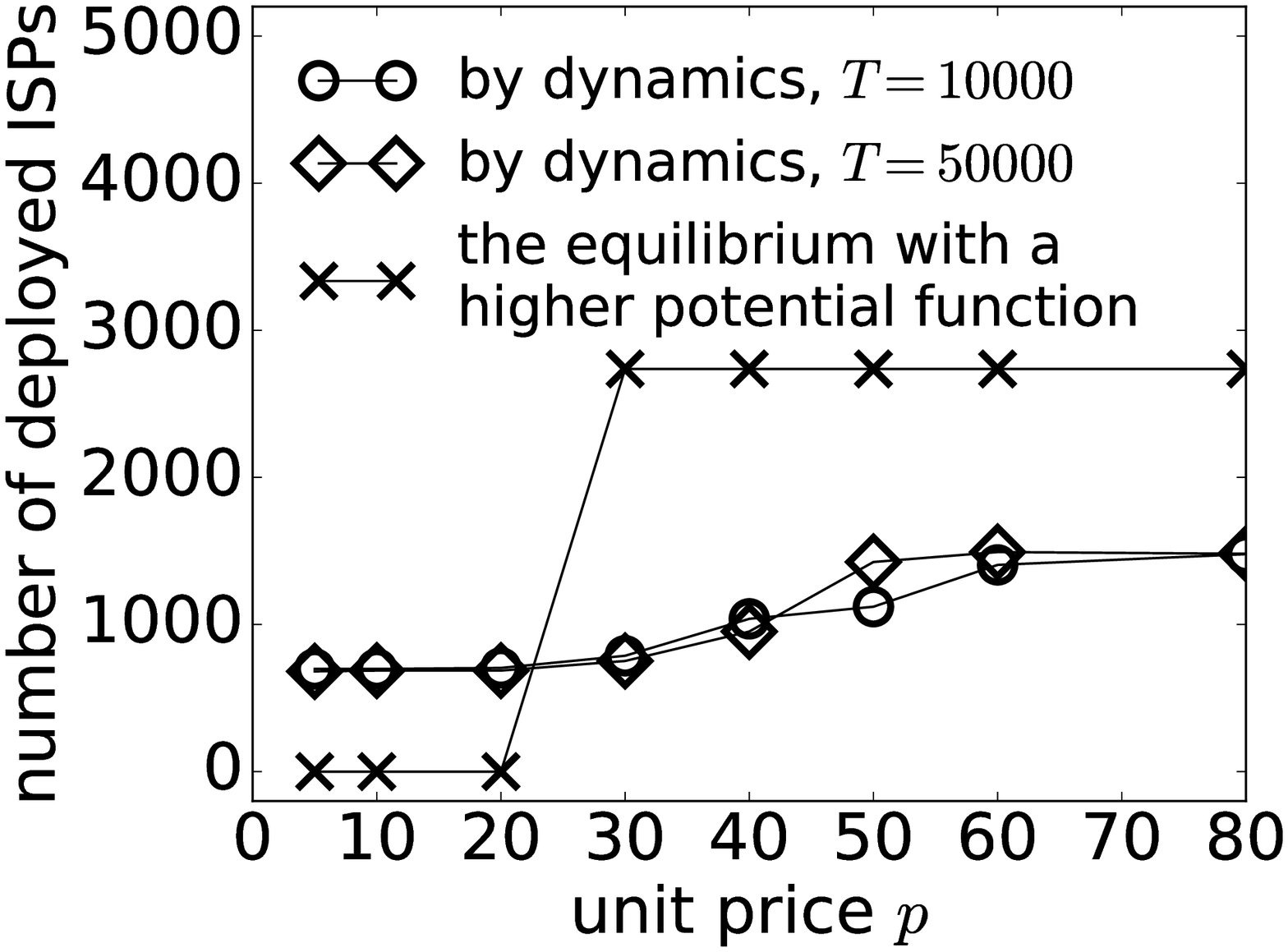}
  \caption{Logit-response dynamics (IPv4-net)}
  \label{fig:logit_response_large}
  \end{minipage}
  \begin{minipage}{0.245\textwidth}
  \centering
  \includegraphics[width=\textwidth]{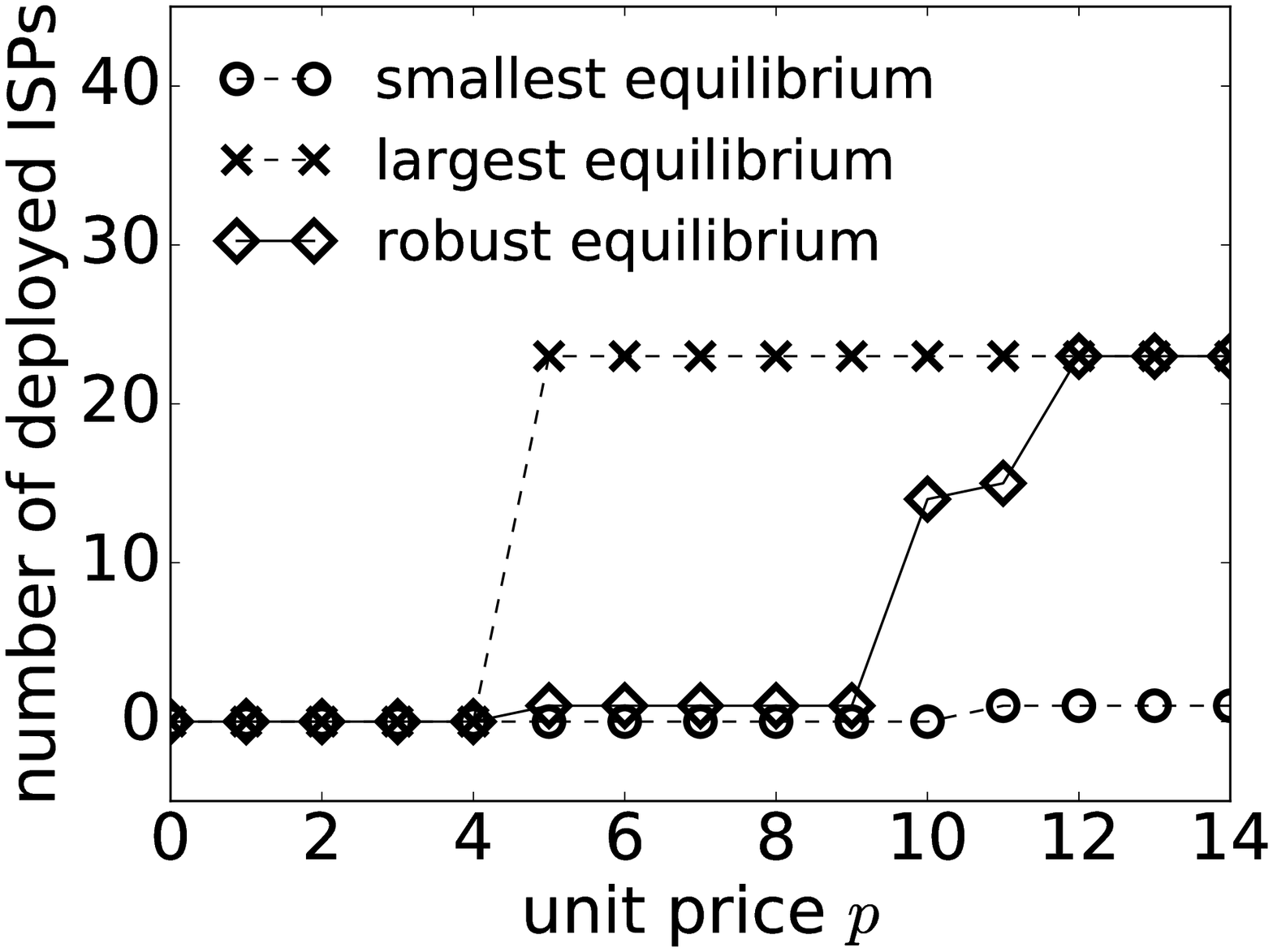}
  \caption{Unit price $p$ and the scale of equilibria (G\'EANT)}
  \label{fig:benefit_scale}
  \end{minipage}
\begin{minipage}{0.245\textwidth}
  \centering
  \includegraphics[width=\textwidth]{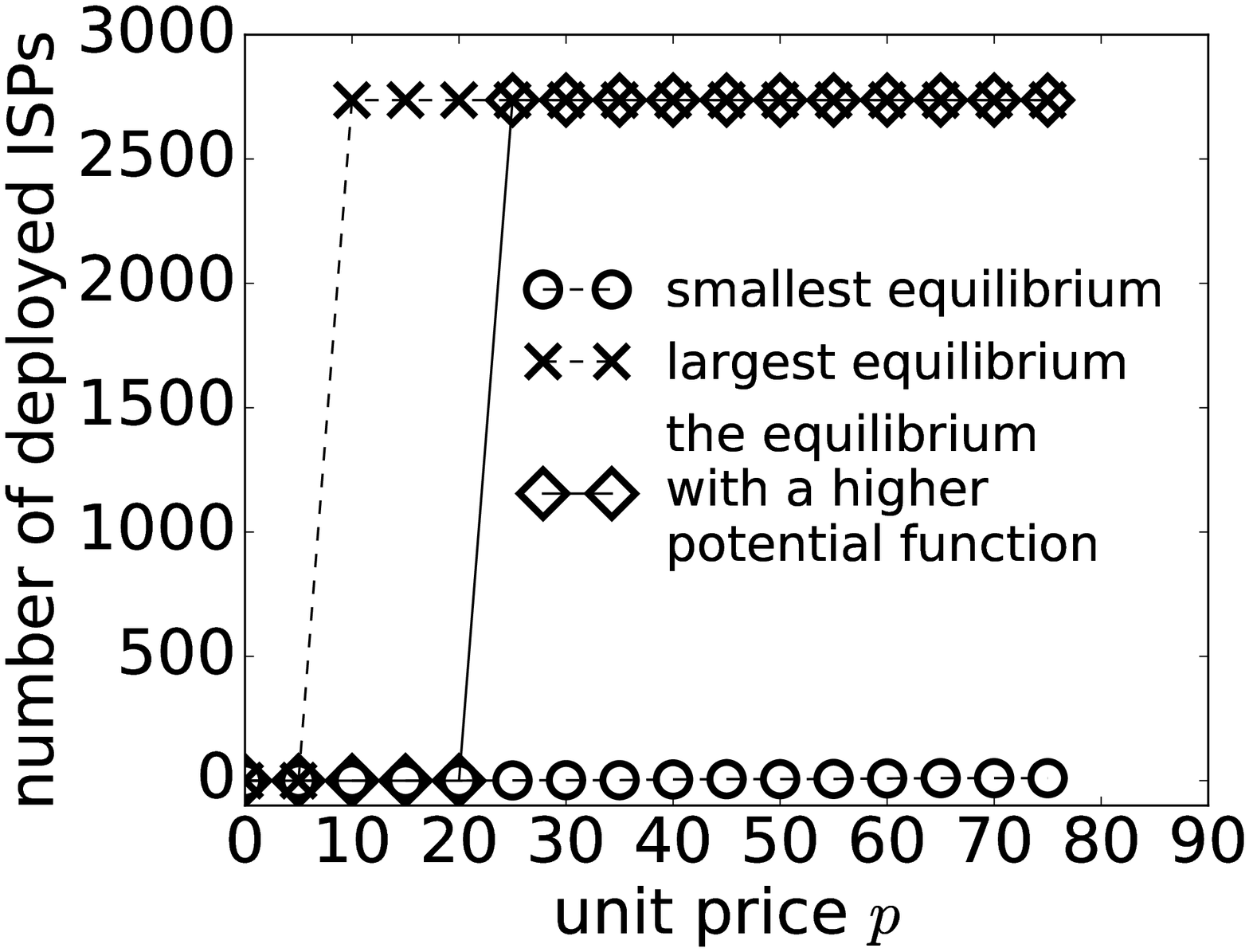}
  \caption{Unit price $p$ and the scale of equilibria (IPv4-net)}
  \label{fig:benefit_scale_large}
  \end{minipage}
\begin{minipage}{0.245\textwidth}
  \centering
  \includegraphics[width=\textwidth]{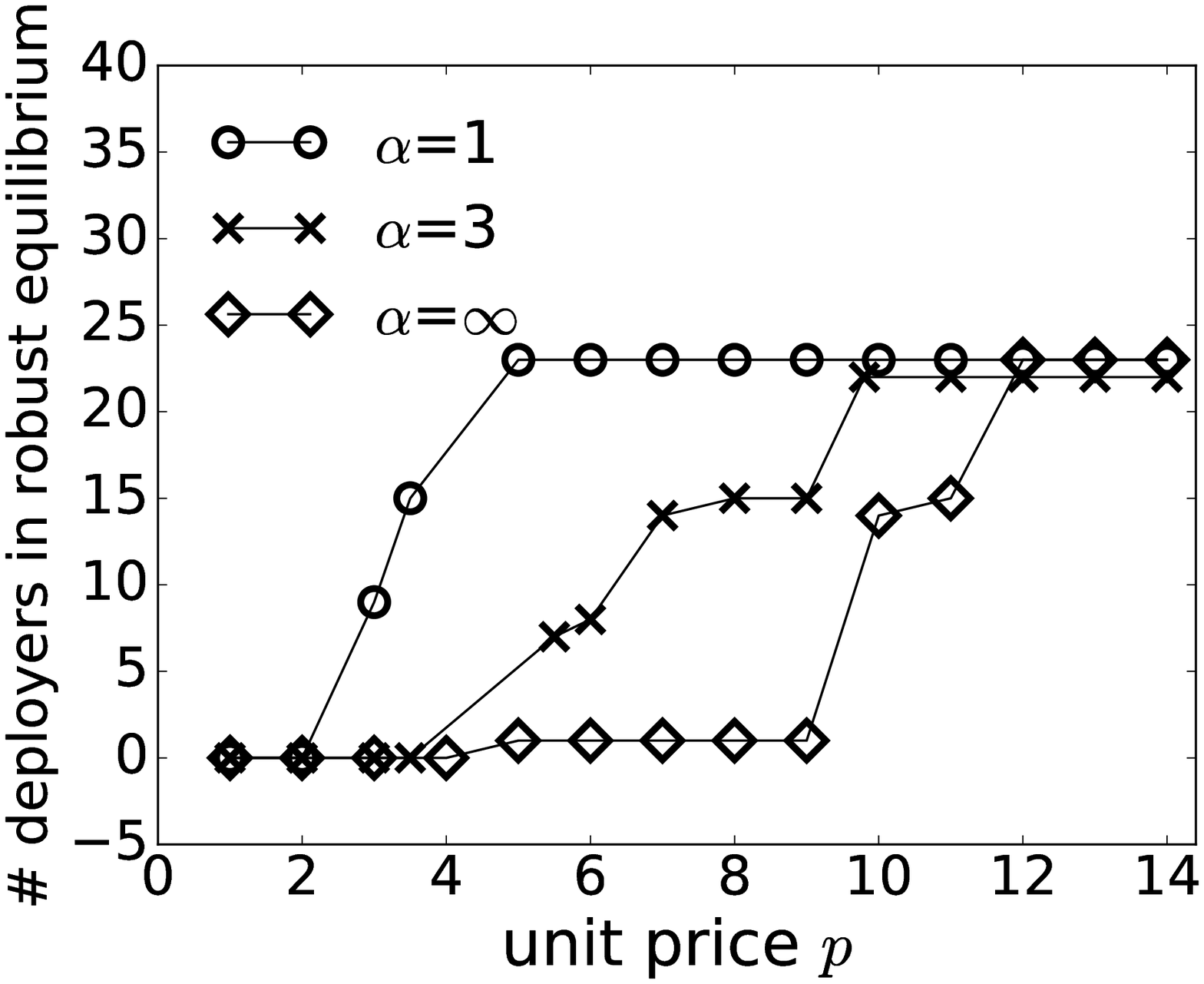}
  \caption{Incremental deployment mechanisms (G\'EANT)}
  \label{fig:incremental_scale}
  \end{minipage}
\end{figure*}
 
\section{\bf Numerical Experiments}
\label{sec:experiments}


\subsection{\bf Experiment Settings}
\noindent
{\bf Datasets.}
The first dataset~\cite{uhlig2006providing} was collected from a European
education network G\'EANT with 23 ASes (Autonomous System).
The data contains a network topology
$G{=}(\CI{,} \mathcal{E})$
and a traffic matrix $\mathbf{T} {\in} \mathbb{R}^{I\times I}_{{\ge}0}$,
where $T_{ij}$ records the traffic volume from the source node $i$
to the destination node $j$.
There are $477$ flows of non-zero traffic volume in this dataset.
The second dataset is the AS-level IPv4 topology collected
by CAIDA in Dec. 2017~\cite{caidatopology}.
The dataset contains a weighted graph of 28,499 ASes
$G{=}(\CI{,} \mathcal{E})$, where each edge $(i{,}j){\in} \mathcal{E}$
can be either a direct or an indirect link from $i$ to $j$.
This dataset does not contain the traffic matrix data.
Thus, we synthesize a traffic matrix based
on the Gravity method~\cite{roughan2005simplifying}.
The idea is that the traffic volume from node $i$ to $j$
is proportional to the {\it repulsive} factor of
the source node $i$ denoted by $T^{out}(i)$, and
the {\it attractive} factor
of the destination node $j$, denoted by $T^{in}(j)$, i.e.
$T_{ij}
{\propto}
T^{out}(i){\times} T^{in}(j).
$
We apply the Clauset-Newman-Moore method~\cite{clauset2004finding}
to extract the largest cluster in the network.
This cluster contains 2,774 nodes,
and we take $T^{in}(i)$ and $T^{out}(j)$ to be i.i.d. exponential
random variables with mean 1. 
From 2774$\times$2774 possible flows,
we randomly select 74,424 (2\%) as the flows with a positive demand of traffic
for the new architecture.

\noindent
{\bf Parameter settings.}
An ISP corresponds to an AS that has its own network policies,
so we regard the ASes in the datasets as the ISPs in our model.
It is known that G\'EANT network uses IS-IS protocol~\cite{uhlig2006providing}
that implements the Dijkstra shortest path algorithm.
For flows with a positive traffic, let
$
\CF {=} \{ \text{traffics from $i$ to $j$}
| i, j \!\in\! \CI, T_{i,j} \!>\! 0 \}
$ be the set of flows, and $\vec{p}_f$ be
the shortest path from the source to the destination for all flows
$f\in \CF$.
As discussed in Section~\ref{sec:competition},
a new functionality such as CDN typically charges customers based on
usage volume.  Thus we assume that the revenue gain from a
flow is proportional to the usage volume, i.e.,
$
\Delta_f(\tilde{\CC})
{=} p {\times} w_f
$, where $p$ is the unit price.
Given a traffic matrix $\mathbf{T}$,
the proportional traffic volume of a flow ${f}$ from source $s$ to destination
$t$ is 
$
w_{f}
{\propto} T_{i,j}.
$
Note that the launching cost of an ISP depends on the workload of the ISP.
Therefore we assume that the launching cost of an ISP $i$ {\MARK is}
proportional to the total amount of traffic through this ISP, i.e.
$
c_i {=} C{\times}
\sum_{{f} \in \CF}
\bm{1}_{\{i\in \vec{p}_f \}} w_f,
$
where $C$ is the launching cost for a unit amount of traffic.
For an architecture deployment game
${\langle} {{\tilde{\mathcal{C}}}{,} \CA{,} \bm{u}} {\rangle}$,
as we scale $p$ and $C$ linearly at the same rate,
the utility function $\bm{u}$ scales linearly as well.
Hence the Nash equilibria and the robust equilibrium will not
be changed.
Without loss of generality, we
set $C{=}1$, and see the impact of $p$
(more generally, $p{/}C$).

{\TONminor 
\noindent
\noindent{\bf Model settings.} In our numerical experiments we
consider fixed routing paths, no partial deployment 
and no revenue loss
caused by the competition between the old and new architectures.  
This is because we focus on the reasons why a new architecture is
difficult to deploy,   
and previously we have seen that other factors (change of routing path, 
partial deployment, and the competition between the old and new
architectures) are generally \emph{not\/} the major barriers to
deployment.
}

\subsection{\bf Equilibrium of the Deployment Game}

\noindent
{\bf Logit-response dynamics of ISPs. }
We simulate ISPs' behaviors by the logit-response dynamics
defined in Sec.~\ref{sec:equilibrium}, where we randomly initiate an ISP
to deploy with probability 0.5.
For the G\'EANT network, we set $\beta_t {=} 8{\times} 10^{-5}{/}t$, and
take the average of 200 runs.
As shown in Fig.~\ref{fig:logit_response}, the number of deployed ISPs is
close to the predictions of the robust equilibrium. When $T{=}50$, each ISP on the average makes two decisions, and the
outcome of dynamics is very close to the robust equilibrium.
As we increase $T{=}500$,
the outcome becomes closer to the robust equilibrium.
{
Similar results are observed for the IPv4 network in Fig.~\ref{fig:logit_response_large}.
}
The logit-response
dynamics lead to the ``robust'' equilibrium, {\MARK as if ISPs are maximizing
  some potential function in the deployment}. 
 
\noindent
{\bf ISPs' inferring process to eliminate dominated strategies.}
Recall that ISPs eliminate the dominated strategies according to the process in
our Sec.~\ref{sec:equilibrium}. After the
$t^{th}$ round of inference, it requires the perceived benefit to be at least 
$\ubar{V}^{(t)}{\triangleq} (1{+}\ubar{\theta}^{(t)})\Delta_f(\CI)$ for
an ISP to decide to adopt, where $\ubar{\theta}^{(t)}$ is the lower bound of the
estimated parameter $\theta$ after the $t^{th}$ round. Similarly, the upper bound of perceived benefit is 
$\bar{V}^{(t)}{\triangleq} (1{-}\bar{\theta}^{(t)})\Delta_f(\CI)$ for an ISP to decide not to
adopt, where $\bar{\theta}^{(t)}$ is the upper bound of the estimation of $\theta$. It means that an ISP will
definitely not adopt when the perceived benefit is below 
$(1+\ubar{\theta}^{(t)})\Delta_f(\CI)$,
 and an ISP will
definitely adopt when the perceived benefit is above $(1+\bar{\theta}^{(t)})\Delta_f(\CI)$. When the
perceived benefit is between 
$(1+\ubar{\theta}^{(t)})\Delta_f(\CI)$ and
$(1+\bar{\theta}^{(t)})\Delta_f(\CI)$, 
an ISP is uncertain about its decision.  

Consider a line-graph of $I$ ISPs ($I=2,3,4$)  
and one flow $f$. Each ISP has a launching
cost of $c=1/I$, and there is no incremental deployment for this new
architecture.  Note that in
this setting, all the ISPs are symmetric and have symmetric strategies.
Fig.~\ref{fig:2ISP_induction},
Fig.~\ref{fig:3ISP_induction} and \ref{fig:4ISP_induction} 
show the induction dynamics of firms.  
One can observe that ISPs become more certain about their decisions as the number of rounds $t$ increases. When the number of induction rounds 
exceeds 60, there is a threshold of perceived benefits for an ISP to
decide whether to deploy the new architecture.   
When the number of ISPs increases, 
the threshold of perceived benefits increases.  
This complements our theoretical results for
$\sigma\rightarrow 0$.  
One can see that an architecture is less deployable when
there are more critical ISPs along the paths. Moreover, when the uncertainty
$\sigma$ is larger, it requires fewer inference rounds to converge.
This is because ISPs' inferring process eliminates {\em what they will not do},
and with more uncertainty, ISPs will more quickly find {\em what they will not do}.
We need to point out that realistic ISPs may not do inferences for a
large number of rounds. 


\noindent
{\bf Lessons learned. } 
The numerical results further validates our theoretical characterizations 
on the equilibrium of ISPs.  

\subsection{\bf Quantifying the Deployability}
\vspace{-0.05in}

\noindent{\bf Benefit-cost ratio.}
The unit price
$p$ determines the benefit-cost ratio of the new architecture.
For G\'EANT network, the total benefit will be
more than the total launching cost when $p{>}3.3$.
Fig.~\ref{fig:benefit_scale} shows the impact of $p$ on the deployability without any incremental deployment mechanism.
When $p{\le} 4$, there is only one equilibrium, namely
``no ISP will deploy'', because the benefit of the new architecture is
not enough to cover ISPs' launching cost.
When $p{\ge} 5$, the largest equilibrium {\ICNP (with the largest set
  of deployers)} is ``all 23 ISPs deploy''.
However, in the robust equilibrium, no more than one ISP will deploy until
$p{\ge} 10$. This is because the
largest equilibrium will not have a positive potential function unless
$p{\ge}9.54$.
For the IPv4 network with 2,774 ISPs,
computing the robust equilibrium is intractable, so in
Fig.~\ref{fig:benefit_scale_large},
Fig.~\ref{fig:incremental_scale_large}, we choose 
one of the smallest/largest equilibria with a higher potential function which is
{\MARK the one that is possible} to be the robust equilibrium.
For the IPv4 network, a new architecture will be profitable when $p{\ge} 4.78$.
But only when $p{\ge} 21.88$, the condition (9) for successful
deployment can be satisfied. Hence, as seen from
Fig.~\ref{fig:benefit_scale_large}, only when $p{\ge}25$, ``all
ISPs to deploy'' is the one that is possible to be a robust equilibrium.
Similar phenomena are observed in both networks.

\begin{figure}[htb]
  \begin{minipage}{0.24\textwidth}
  \centering
  \includegraphics[width=\textwidth]{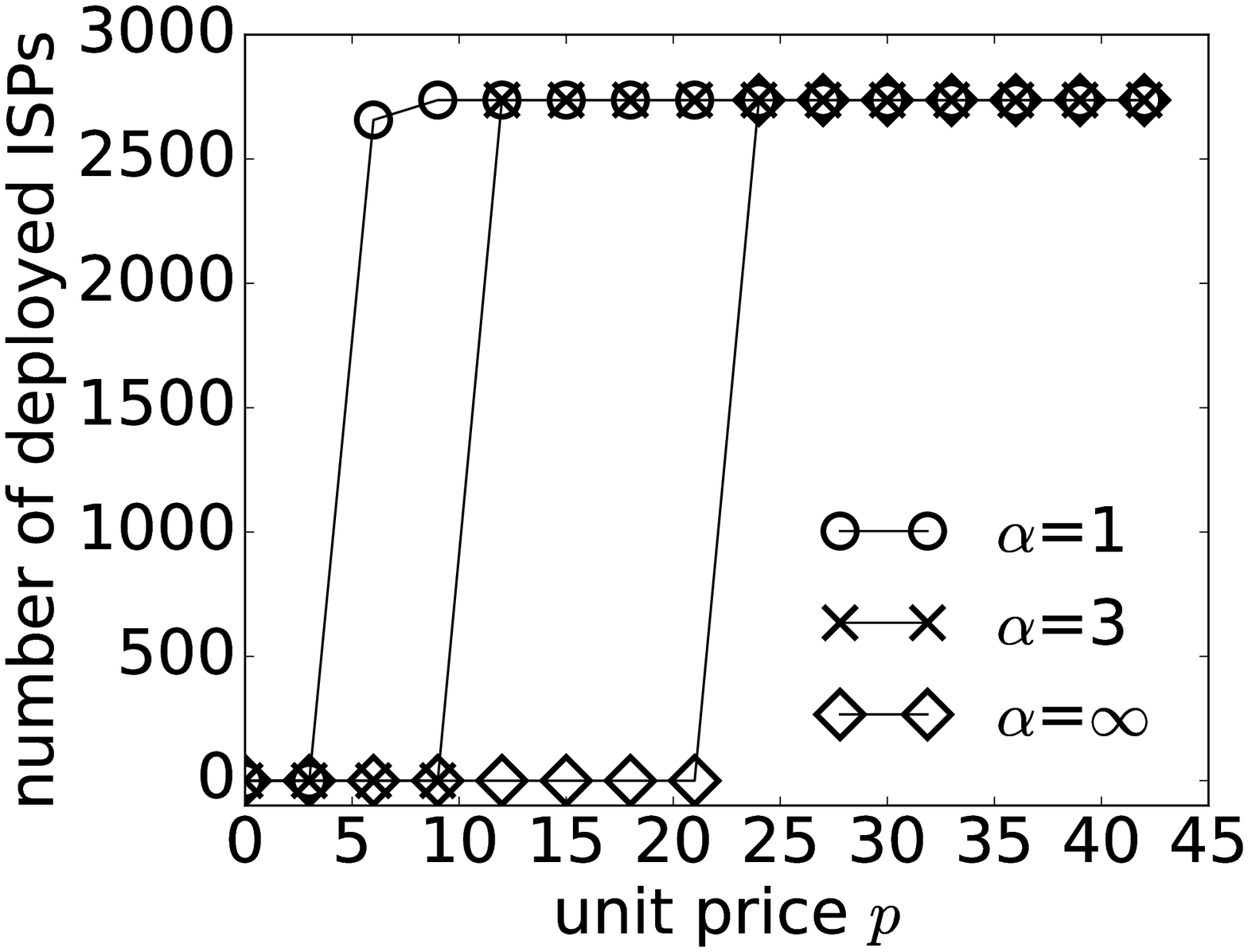}
  \caption{Incremental deployment mechanisms (IPv4-net)}
  \label{fig:incremental_scale_large}
  \end{minipage}
\begin{minipage}{0.24\textwidth}
  \centering
  \includegraphics[width=\textwidth]{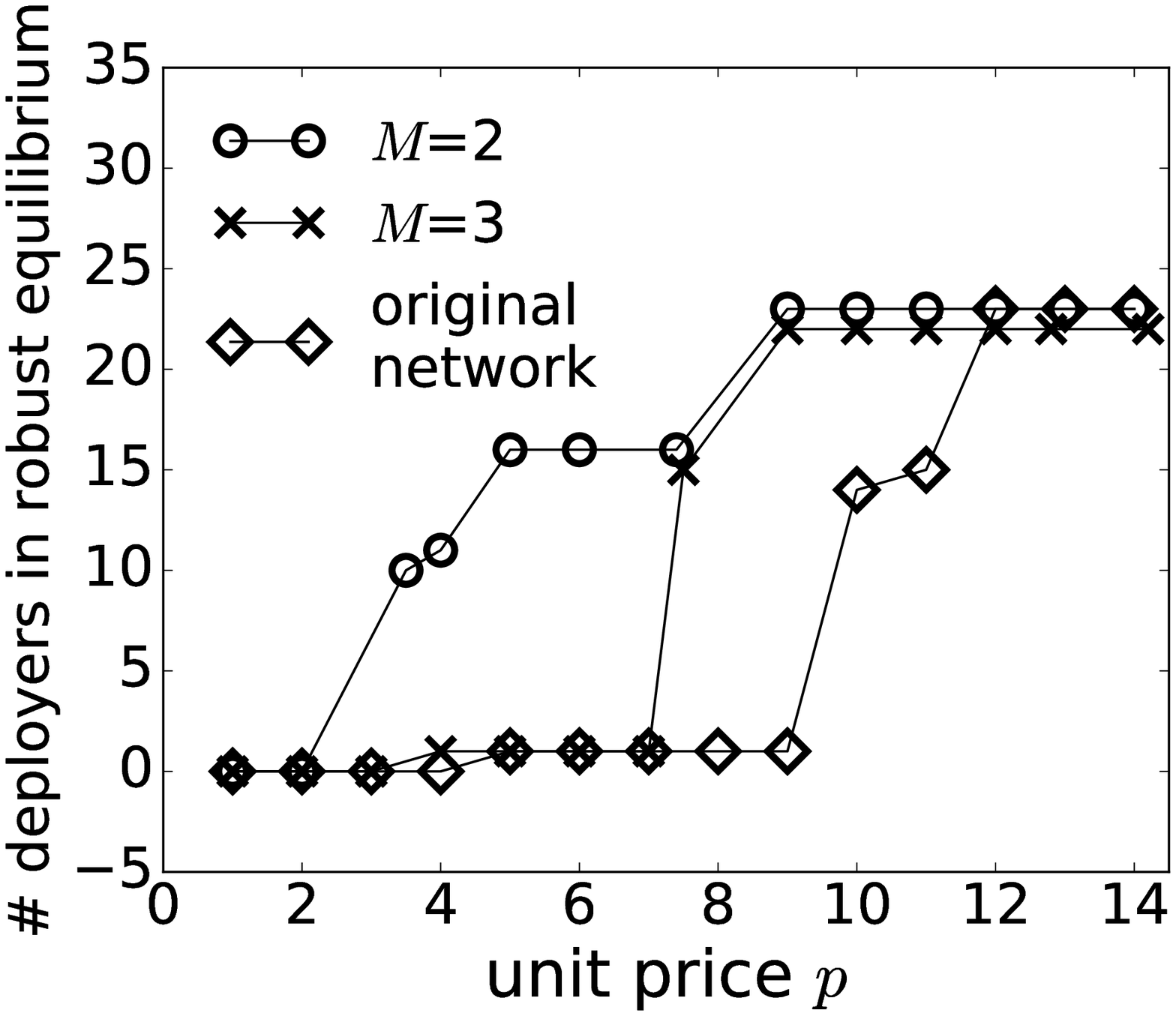}
  \caption{Impact of the flattening of network (G\'EANT)}
  \label{fig:flatten_robust_large}
  \end{minipage}
\end{figure}

\noindent{\bf Incremental deployment mechanisms.}
We set the incremental benefit for flow ${f}$ as
$ \Delta_f(\CS){=} \left({n_f(\CS)} / {|\CC(\vec{p}_{f})|}\right)^\alpha
p w_f$, when a set $\CS{\subseteq}\tilde{\CC}$ of ISPs deploy. The parameter
$\alpha{\ge} 1$ represents performances of incremental deployment mechanisms,
where a smaller $\alpha$ indicates better performances.
When $\alpha{=}{+}\infty$, there is no incremental benefit. In this case, as
depicted in Fig.~\ref{fig:incremental_scale}, the new architecture
will not be deployed in the G\'EANT network until $p{\ge} 10$.
In contrast, when $\alpha{=}1$,
the architecture will be immediately deployed by 7 ISPs when $p{=} 3$ and will be fully
deployed by all ISPs when $p{\ge} 5$. As $\alpha$ decreases, the new
architecture gets deployed by more ISPs for a fixed $p$, in help of better
incremental mechanisms.
As seen in Fig.~\ref{fig:incremental_scale_large},
better incremental mechanisms also improve the deployability of new architectures in the IPv4 network.

\noindent{\bf Internet flattening phenomenon.} To see the impact of a flattening
Internet, we shrink the paths of the original flows to have a maximum length of $M$. For a
original flow $(i{,}v_1{,} {\ldots}{,} v_L{,} j)$ with a path length $L{+}2{>}M$, the
flattened flow will be $(i{,}v_{L{-}M{+}3}{,} {\ldots}{,} v_L{,} j)$ that contains $i,j$ and
$M{-}2$ ISPs which are nearest the destination $j$.
This setting emulates that the sender uses data centers near the receiver.
 For G\'EANT, Fig.~\ref{fig:flatten_robust_large} shows that when the maximum
 path length is shortened to $M{=}2$ (i.e. only the sender \& receiver are in the flow), more than 10 ISPs will deploy when $p{\ge} 3.5$. Meanwhile,
 in the original network, ISPs will deploy the new architecture only when
 $p{\ge} 10$. Generally, the new architecture will be deployed by more ISPs in a
 more flattened network for a fixed $p$. 
{\MARK One may observe that when content providers use
   data centers which are close to end users, they can pay a lower unit price $p$ to
   the ISPs so to enjoy the deployed new architectures/technologies.}

\noindent
{\bf Lessons learned.}
A profitable new architecture may not be
deployed. Also (and unsurprisingly), higher benefit-cost ratio means
higher deployability.
The enhancement of incremental deployment mechanisms and the
flattening Internet both 
improve the deployability of the new architectures.

We also conduct experiments to show the benefits of our 
coordination mechanism
in Section~\ref{sec:mechanism}.
Please refer to the supplementary material or our
technical report\cite{arXiv_version} for details.

\section{\bf Related Works}

\label{sec:relatedwork}

Designing future Internet architecture has been on the agenda 
since the early ages of the Internet~\cite{shenker1995fundamental}.  
A variety of future architectures were
proposed~\cite{diff_serv,ndn,mobilityfirst,xia} 
to improve IPv4.
Unfortunately, most of these proposals fail to deploy at scale.
To make the Internet architectures evolvable, 
incremental deployment mechanisms 
are developed to enable universal access of 
IPv6~\cite{gilligan2005basic,despres2010ipv6,mukerjee2013tradeoffs}.  
While an evolvable new architecture should be compatible with old
architectures,  
our work shows {\ICNP via economic models} that an evolvable
architecture should also provide incremental {\it ``benefits''} to ISPs.
{\SHEEP Internet flattening phenomenon was studied in
 paper~\cite{Gill:2008:FIT,chiu2015we}, and our work formalizes their
 observations.}
{\SHEEP A recent work~\cite{kirilin2018protocol} studied the incremental
  deployment of routing protocols, and suggested a coordinated adoption of a large number of ISPs}.

Economics issues with the future Internet architectures have also been noticed.
Wolf et al. developed ChoiceNet~\cite{Wolf:2014:CTE} to provide 
an economics plane to the Internet and a clear economics incentive for ISPs.
Our work also points out that a new architecture may not be deployed
even if it could be profitable for all ISPs.  
Along this direction, Ratnasamy et al.~\cite{Ratnasamy:2005:TEI}
has a similar \textit{``chicken-and-egg''} argument.
Our economics analysis strengthens these arguments and quantitatively analyze
the difficulty of coordination among decentralized ISPs.
{\ICNP Some works studied the adoptability of BGP security
  protocols~\cite{Chan:2006:MAS,Gill:2011:LMD}. They conduct simulations, while we provide game-theoretic
  analysis to reveal key factors for the deployability, e.g. the coordination
  of ISPs. How to select some seeding ISPs to stimulate the deployment was
  studied~\cite{Goldberg:2013:DNT}, but the incentives for the seeding
  ISPs remain 
  a problem. Our economic mechanism considers the launching cost of the ISPs and requires the coordinator to
  invest nothing.}

The \textit{``coordination failure''} phenomenon was also studied
in economics~\cite{cooper1988coordinating}.
Monderer et al.~\cite{monderer1996potential}
found that the equilibrium that maximizes a potential function 
accurately predicts Huyck's experiments~\cite{van1990tacit}. Then Morris et
al.~\cite{morris2001global} give reasons via the \textit{``global game''},
which is used in our analysis.

\section{\bf Conclusion}
\label{sec:conclusion}

This paper studies the deployability \& evolvability of new
architectures/protocols from an economic perspective.
Our economic model shows that:
(1) Due to coordination difficulty,
being profitable is not sufficient to guarantee a new architecture
to be widely deployed;
(2) A superior architecture may lose
to another competing architecture which requires less coordination.
Our model explains why IPv4 is hard to be replaced,
why IPv6, DiffServ, CDN have different deployment difficulties,
{\color{black}
and why we
observe the ``Internet flattening phenomenon''.
  Our model suggests that by changing the routing path, a new architecture
  becomes easier to deploy.
  In addition, the new architecture will be more deployable when the
  competition from the new architecture cause the revenue from the old
  functionality to decline, provided that each ISP participates in a
  small fraction of traffic volumes in the network. 
Our model also quantifies the importance of incremental deployment
mechanisms for the deployment of new Internet architectures.}
For architectures like DiffServ with which incremental deployment mechanisms
are not available, people may consider a centralized mechanism to help
the deployment. 
The designers of new architectures like NDN and XIA
can also use our model to evaluate and improve the deployability of their
design. {\SHEEP Our model predicts that the current design of NDN and XIA are
  difficult to deploy, and MPQUIC will win over MPTCP.}

\section*{\bf Acknowledgment}
The work by John C.S. Lui was supported in part by the RGC R4032-18
RIF funding.  The work of Kenneth L. Calvert was supported by the
U.S.\ National Science Foundation during his temporary assignment there.

\bibliographystyle{IEEEtran}
\bibliography{bib}

\vspace{-0.3in}
\begin{IEEEbiography}[{\includegraphics[width=1in,height=1.25in,clip,keepaspectratio]{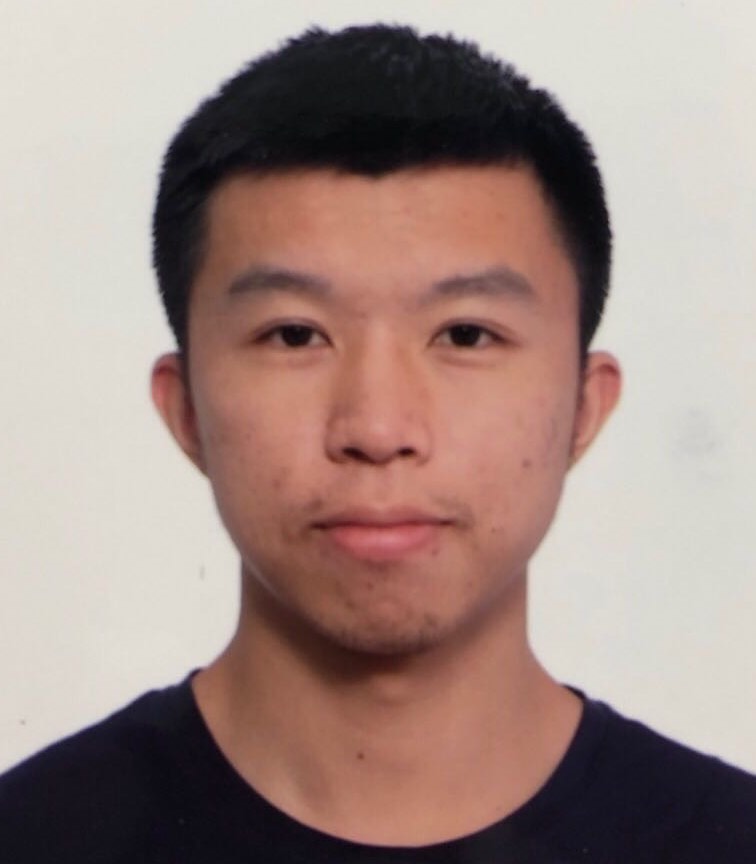}}]{Li Ye}
received the B.Eng. degree from the School of Computer Science and Technology,
University of Science and Technology of China, in 2016. He is currently a Ph.D
student in the Department of Computer Science and Engineering, The Chinese
University of Hong Kong, under the supervision of Prof. John. C.S. Lui. His research interests include network economics, data
driven decision, and stochastic modeling. He is a member of the IEEE.
\end{IEEEbiography} 
\vspace{-0.3in}

\begin{IEEEbiography}[{\includegraphics[width=1in,height=1.25in,clip,keepaspectratio]{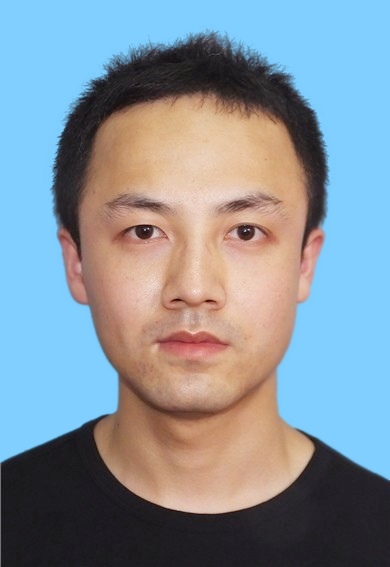}}]{Hong Xie}
received the B.Eng. degree from the School of Computer Science and
Technology, University of Science and Technology of China, in 2010, and the
Ph.D degree from the Department of Computer Science and Engineering, The
Chinese University of Hong Kong, in 2015, under the supervision of Prof. John.
C.S. Lui. He is currently a Professor in the College of
Computer Science, Chongqing University. His
research interests include online learning algorithm and applications. 
He is a member of the IEEE.
\end{IEEEbiography} 
\vspace{-0.3in}

\begin{IEEEbiography} [{\includegraphics[width=1in,height=1.25in,clip,keepaspectratio]{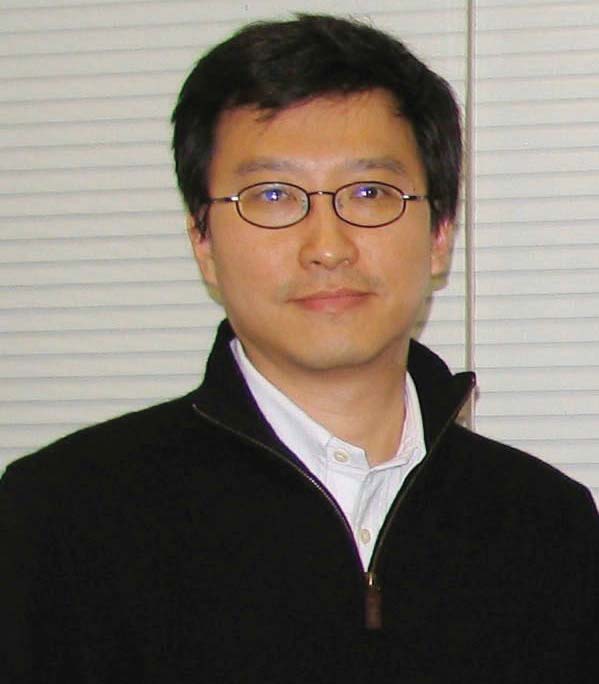}}]{John C.S. Lui}
received the Ph.D degree in computer science from the University of California
at Los Angeles. He
is currently the Choh-Ming Li Chair Professor with the Department of Computer
Science and Engineering, The Chinese University of Hong Kong. His current
research interests include Machine learning, online learning (e.g., multi-armed
bandit, reinforcement learning), Network Science, Future Internet Architectures
and Protocols, Network Economics, Network/System Security, Large Scale Storage
Systems.
He is an elected member of the IFIP WG 7.3, Fellow of ACM, Fellow of IEEE, Senior Research Fellow of the Croucher Foundation and was the past chair of the ACM SIGMETRICS (2011-2015). 
He received various departmental teaching awards and the CUHK Vice-Chancellor's Exemplary Teaching Award. 
John is a co-recipient of the best paper award in the IFIP WG 7.3 Performance 2005, IEEE/IFIP NOMS 2006, SIMPLEX 2013, and ACM RecSys 2017.
\end{IEEEbiography}

\vspace{-0.3in}

\begin{IEEEbiography}[{\includegraphics[width=1in,height=1.25in,clip,keepaspectratio]{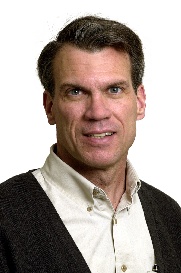}}]{Kenneth L. Calvert}
is Gartner Group Professor in Network Engineering 
at the University of Kentucky.  His research deals with the
design and implementation of advanced network protocols and services.
He has been an associate editor of IEEE/ACM
Transactions on Networking, a faculty member at Georgia Tech, and a
Member of Technical Staff at Bell Telephone Laboratories in Holmdel,
NJ.
During 2016-2019 he served as Division Director for Computer and
Network Systems in the Computer and Information Science and
Engineering Directorate at the National Science Foundation of the US.
He holds degrees from MIT, Stanford, and the University
of Texas at Austin.
\end{IEEEbiography}

\includepdf[pages=-]{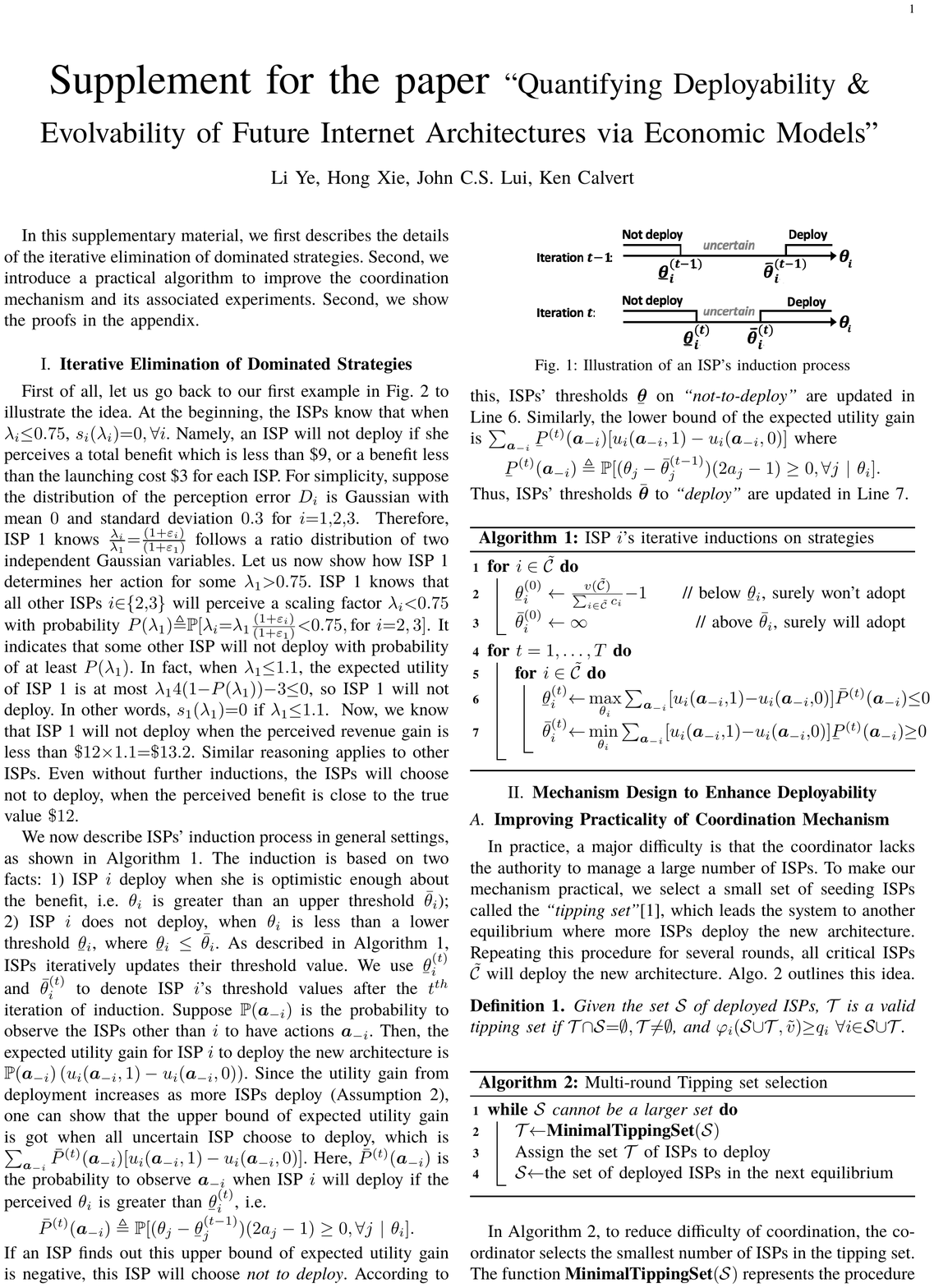}

\end{document}


\setlength{\textfloatsep}{0.15\textfloatsep}
  \setlength{\dbltextfloatsep}{0.15\dbltextfloatsep}
  \setlength{\floatsep}{0.15\floatsep}
  \setlength{\dblfloatsep}{0.15\dblfloatsep}
  \setlength{\belowdisplayskip}{0.15\baselineskip}
  \setlength{\abovedisplayskip}{0.15\baselineskip}
  \captionsetup[subfloat]{captionskip=2pt}  
\maketitle

In this supplementary material, we first describes the details of the iterative
elimination of dominated strategies. Second, we introduce a practical algorithm to
improve the coordination mechanism and its associated experiments. Second, we show the proofs in the appendix.

\section{\bf Iterative Elimination of Dominated Strategies}
First of all, let us go back to our first example in Fig.~\ref{fig:example} to
illustrate the idea.
{At the beginning, the ISPs know that when $\lambda_i{\le} 0.75$,
$s_i(\lambda_i){=}0, \forall i$. Namely, an ISP will not deploy if she
perceives a total benefit which is less than \$9, or a benefit less than the launching
cost \$3 for each ISP.} For simplicity, suppose the distribution of the perception
error $D_i$ is Gaussian with mean $0$ and standard deviation $0.3$ for $i{=}1{,}2{,}3$.
{
  Therefore, ISP 1 knows $\frac{\lambda_i}{\lambda_1}{=}
  \frac{(1{+}\varepsilon_i)}{(1{+}\varepsilon_1)}$ follows a ratio distribution of two
  independent Gaussian variables.
Let us now show how ISP 1 determines her action for some $\lambda_1{>}0.75$.
ISP 1 knows that all other ISPs $i{\in}\{2{,}3\}$ will perceive a scaling factor
$\lambda_i{<}0.75$ with probability $P(\lambda_1){\triangleq}
\mathbb{P} [\lambda_i{=}\lambda_1\frac{(1+\varepsilon_i)}{(1+\varepsilon_1)}{<}0.75,\text{for }i{=}2,3]$. It indicates that some
other ISP will not deploy with probability of at least $P(\lambda_1)$.
In fact, when $\lambda_1{\le} 1.1$,
the expected utility of ISP 1 is at most $\lambda_1 4(1{-}P(\lambda_1)) {-} 3{\le} 0$, so ISP 1 will not
 deploy. In other words, $s_1(\lambda_1){=}0$ if $\lambda_1{\le} 1.1$. }
Now, we know that ISP 1 will not deploy
when the perceived revenue gain is less than
$\$12{\times} 1.1{=}\$13.2$.
Similar reasoning applies to other ISPs.
Even without further inductions, the ISPs will choose not to deploy,
when the perceived benefit is close to the true value $\$12$.

{\TON
We now describe ISPs' induction process in general settings, 
as shown in Algorithm~\ref{alg:global_game}.
The induction is based on two facts: 
1) ISP $i$ deploy when she is optimistic enough about the benefit, 
i.e. $\theta_i$ is greater than an upper threshold $\bar{\theta}_{i}$);  
2) ISP $i$ does not deploy, when $\theta_i$ is less than 
a lower threshold $\ubar{\theta}_{i}$, where
$\ubar{\theta}_i\le \bar{\theta}_i$.  
As described in Algorithm~\ref{alg:global_game}, ISPs iteratively 
updates their threshold value.   
We use $\ubar{\theta}_{i}^{(t)}$ and $\bar{\theta}_{i}^{(t)}$ to denote ISP
$i$'s threshold values after the $t^{th}$ iteration of induction.  
Suppose $\mathbb{P}(\bm{a}_{-i})$ is the probability to observe the ISPs other
than $i$ to have
actions $\bm{a}_{-i}$. Then, the expected utility gain for ISP $i$ to deploy the
new architecture is $\mathbb{P}(\bm{a}_{-i})\left( u_i(\bm{a}_{-i},1)-u_i(\bm{a}_{-i},0) \right)$.
Since the utility gain from deployment increases as more ISPs deploy
(Assumption~\ref{asmp:supermodular}), one can show that the upper bound of
expected utility gain is got when all uncertain ISP choose to deploy, which is
$\sum_{\bm{a}_{-i}}\bar{P}^{(t)}(\bm{a}_{-i})[u_i(\bm{a}_{-i},1)-u_i(\bm{a}_{-i},0)]$.
Here, $\bar{P}^{(t)}(\bm{a}_{-i})$ is the probability to observe $\bm{a}_{-i}$
when ISP $i$ will deploy if the perceived $\theta_i$ is greater than
$\ubar{\theta}_{i}^{(t)}$, i.e.
\[
    \bar{P}^{(t)}(\bm{a}_{-i}) \triangleq
   \mathbb{P}[(\theta_j - \ubar{\theta}_{j}^{(t-1)})(2a_j-1)\ge 0, \forall j ~|~ \theta_i].
\]
If an ISP finds out this upper bound of expected utility gain is negative, this
ISP will choose {\em not to deploy}. According to this, ISPs' thresholds
$\ubar{\bm{\theta}}$ on {\em``not-to-deploy''} are updated in Line~\ref{line:update_lower_bound}.
Similarly, the lower bound of the expected utility gain is
$\sum_{\bm{a}_{-i}}\ubar{P}^{(t)}(\bm{a}_{-i})[u_i(\bm{a}_{-i},1)-u_i(\bm{a}_{-i},0)]$
where
\[
    \ubar{P}^{(t)}(\bm{a}_{-i}) \triangleq
    \mathbb{P}[(\theta_j - \bar{\theta}_{j}^{(t-1)})(2a_j-1)\ge 0, \forall j ~|~ \theta_i].
\]
Thus, ISPs' thresholds $\bar{\bm{\theta}}$ to {\em``deploy''} are updated in Line~\ref{line:update_upper_bound}.
}

\begin{algorithm}
  \caption{\TON ISP $i$'s iterative inductions on strategies}\label{alg:global_game}
  \For{$i\in \tilde{\CC}$}{
    {$\ubar{\theta}_{i}^{(0)}\gets \frac{v(\tilde{\CC})}{\sum_{i\in
          \tilde{\CC}}c_i}{-}1$}\tcp*{below $\ubar{\theta}_{i}$, surely won't adopt}\label{line:not_deploy}
    {$\bar{\theta}_{i}^{(0)}\gets \infty$}\tcp*{above $\bar{\theta}_{i}$, surely will adopt}
  }
  \For{$t=1,\ldots,T$}{
    \For{$i\in \tilde{\CC}$}{
      $\ubar{\theta}_{i}^{(t)}{\gets}
      \max\limits_{\theta_i}
    \sum\nolimits_{\bm{a}_{-i}} [u_i(\bm{a}_{-i}{,} 1){-}u_i(\bm{a}_{-i}{,}0)]
    \bar{P}^{(t)}(\bm{a}_{-i})
    {\le} 0$\label{line:update_lower_bound} \\
    $\bar{\theta}_{i}^{(t)}{\gets}
      \min\limits_{\theta_i} 
    \sum_{\bm{a}_{-i}} [u_i(\bm{a}_{-i}{,} 1){-}u_i(\bm{a}_{-i}{,}0)]
    \ubar{P}^{(t)}(\bm{a}_{-i})
    {\ge} 0$\label{line:update_upper_bound}
    }
  }
\end{algorithm}
\vspace{-0.3in}

\begin{figure}
  \centering
  \includegraphics[width=0.35\textwidth]{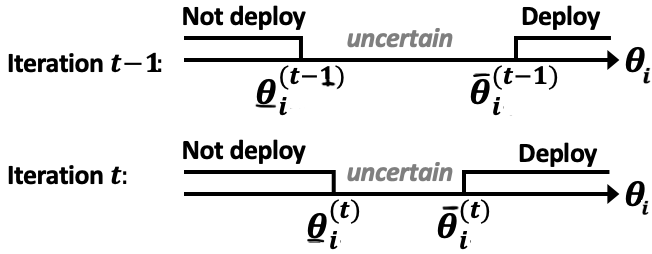}
  \caption{Illustration of an ISP's induction process}
  \label{fig:induction_global_game}
\end{figure}
 
\section{\bf Mechanism Design to Enhance Deployability}

\subsection{\bf Improving Practicality of Coordination Mechanism}

In practice, a major difficulty is that the coordinator lacks the authority to
manage a large number of ISPs.
To make our mechanism practical, 
we select a small set of seeding ISPs
called the {\it``tipping set''}\cite{heal2006supermodularity}, 
 which leads the system to another equilibrium where 
more ISPs deploy the new architecture.
Repeating this procedure for several rounds, all critical ISPs $\tilde{\CC}$
will deploy the new architecture.  
Algo.~\ref{alg:improved_selection} outlines this idea.  

\begin{definition}
Given the set $\CS$ of deployed ISPs, $\CT$ is a valid tipping
set if $\CT{\cap} \CS{=}\emptyset, \CT{\ne} \emptyset$, and $\varphi_i(\CS{\cup}
\CT, \tilde{v}){\ge} q_i$~$\forall i{\in} \CS{\cup}\CT$.
\end{definition} 

\begin{algorithm}[h]
  \caption{Multi-round Tipping set selection}\label{alg:improved_selection}
  \While{$\CS$ cannot be a larger set}{
    {$\CT {\gets} \textbf{MinimalTippingSet}(\CS)$}\\
    {Assign the set $\CT$ of ISPs to deploy}\\
    {$\CS{\gets}$the set of deployed ISPs in the next equilibrium}
  }
\end{algorithm}

In Algorithm~\ref{alg:improved_selection}, to reduce difficulty of coordination, 
the coordinator selects the smallest number of ISPs in the tipping set. 
The function $\textbf{MinimalTippingSet}(\CS)$ represents 
the procedure to find the minimal valid tipping set, 
when currently a set $\CS$ of ISPs deploy.   
Formally, we find the minimal valid tipping set via the following optimization: 
\begin{align}
& 
\underset{\CT\ne \emptyset, \CT\cap \CS=\emptyset}{\text{minimize}}
&&  |\CT|, \nonumber\\
& \text{subject to} 
&& \varphi_i(\CS{\cup} \CT, \tilde{v}){\ge} q_i, \forall i\in \CS{\cup}\CT.
  \label{eq:problem_coordinator}
\end{align}

This problem is NP-hard even when $\varphi_i(\cdot,\tilde{v})$ is a supermodular function (because it
is equivalent to the NP-hard submodular maximization problem). 
We propose a heuristic greedy algorithm (Algorithm \ref{alg:greedy_selection}) to address 
this computational challenge.  
We aim to select the minimal number of ISPs so that constraints 
(\ref{eq:problem_coordinator}) is satisfied.  
Note that the gap between ISP $i$'s cost and ISP $i$'s distributed benefits is
$\max\{c_i-\varphi_i(\CS, \tilde{v}), 0\}$, which characterizes the violation of
the constraint (\ref{eq:problem_coordinator}) for an ISP $i$.
For a set $\CS$ of ISPs, we
define the total gap as $\texttt{gap}(\CS) {\triangleq} \sum_{i\in \CS} \max\{c_i-\varphi_i(\CS, \tilde{v}), 0\}$. When $\texttt{gap}(\CS\cup \CT)=0$, all the constraints in
(\ref{eq:problem_coordinator}) are satisfied.
In Line 3-7 of Algorithm~\ref{alg:greedy_selection}, we greedily select ISPs to minimize the total gap, until the total gap reduces to 0.
Before the execution of Line~8,
$\texttt{gap}(\CS\cup \CT)=0$ and all the constraints in
(\ref{eq:problem_coordinator}) are satisfied, hence we get a feasible solution.
But the number of selected ISPs could be further reduced,
by removing the redundant ISPs whose removal will not violate the constraints (Line 7-8).

\vspace{-0.1in}
\begin{algorithm}[h]
 \caption{$\textbf{MinimalTippingSet}(\CS)$}\label{alg:greedy_selection}
 \SetKwFunction{FMinimalTippingSet}{MinimalTippingSet}
 \SetKwProg{Fn}{Function}{:}{}
 \Fn{\FMinimalTippingSet{$\CS$}}{
 $\CT \gets \{j\}$ ($j$ is randomly chosen from $\tilde{\mathcal{C}}-\CS$)\\
 \While{$\texttt{gap}(\CS \cup \CT) \ne 0$}{
   \If{$\CS\cup \CT = \tilde{\mathcal{C}}$}{
      {$\CT\gets \CS_{\overline{\bm{a}}^\ast} - \CS$}, \Break
   }
   $i\gets \arg\min_{i\in (\tilde{\mathcal{C}} -\CT-\CS)} \texttt{gap}(\CS \cup \CT \cup \{i\})$
   \label{alg:greedy_increase}\\
   $\CT\gets \CT\cup \{i\}$
 }
         \While{$\exists j\in \CS$, such that $\texttt{gap}(\CS\cup\CT - \{j\}) = 0$}{
     (randomly choose such $j$),  $\CT\gets \CT-\{j\}$
 }
 \Return{$\CT$}
 }
\end{algorithm}
\vspace{-0.1in}

\begin{theorem}
 \label{thm:tipping_algorithm}
 Algorithm \ref{alg:improved_selection} will terminate.
 After the termination, the largest equilibrium $\overline{\bm{a}}^\ast$ in
 Lemma~\ref{lemma:super_eq} will be reached.
\end{theorem}

\subsection{\bf Experiments on the Coordination Mechanism's Benefits}
\label{sec:exp_mechanism}

\begin{figure}
  \centering
                \begin{minipage}{0.4\textwidth}
  \centering
  \includegraphics[width=0.7\textwidth]{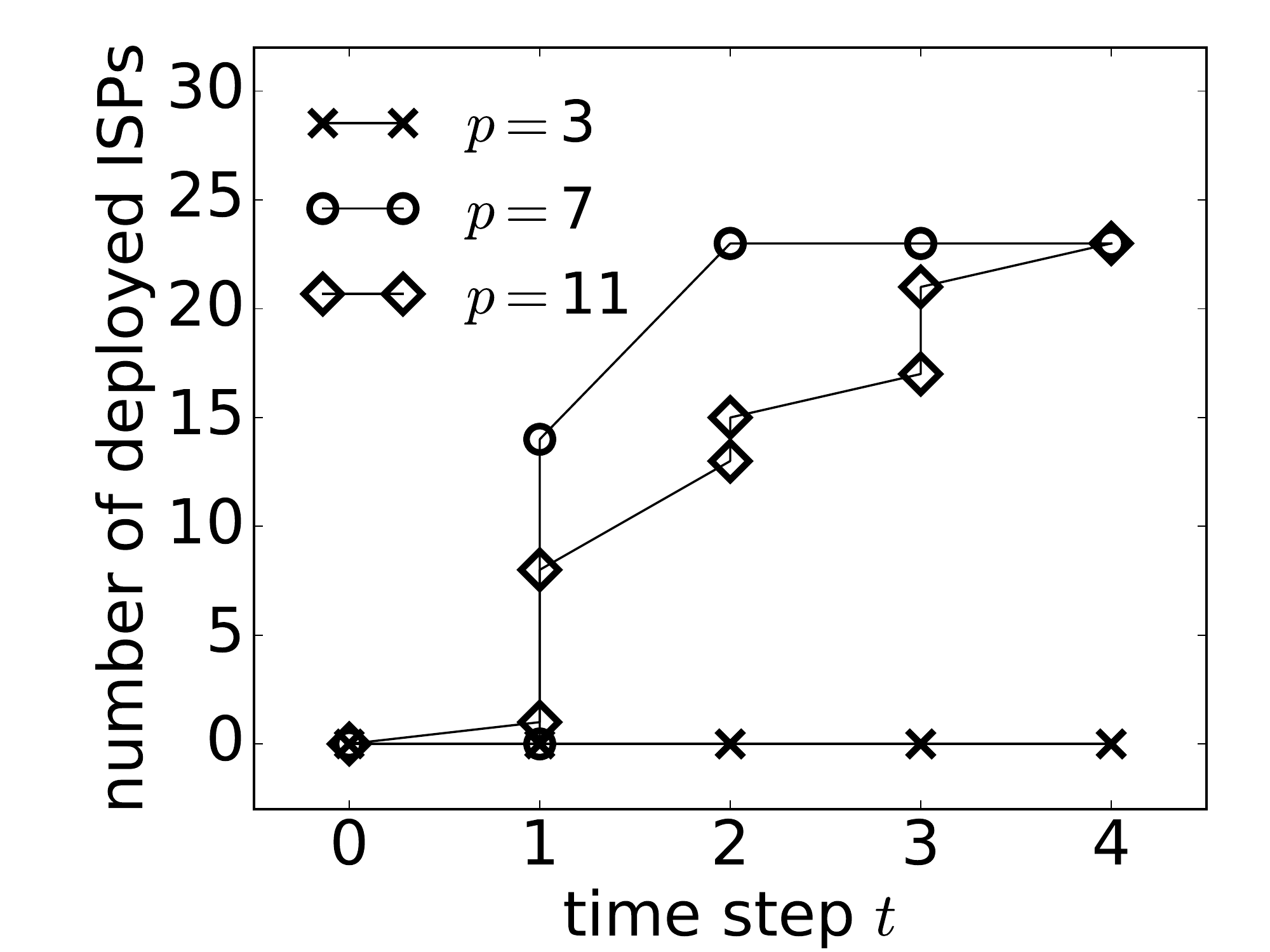}
  \caption{Centralized economics mechanism (G\'EANT)}
  \label{fig:mechanism_exp}
  \end{minipage}
\end{figure}

How the centralized coordinator can select the small ``tipping set'' of ISPs in the G\'EANT network is shown in Fig.~\ref{fig:mechanism_exp}.
At each time step $t{=}1,2,{\cdots}$, the coordinator selects some ISPs,
where we see a vertical increment for the number of deployers. 
Between time steps $t$ \& $t{+}1$, some ISPs will deploy if they have higher
utilities, where we see an increasing slope.
For a small $p{=}3$, the economics mechanism cannot help and ``all ISPs do not deploy''
is the only equilibrium.
When $p{=}7$, originally the new architecture
cannot be successfully deployed as shown in Fig.~\ref{fig:benefit_scale}. With the help of economics mechanism, by
selecting $15$ ISPs as the tipping set, all ISPs will finally deploy. When $p$ further increases
to $11$,
the coordinator only needs to select 7,2,4 ISPs in three steps respectively.

For the IPv4 network, when $p{=}5$, no ISP will deploy even with the existence
of the coordinator.
When $p{=}10$ and $20$, after the coordinator selects 1,970 and 1,350 tipping
ISPs respectively, all ISPs will finally deploy. Note that 1,350<1,970.

\noindent
{\bf Lessons learned. }
As the benefit-cost ratio of architectures increases, the coordinator selects a smaller number of ISPs.

\appendix
\subsection{\bf Closed-form of Shapley value and the potential function}

{\TON
\begin{proof}[\bf Proof of Lemma~\ref{lem:EquilibriumSecondStage}]
 From Hart's paper\cite{hart1996bargaining}, we know Shapley value distribution is
 the unique stationary subgame perfect equilibrium (SSPE) in the case where ISPs do not
 have revenue loss when they do not deploy the new architecture, i.e. 
 $\delta_{i,f}(\CS){=}0$ for $i{\not\in}\CS$. We will first show that a SSPE in
 their game is still a SSPE in ours.
The only modification we need to make on the proofs of Hart's
paper\cite{hart1996bargaining} is ISPs' payoffs when they quit the bargaining
(or when they do not deploy the new architecture).
Instead of the utility $0$ in Hart's paper\cite{hart1996bargaining}, an ISP $i$ will get $\sum_{f\in\CF}\delta_{i,f}(\CS\backslash\{i\})$
if it chooses to quit bargaining.

Let us define $\bm{\phi}^{(i)}(\CS,{v})=(\phi_j^{(i)}(\CS,{v})_{j\in\CS}$ as the distribution mechanism proposed by ISP $i$ when a set
$\CS$ of ISPs remain in the bargaining process. We also define
$\bar{\bm{\phi}}(\CS,{v}){=}(\bar{\phi}_j(\CS,{v}))_{j\in\CS}{\triangleq} ({1}/{|\CS|})\sum_{i\in\CS}[(\phi_j^{(i)}(\CS,{v}))_{j\in\CS}]$ as ISPs'
expected distribution mechanism from a random proposer.

In the Proposition 1 of the paper\cite{hart1996bargaining}, they claim that the
proposals corresponding to an SSPE are always accepted.
We define
\[
  V(\CS){=}\{(\bm{\phi}(\CS,v))|\phi_i(\CS,v)
  {\ge}
\delta_{i,f}(\CS\backslash\{i\}), \sum_{i\in\CS}\phi_i(\CS,v){\le} v(\CS) \}\] as the
set of feasible distribution mechanisms when ISPs in $\CS$ deploy the new architecture.
Moreover, $\partial V(\CS)$ is the boundary of the set $V(\CS)$.
According to the paper\cite{hart1996bargaining}, the equilibriums are characterized by
\begin{itemize}
\item[(C1)] $\bm{\phi}^{(i)}(\CS,v)\in \partial V(\CS)$ for all $i\in\CS\subseteq \CI$;
\item[(C2)] $\phi_{j}^{(i)}(\CS,v){=}\rho \bar{\phi}_{j}(\CS,v) {+} (1{-}\rho)\bar{\phi}_j(\CS\backslash\{i\},v)$ for all
$i,j{\in} \CS{\subseteq} \CI$ with $i{\ne} j$.
\end{itemize}

\noindent
We let
$\varphi_{i}(\CS,\tilde{v})=\phi_{i}(\CS,v)-\sum_{f\in\CF}\delta_{i,f}(\CS\backslash\{i\})$,
and
\[
  \mathcal{V}(\CS)
  {=}
  \{(\varphi_i(\CS,\tilde{v}))_{i\in\CS}|\varphi_i(\CS,\tilde{v}){\ge}0,
\sum_{i\in\CS}\varphi_i(\CS,\tilde{v}){\le} \tilde{v}(\CS)\}.
\]

Similarly, we define
$\bm{\varphi}^{(i)}(\CS,\tilde{v}){=}(\varphi^{(i)}(\CS,\tilde{v}))_{j\in\CS}$, 
and 
$\bar{\varphi}_i(\CS,\tilde{v}){=}(1/|\CS|)\sum_{i\in\CS} \phi_{\CS,i}$.
Then, conditions (C1) and (C2) are equivalent to the following two conditions (C3)
and (C4)
\begin{itemize}
\item[(C3)] $\bm{\varphi}^{(i)}(\CS,\tilde{v})\in \partial \mathcal{V}(\CS)$ for all $i\in \CS \subset\CI$;
\item[(C4)] $\varphi_j^{(i)}(\CS,\tilde{v}){=}\rho \bar{\varphi}_j(\CS,\tilde{v})
  {+} (1{-}\rho) \bar{\varphi}_j(\CS\backslash \{i\}, \tilde{v})$
for $i,j{\in}\CS{\subseteq} \CI$ with $i{\ne} j$.
\end{itemize}

Our proof follows the arguments in the proof of Theorem 2 (page 7) of Hart's
paper\cite{hart1996bargaining}. 
Hart's paper\cite{hart1996bargaining} has shown that  
the payoff configurations $\{\bm{\varphi}^{(i)}(\CS,\tilde{v})\}_{i\in\CS}$ that
solve the equation system defined by conditions (C3)-(C4) satisfy the four axioms of Shapley value
defined in Shapley's paper\cite{shapley1953value}.
In particular, the solving configurations
are {\em ``linear''} in $\tilde{v}$ because both (C3) and (C4) are linear, {\em ``symmetric''} relative to the labels of the players, and {\em ``Pareto
  efficient''} because the distribution lies on the border of $\mathcal{V}$.
Also, the solving configurations satisfy the {\em ``null player''} axiom because
$\varphi_j^{(i)}(\CS\backslash\{j\},\tilde{v})=0$.
Therefore, the $\varphi_j^{(i)}(\CS,\tilde(v))$ that satisfy conditions (C3) and (C4) is the
Shapley value of the ISP $j$ which corresponds to
$\varphi_i(\CS,\tilde{v})$ defined in Equation (\ref{eq:sharplyvaluedef}).
\end{proof}
}

\begin{proof}[{\bf Proof of Theorem~\ref{thm:Shapley_value}}]
{\TON
  Shapley value satisfy the four properties listed in this paper\cite{Ma:2010:IEU}.
}
First, we show the closed-form of the Shapley value. Recall that we define the revenue gain function as $\Delta_{f}(\CS)$ for a flow
$f\in\CF$. 
When Eq.~(\ref{eq:CondRoutPath}) and (\ref{eq:RevLoss}) hold, we do not have
change of path and revenue loss from flows that use the old architecture.
According to Proposition~\ref{Assum:RevDis}, for both $f\in \CF^{old}(\CS)$ or
$f\in\CF^{new}(\CS)$, we have $\delta_{i,f}(\CS)=0$.
Therefore,
$v(\CS)=\sum_{f\in \CF}\Delta_{f}(\CS)$. Consider the Shapley value
$\varphi_i(\CS,\Delta_f)$ w.r.t. the function $\Delta_f$. According to Property
{\em``dummy''} (a.k.a. {\em``null player''}),   any $i{\not\in} \CC(\vec{p}_f)$ is
  dummy for the flow $f$ and $\phi_i(\CS, \Delta_f){=}0$ for $i{\not\in} \CC(\vec{p}_f)$. Furthermore, for
  $i,j\in \CS\cap \CC(\vec{p}_f)$, we have $\Delta_{f}(\CT\cup\{i\})=\Delta_{f}(\CT\cup\{j\})$ for all
  $\CT\subseteq \CS\setminus \{i,j\}$ because of Assumption \ref{asump:revenue_gain}.
  Therefore, by Property {\em ``symmetry''}\cite{Ma:2010:IEU}, we have $\varphi_i(\CS, \Delta_f)=\varphi_j(\CS,\Delta_f)$ for any $i,j\in \CC(\vec{p}_f)\cap \CS$, which means
  that two critical ISPs for a flow $f$ should share the same revenue gain if
  they deploy the new architecture. Now, by Property {\em
    ``efficiency''}\cite{Ma:2010:IEU} (a.k.a. {\em``Pareto efficient''}), we know
  $\sum_{i\in \CS} \varphi_i(\CS, \Delta_f)=\Delta_f(\CS)$, so each critical ISP $i\in
  \CC(\vec{p}_f)$ that deploys the new architecture has an equal share
  $\varphi_i(\CS,\Delta_f)=\Delta_f(\CS)/n_f(\CS)$. According to
  Property {``additivity''}\cite{Ma:2010:IEU},
  \begin{align*}
    &\varphi_i(\CS, v)=\varphi_i(\CS, \sum_{f\in \CF}\Delta_f)=\sum_{f\in \CF} \varphi_i(\CS, \Delta_f)\\
    =&\sum\nolimits_{f\in \CF}  \mathbbm{1}_{\{i\in \CC(\vec{p}_f)\}} \frac{\Delta_f(\CS)}{n_f(\CS)} \\
  \end{align*}
 Moreover, when conditions (\ref{eq:CondRoutPath}) and (\ref{eq:RevLoss}) hold,
 we have $\delta_{i,f}(\CS)=0$ for $\forall i{\not\in\CS}$ and $f{\in}\CF$, thus
 $\varphi_i(\CS,\tilde{v}){=}\phi_i(\CS,v)$. Then, the above equation we have proved is exactly Eq.~(\ref{eq:Shapley}).
\end{proof}

Then, we show the closed-form of potential function $\Phi$.
\begin{proof}[{\bf Proof of Theorem~\ref{thm:potential_game}}]
Under the same condition of Theorem~\ref{thm:Shapley_value}, we have $u_i(0, \bm{a}_{-i})=0$, then
\[u_i(1, \bm{a}_{-i}) {-} u_i(0, \bm{a}_{-i}) 
{=}
\sum_{f\in\CF} \mathbbm{1}_{\{i{\in} \CC(\vec{p}_f)\}}
\frac{\tilde{\Delta}_f\left( n_f(\CS_{\bm{a}_{-i}}){+}1\right)}{n_f(\CS_{\bm{a}_{-i}}){+}1}{-}c_i,
\]
where we recall that $\CS_{\bm{a}}=\{i|a_i=1\}$.
Now, consider $\Phi(1, \bm{a}_{-i}) - \Phi(0, \bm{a}_{-i})$. First,
\begin{align*}
  \Phi(1, \bm{a}_{-i}) {=} \sum_{{f} \in \CF}
\sum_{m=1}^{n_f(\CS_{\bm{a}_{-i}}) + \mathbbm{1}_{\{i\in \CC(\vec{p}_f)\}}}
\frac{\tilde{\Delta}_f(m)}{m}
{-}
\sum_{j\ne i} a_j c_j {-} c_i.
\end{align*}
Second,
\begin{align*}
  \Phi(0, \bm{a}_{-i}) 
=
 \sum_{{f} \in \CF}
\sum\nolimits_{m=1}^{n_f(\CS_{\bm{a}_{-i}}) }
\frac{\tilde{\Delta}_f(m)}{m}
-
\sum\nolimits_{j\ne i} a_j c_j 
\end{align*}
Therefore, we have 
\[
  \Phi(1, a_{-i}) - \Phi(0, a_{-i}){=} \sum_{f\in\CF} \mathbbm{1}_{\{i{\in} \CC(\vec{p}_f)\}}\frac{\tilde{\Delta}_f\left( n_f(\CS_{\bm{a}_{-i}}){+}1\right)}{n_f(\CS_{\bm{a}_{-i}}){+}1}{-}c_i.
\]
We could see that $u_i(1, \bm{a}_{-i}) - u_i(0, \bm{a}_{-i})=\Phi(1,
\bm{a}_{-i}) - \Phi(0,\bm{a}_{-i})$ for any $\bm{a}_{-i}$, which satisfies the
definition of a potential game in Eq.~(\ref{eq:potential}). Hence, we have
proved that a function $\Phi$ in the form of Eq.~(\ref{eq:potential_func}) is a
valid potential function and our game $G$ is a potential game by definition.
\end{proof}

\subsection{\bf Architecture Deployment Game}
In this section, we will show that our game $G$ is a supermodular game. We will also
guide the readers to the theory of the ``global game'' which uses the idea of
iterative elimination of dominated strategies.
Furthermore, we will prove our corollaries about the game.

We next show that under the complementarity effect as stated in Assumption \ref{asmp:supermodular},
an ISP will have more incentive to deploy
the new architecture if more ISPs have already deployed it, as follows:
\begin{theorem}
Suppose $\varphi$ satisfies (\ref{eq:sharplyvaluedef}),
and Assumption \ref{asmp:supermodular} holds,
then the game $G$ is a supermodular game, i.e.,
for all $i\in{\tilde{\CC}}$, 
 \[
   u_i(1, \bm{a}_{-i}^\prime) - u_i(0, \bm{a}_{-i}^\prime) \ge u_i(1, \bm{a}_{-i}) -  u_i(0, \bm{a}_{-i}),
 \]
   whenever $\bm{a}_{-i}^\prime\ge \bm{a}_{-i}$ holds
   component-wisely.
\label{thm:supermodulargame}
\end{theorem}
\begin{proof}[{\bf Proof of Theorem~\ref{thm:supermodulargame}}]
  In Section 5.2 of the paper\cite{narahari2012game}, it is shown that for some $i\in
  \CS\subseteq \CT$, we have
  \begin{align}
    \label{eq:Shapley_monotone}
\varphi_i(\CS, \tilde{v})\le \varphi_i(\CT, \tilde{v}).
  \end{align}
  This is because
  the marginal improvement of ISPs' total revenue by adding ISP $i$ will be
  higher as more other ISPs have deployed the new architecture.
  Since $\bm{a}_{-i}\le
  \bm{a}_{-i}^\prime$, we have $\CS_{a_{-i}}{\cup} \{i\} \subseteq
  \CS_{a_{-i}^\prime}{\cup} \{i\}$. Thus,
  \begin{align}
    \label{eq:using_Shapley_monotone}
    \varphi_i(\CS_{a_{-i}}{\cup} \{i\}, \tilde{v})\le \varphi_i(\CS_{a_{-i}^\prime}{\cup}
    \{i\}, \tilde{v}).
  \end{align}

We also note that
  \begin{align}
    \label{eq:utilities}
    u_i(1,\bm{a}_{-i}^\prime)-u_i(0,\bm{a}_{-i}^\prime)&=\varphi_i(\CS_{a_{-i}^\prime}{\cup} \{i\}, \tilde{v})-c_i, \nonumber\\
    u_i(1,\bm{a}_{-i})-u_i(0,\bm{a}_{-i})&=\varphi_i(\CS_{a_{-i}}{\cup} \{i\}, \tilde{v})-c_i.
  \end{align}
  Combine (\ref{eq:utilities}) and (\ref{eq:using_Shapley_monotone}), we have
  \[u_i(1, \bm{a}_{-i}^\prime) {-} u_i(0, \bm{a}_{-i}^\prime)
    {\ge} u_i(1, \bm{a}_{-i}) {-}  u_i(0, \bm{a}_{-i}),\]
  which concludes our proof.
\end{proof}

Recall that we define the ``robust
equilibrium'' in Sec.~\ref{sec:equilibrium} and use it in the analysis. Now, we
show that it can be located in polynomial time. This is because the set
function $\tilde{\Phi}$ defined by $\tilde{\Phi}(\CS_{\bm{a}}){\triangleq}
\Phi(\bm{a})$ is a supermodular set function as we will show, and the global maximizer of a
supermodular function can be found in strongly polynomial time\cite{iwata2001combinatorial}.

\begin{proof}[{\bf Proof of supermodularity of $\tilde{\Phi}(\cdot)$}]
 Because a supermodular game has increasing differences of ISPs' actions, i.e. 
 \[
   u_i(a_i^\prime, \bm{a}_{-i}^\prime) - u_i(a_i, \bm{a}_{-i}^\prime) \ge u_i(a_i^\prime, \bm{a}_{-i}) -  u_i(a_i, \bm{a}_{-i}),
 \]
   whenever $a_i^\prime\ge a_i$ and $\bm{a}_{-i}^\prime\ge \bm{a}_{-i}$ 
   component-wisely.  
   In a potential game, it also means that
 \begin{align}
   \label{eq:potential_supermodular}
   \Phi(a_i^\prime, \bm{a}_{-i}^\prime) - \Phi(a_i, \bm{a}_{-i}^\prime) \ge \Phi(a_i^\prime, \bm{a}_{-i}) -  \Phi(a_i, \bm{a}_{-i})
 \end{align}
 Let $\CS_1{\triangleq}\CS_{\bm{a}_{-i}}$ and
 $\CS_2{\triangleq}\CS_{\bm{a}_{-i}^\prime}$, where $\CS_1$ and $\CS_2$ can be
 any sets as long as $\CS_1{\subseteq} \CS_2$. By Eq.~(\ref{eq:potential_supermodular}),
 using the set function $\tilde{\Phi}$, we have $\tilde{\Phi}(\CS_2\cup \{i\})-\tilde{\Phi}(\CS_2){\ge} \tilde{\Phi}(\CS_1\cup \{i\}){-}\tilde{\Phi}(\CS_1)$. Because $\CS_1$ and $\CS_2$
 can be any sets, we proved that $\tilde{\Phi}$ is a supermodular function\cite{iwata2001combinatorial}.
\end{proof}

\begin{proof}[{\bf Proof of Corollary \ref{cor:EquiFailSucc}}]
  It is easy to see that $\underline{\bm{a}}^*=(0,\ldots, 0)$ is a pure Nash
  equilibrium, if $v(\{i\}) \leq c_i, \forall i \in \tilde{\CC}$. For any $i{\in}\tilde{\CC}$, we have $u_i(1,\bm{0}_{-i}){=}v(\{i\}){-}c_i{\le}
  0 {=} u_i(0,\bm{0}_{-i})$. By Definition \ref{def:nash_eq},
  $\underline{\bm{a}}^*$ is a pure strategy Nash equilibrium. Moreover, there is no smaller equilibrium than $(0,\ldots, 0)$.
  Similarly, we can see that the action profile
  $\overline{\bm{a}}^\ast=(1,\ldots, 1)$ is also a pure strategy Nash
  equilibrium, if $\varphi_i(\tilde{\CC},\tilde{v}) \geq c_i, \forall i \in \tilde{\CC}$.
  This is because for any $i{\in}\tilde{\CC}$,
  $u_i(1,\mathbbm{1}_{-i})-u_i(0,\mathbbm{1}_{-i}){=}\varphi_i(\tilde{\CC},\tilde{v}){-}c_i{\ge}
  0$. Also, there is no equilibrium that is larger than $(1,\ldots, 1)$ point-wisely. 
\end{proof}

\begin{proof}[{\bf Remarks on Lemma \ref{lemma:equiGlobalGame}}]
This Theorem is a result of the paper by Frankel, Morris et al. --- {\em``Equilibrium
  Selection in Global Games with Strategic Complementarities''}\cite{frankel2003complementarity}. We apply the
``global game'' framework. The idea of the ``global game'' is that the ISPs have
incomplete information and ISPs have a small perception error on the payoff,
just as we stated in Sec.~\ref{sec:equilibrium}-{\bf (2) Iterative elimination of
dominated strategies}. In Theorem 1 of
that paper\cite{frankel2003complementarity},
they show the strategic profile $(s^\ast_i (\cdot))_{i\in\tilde{\CC}}$ of the
players is unique as the error distributions $\CD_i$'s concentrate around zero,
if our original game without perception error is a supermodular game. Theorem 4
in that paper states that the unique strategy profile will be the ``Local Potential(LP) maximizer'' (in our case, the maximizer of the potential function
$\Phi(\cdot)$ is their ``LP-maximizer'') as the perception error becomes 0, i.e.
$\varepsilon_i(\bm{a})=0$, or $\theta_i=1$.

Note that the quasi-concavity in their assumption could be extended to discrete
cases, and our case with only two actions automatically satisfies the
quasi-concavity since it is indeed linear. Moreover, we have a slightly
different way to model the perception error, although both ways have the same
physical interpretation that is an ISP will possibly perceive a higher or lower
payoff compared to its truth value. The key question is how an ISP (say $i$) can
infer the probability for some other ISP (say $j$) to deploy or not. In their model, the perception error is additive,
therefore the ISP $j$'s perceived signal $\theta_j$ follows a posterior distribution $\theta_i{+}(\CD_j{-}\CD_i)$ where $\CD_i$ and
$\CD_j$ are the distributions of $\varepsilon_i$ and $\varepsilon_j$. In our
model, ISP $j$'s perceived signal $\theta_j$ follows a ratio distribution of
the random variable $\theta_i\frac{1+\varepsilon_j(\bm{a})}{1+\varepsilon_i(\bm{a})} = \theta_i {+}
\frac{\varepsilon_j(\bm{a})-\varepsilon_i(\bm{a})}{1+\varepsilon_i(\bm{a})}\theta_i$. Their proof
only requires that the perception deviation from $j$ to $i$ that is
$(\CD_j{-}\CD_i)$ to have both positive and negative support with no restriction on the specific
distribution $\CD_i, \CD_j$. In fact, our perception deviation
$\frac{\varepsilon_j(\bm{a})-\varepsilon_i(\bm{a})}{1+\varepsilon_i(\bm{a})}\theta_i$ does have support on
both positive and negative values, hence their proofs still apply.
\end{proof}

\begin{proof}[{\bf Proof of Corollary \ref{corollary:necessary_condition}}]
 If all critical ISPs $\tilde{\CC}$ deploy the new architecture in the robust
 equilibrium, then $\Phi((1,\ldots, 1))$ should be the maximum value of the potential
 function $\Phi(\cdot)$. Therefore, at least $\Phi((1,\ldots, 1))\ge
 \Phi((0,\ldots, 0))$. 

Under conditions (\ref{eq:CondRoutPath}) and (\ref{eq:RevLoss}), we have $\varphi_i(\CS,\tilde{v})=\phi_i(\CS,v)$. Then the potential function has the form of (\ref{eq:potential_func}). It directly leads to condition
 (\ref{eq:necessary_condition}).

 Next, we show $B(\tilde{\CC})\le v(\tilde{\CC})$.
 We consider a special sequence of deployers $\tilde{\CC}=\{1,2,\ldots,
 |\tilde{\CC}|\}$. Namely, the first deployer is ISP 1, the second deployer
 is ISP 2, and so on.
 The total immediate benefits is now
 \begin{align}
   \label{eq:B}
   B(\tilde{\CC})=\phi_1(\{1\},v)+\phi_2(\{1,2\},v)+\ldots+\phi_{|\tilde{\CC}|}(\tilde{\CC},
   v).
 \end{align}
 Note that the total immediate benefits do not depend on the specific sequence
 of deployers because it is a potential game. Also,
 \begin{align}
   \label{eq:v}
   v(\tilde{\CC}) =
   \phi_1(\tilde{\CC},v)+\phi_2(\tilde{\CC},v)+\ldots+\phi_{|\tilde{\CC}|}(\tilde{\CC},
   v).
 \end{align}
 We could see that $\{i: i{\le} m\}\subseteq \tilde{\CC}$ when $m{\le}
 |\tilde{\CC}|$. Therefore, according to (\ref{eq:Shapley_monotone}),
 $\phi_m(\{i: i{\le} m\},v)\le \phi_m(\tilde{\CC},v)$ for any $m{\le}
 |\tilde{\CC}|$. Immediately, from (\ref{eq:B}) and (\ref{eq:v}), we know
 $B(\tilde{\CC})\le v(\tilde{\CC})$.
 
Therefore, condition (\ref{eq:necessary_condition}) implies
(\ref{eq:profitable}). But the reverse is not true.
\end{proof}

\begin{proof}[{\bf Proof of Corollary \ref{corollary:sufficient}}]
  We discuss two cases. The first case is that
\[
  \Phi((1,{\ldots},1))> \Phi((0,\ldots,0)).
\]
Then, for some maximizer $\bm{a}^*{\in} \arg\max_{\bm{a}} \Phi(\bm{a})$, $\Phi(\bm{a}^*){>}0$. Moreover, $\bm{a}^*{\ne} \bm{0}$. Therefore, in the robust
equilibrium $\bm{a}^*$, there exist a non-empty set of ISPs to deploy the new
architecture.

The second case is that $\Phi((1,{\ldots},1))= \Phi((0,\ldots,0))$. If there exist some
$\bm{a}\ne \bm{0}$ such that $\Phi(\bm{a})> 0$, then $\Phi(\bm{a}^*){>}0$ and
$\bm{a}^*{\ne} \bm{0}$, so we are done. If
there is no such $\bm{a}$, then we have $\max_{\bm{a}} \Phi(\bm{a}){=}0$ and
$(1,\ldots,1){\in} \arg\max_{\bm{a}} \Phi(\bm{a})$. Therefore, ``all ISPs to
deploy'' will be a robust equilibrium and in this robust equilibrium, more than one ISPs will deploy.
\end{proof}

{\TON
\subsection{\bf Impacts of Change of Routing Paths}
\begin{proof}[\bf Proof of Lemma~\ref{lemma:shapley_routing_path}]
First we show the decomposition of the part
$v_g$.
Note that when flow $f$ can use the native new architecture, the revenue
gain from this flow is $\Delta_f(\tilde{\CC})$, which is the maximum
revenue gain from this flow. 

Here, $v_g(\CS)=\sum_{f\in\CF^{new}(\CS)} \Delta_f(\tilde{\CC})$. In other words, ISPs have
revenue gain from flow $f$ when it can use the new architecture. Let us use $E(\vec{p})$ to denote the event that all critical
ISPs for routing path $\vec{p}$ deploy the new architecture, i.e.
$\CC(\vec{p})\subseteq\CS$. Then, ``flow $f$ can use the new architecture'' is
equivalent to the event ``$\bigcup_{\vec{p}\in \CP_f} E(\vec{p})$''. By Venn graph or
by the inclusion-exclusion principle for set operations, we will get the
following equation
\[
\MBBP[\bigcup_{\vec{p}\in \CP_f} E(\vec{p})] = \sum_{\CP\subseteq \CP_f}
(-1)^{|\CP|-1}\MBBP[\bigcap_{\vec{p}\in \CP} E(\vec{p})].
\]
As an illustrative example, when
$\CP_f=\{\vec{p}_1,\vec{p}_2,\vec{p}_3\}$, we have
\begin{align*}
 &\MBBP[ E(\vec{p}_1){\cup} E(\vec{p}_2){\cup} E(\vec{p}_3)] {=}
 \MBBP[E(\vec{p}_1)]{+}\MBBP[E(\vec{p}_2)]{+}\MBBP[E(\vec{p}_3)] \\
 & -\MBBP[E(\vec{p}_1){\cap} E(\vec{p}_2)] {-} \MBBP[E(\vec{p}_2){\cap} E(\vec{p}_3)] {-} \MBBP[E(\vec{p}_1){\cap} E(\vec{p}_3)] \\
 & +\MBBP[E(\vec{p}_1){\cap} E(\vec{p}_2){\cap} E(\vec{p}_3)].
\end{align*}

Since in the deployment process, an event $E(\vec{p})$ can either happen or not, we can replace the probabilistic
notation $\MBBP$ with the deterministic notation $\mathbbm{1}$.
Therefore, we have the following decomposition of the revenue gain function: 
\begin{align}
  \label{eq:combination}
v_g(\CS){=}\sum_{f\in\CF}\Delta_f(\tilde{\CC}) \sum_{\CP \subseteq \CP_f} 
  (-1)^{|\CP|-1}\mathbbm{1}_{\{ \bigcup_{\vec{p}\in\CP}\CC(\vec{p}_f) \subseteq \CS\}}.
\end{align}
For a set $\CT$, we say the game for ISPs $\CS$ to distribute the value $v_\CT(\CS)$ is a ``unanimity
game'', if the value
$v_\CT(\CS)$ is non-zero only when $\CT\subseteq \CS$.
In Equation~(\ref{eq:combination}),
for each $f\in\CF$ and $\CP\subseteq \CP_f$, we have a unanimity game to
distribute
$(-1)^{|\CP|-1}
\Delta_f(\tilde{\CC})\mathbbm{1}_{\{ \bigcup_{\vec{p}\in\CP}\CC(\vec{p}) \subseteq \CS\}}$,
where all the ISPs in the set
$\bigcup_{\vec{p}\in\CP}\CC(\vec{p})$ should deploy the new architecture to get
non-zero valuation. According to Lemma 3 (page 10) of paper\cite{ui2000shapley},
all the ISPs in the set $\bigcup_{\vec{p}\in\CP}\CC(\vec{p})$ will have the same
share $(-1)^{|\CP|-1}\Delta_f(\tilde{\CC})/|\bigcup_{\vec{p}\in\CP}\CC(\vec{p})|$
of the total utility of this unanimity game.
Also note that $\sum_{f\in\CF^{old}(\CS)}\Delta_f(\CS){=}0$, thus we only need
to count the revenue increment of the flow in $\CF^{new}(\CS)$.
According to the additive property of Shapley value, ISP $i$'s Shapley value is
the sum of shared utilities from all unanimity games related to ISP $i$.
\begin{align*}
 & \varphi_i(\CS,v_g){=}\\
 & \sum_{f\in\CF}\sum_{\CP \subseteq \CP_f} 
\frac{\Delta_f(\tilde{\CC}) (-1)^{|\CP|-1}}{|\bigcup_{\vec{p}\in \CP}\CC(\vec{p})|} 
\mathbbm{1}_{\{i\in \bigcup_{\vec{p}\in \CP}\CC(\vec{p})\}}
\mathbbm{1}_{\{\bigcup_{\vec{p}\in \CP}\CC(\vec{p})\subseteq \CS\}}.
\end{align*}

Second, we show that $\varphi_i(\CS,v_l){\ge} 0$.
In fact, we have $v_l(\CS){\ge} 0$
for $\forall \CS\subseteq \CI$. 
According to Proposition~\ref{Assum:RevDis}, we have the claim that $\sum_{i\not\in
  \CS}\delta_{i,f}(\CS)\le 0$, i.e. ISPs who do not deploy the new architecture
have non-zero revenue loss.
This is because an ISP that does not deploy the new architecture will not be in
the routing path $\vec{P}_f(\CS)$, if it is originally not in the routing path
$\vec{p}_f$.
In addition, because $i\not\in \CS\backslash\{i\}$, we have
$\delta_{i,f}(\CS\backslash\{i\})\le 0$. Altogether, we have $v_l(\CS)\ge 0$ for
$\forall \CS$.
Then, from the definition of Shapley value (\ref{eq:sharplyvaluedef}),
we have $\phi_i(\CS,v_l)\ge 0$.
\end{proof}

Let $\Phi_g(\bm{a})$ and $\Phi_l(\bm{a})$ be two functions such that
$
    \Phi_g(1,\bm{a}_{-i}) - \Phi_g(0,\bm{a}_{-i}) = \varphi_i(\CS_{\bm{a}},v_g), 
$ 
and 
$
    \Phi_l(1,\bm{a}_{-i}) - \Phi_l(0,\bm{a}_{-i}) = \varphi_i(\CS_{\bm{a}},v_l).
$
Then, we characterize the potential function of $G$.  
\begin{lem}
\label{lemma:potential_routing_path}
One potential function that satisfies Eq.~(\ref{eq:potential}) can be 
$\Phi(\bm{a}){=} \Phi_g(\bm{a}){+}\Phi_l(\bm{a}){+}\sum_{i{\in}\tilde{\CC}}a_ic_i$.
Given the same condition as Lemma \ref{lemma:shapley_routing_path}, 
we have $\Phi_l(\bm{a}){-}\Phi_l(\bm{0}){\ge} 0, \forall \bm{a}{\in}\{0,1\}^{|\tilde{\CC}|}$,
and $\Phi_g(\bm{a})$ satisfies 
\begin{align}
  \Phi_g(\bm{a})
  {=} 
\sum_{f\in\CF}\sum_{\CP \subseteq \CP_f} 
\frac{\Delta_f(\tilde{\CC})(-1)^{|\CP|-1}}{|\bigcup_{\vec{p}\in \CP}\CC(\vec{p})|}
 \mathbbm{1}_{\{ \bigcup_{\vec{p}\in \CP} \CC(\vec{p}) \subseteq \CS_{\bm{a}}\}}.
  \label{eq:potential_routing_path}
\end{align}
\end{lem}
\begin{proof}[\bf Proof of Lemma~\ref{lemma:potential_routing_path}]
  
When $\Phi(\bm{a}){=}\Phi_g(\bm{a}){+}\Phi_l(\bm{a}){+}\sum_{i{\in}\tilde{\CC}}a_iC_i$,
we have
\begin{align*}
  \Phi(1,\bm{a}_{-i})&{-}\Phi(0,\bm{a}_{-i}){=}\varphi_i(\CS_{\bm{a}},v_g){+}\varphi_i(\CS_{\bm{a}},v_l){+}C_i\\
 & =\varphi_i(\CS_{\bm{a}},\tilde{v})+C_i = u_i(1,\bm{a}_{-i})-u_i(0,\bm{a}_{-i}).
\end{align*}
Then $\Phi(\bm{a})$ is a potential function of $G$ based on the definition.
  
Recall that (\ref{eq:combination}) gives us a decomposition of the valuation
function $v_g(\CS)$, and it also lets us to decompose the original game into
multiple unanimity games. According to Lemma~3 (page 10) of
paper\cite{ui2000shapley}, each unanimity game corresponding to $f\in\CF$ and
$\CP{\subseteq} \CP_f$ has a potential function
$\frac{\Delta_f(\tilde{\CC})(-1)^{|\CP|-1}}{|\bigcup_{\vec{p}\in \CP}\CC(\vec{p})|}$.
Then, we can get the form of $\Phi_g(\bm{a})$ as shown in
(\ref{eq:potential_routing_path}).

Now, we turn to prove $\Phi_l(\bm{a}){-}\Phi_l(\bm{0}){\ge} 0$. Suppose we order
the elements in $\CS{=}\{i|a_i{=}1\}$ into a sequence $(i_1,i_2,\ldots,i_{|\CS|})$. Then,
$\Phi_l(\bm{a})-\Phi_l(\bm{0})=\sum_{k=1}^{|\CS|}\varphi_{i_k}(\{i_1,\ldots,i_k\},v_l)$.
Because $\varphi_i(\CS,v_l)\ge 0$ for $\forall i$ and $\forall \CS$ according to
Lemma~\ref{lemma:shapley_routing_path}, we have $\Phi_l(\bm{a})-\Phi_l(\bm{0})\ge 0$.
\end{proof}

Based on Lemma \ref{lemma:potential_routing_path}, we can analyze the
deployability of a new architecture via
Theorem~\ref{theorem:deployability_routing_path}. 

\begin{proof}[\bf Proof of Theorem~\ref{theorem:deployability_routing_path}]
The claim in the Theorem is reduced to the characterization of the potential
function $\Phi(\bm{1})-\Phi(\bm{0})$. Without loss of generality, we let $\Phi(\bm{0})=0$.
The intuition behind the proof is that if an ISP participates in more routing
paths, then this ISP will share a larger Shapley value.

Since ISPs' revenues can be decomposed by flows, the potential function
$\Phi(\CS)$ can also be decomposed by flows. Therefore, we denote
$\Phi_{g}^{f,\CP_f}(\CS)$ as the potential function $\Phi_g(\bm{a}_{\CS})$
(where $\bm{a}_{\CS}{\triangleq}(\mathbbm{1}_{\{i\in\CS\}})_{i\in\CI}$) when we only have one flow $\CF=\{f\}$, and when the set of
alternative paths of flow $f$ is $\CP_f$. Formally, we define
\begin{align*}
  \Phi_{g}^{f,\CP_f}(\CS) \triangleq
\sum\nolimits_{\CP \subseteq \CP_f} 
\frac{\Delta_f(\tilde{\CC})(-1)^{|\CP|-1}}{|\bigcup_{\vec{p}\in \CP}\CC(\vec{p})|}
 \mathbbm{1}_{\{ \bigcup_{\VBp\in \CP} \CC(\vec{p}) \subseteq \CS\}}
\end{align*}
Note that we use the set $\CS$ instead of the vector $\bm{a}$ in the function
$\Phi_{g}^{f,\CP_f}(\CS)$ for notational convenience.
When the set of alternative routing paths of flow $f$ is $\CP_f$, the potential function
$\Phi_g(\bm{a}){=}\sum_{f\in\CF}\Phi_{g}^{f,\CP_f}(\CS_{\bm{a}})$.

Similarly, we also define $v_{g}^{f,\CP_f}(\CS)$ as the revenue gain $v_g(\CS)$ when we
only have one flow $f$, and when flow $f$'s alternative paths is $\CP_f$.
\begin{align*}
  v_{g}^{f,\CP_f}(\CS)\triangleq 
  \sum\nolimits_{\CP\subseteq \CP_f}
  \Delta_f(\tilde{\CC})(-1)^{|\CP|-1}
  \mathbbm{1}_{\{ \bigcup_{\vec{p}\in \CP} \CC(\vec{p}) \subseteq \CS\}}.
\end{align*}

We now prove by induction that when all critical ISPs deploy the new architecture, the potential function 
increases as we have more routing paths in $\CP_f$, i.e.
$\Phi_{g}^{f,\CP_f}(\bm{1}){\le}\Phi_{g}^{f,\CP_h}(\bm{1})$ when $\CP_f {\subseteq} \CP_h$.

We compare the potential functions in situations where the set of routing paths is $\CP_f$ or
$\CP_f\cup \{\vec{p}\}$. Suppose ISP $i$ is a critical ISP of routing path $\vec{p}$,
i.e. $i\in\CC(\vec{p})$.
For all the critical ISPs in $\tilde{\CC}$, we consider the deploying sequence
$(j_1,\ldots,j_{|\tilde{\CC}|-1},i)$ where ISP $i$ is the last ISP to deploy the
new architecture.
Now, we consider the potential functions
$\Phi_{g}^{f,\CP_f}(\tilde{\CC}\backslash\{i\})$ and $\Phi_{g}^{f,\CP_f\cup\{\vec{p}\}}(\tilde{\CC}\backslash\{i\})$.
When ISP $i$ does not deploy the new architecture, routing path $\vec{p}$ cannot
use the new architecture, because here we do not consider incremental deployment
mechanism. Thus $v_{g}^{f,\CP_f}(\CS)=v_{g}^{f,\CP_f\cup \{\vec{p}\}}(\CS)$
when $i\not\in \CS$.
Therefore,
\begin{align}
\Phi_{g}^{f,\CP_f}(\tilde{\CC}\backslash\{i\})
  = \sum_{k=1}^{|\tilde{\CC}|-1} \varphi_{j_k}(\{j_1,j_2,\ldots,j_k\}, v_{g}^{f,\CP_f}) \nonumber\\
  = \sum_{k=1}^{|\tilde{\CC}|-1} \varphi_{j_k}(\{j_1,j_2,\ldots,j_k\}, v_{g}^{f,\CP_f\cup\{\VBp\}}) 
  = \Phi_{g}^{f,\CP_f\cup \{\VBp\}}(\tilde{\CC}\backslash\{i\}).
  \label{eq:potential_1}
\end{align}
When we add ISP $i$ to deploy the new architecture, the potential function
\begin{align}
  \Phi_{g}^{f,\CP_f}(\tilde{\CC}) =
  \Phi_{g}^{f,\CP_f}(\tilde{\CC}\backslash\{i\})+\varphi_i(\tilde{\CC},v_{g}^{f,\CP_f}). \nonumber\\
  \Phi_{g}^{f,\CP_f\cup\{\VBp\}}(\tilde{\CC}) {=}
  \Phi_{g}^{f,\CP_f\cup\{\VBp\}}(\tilde{\CC}\backslash\{i\}){+}\varphi_i(\tilde{\CC}{,} v_{g}^{f,\CP_f\cup\{\VBp\}}).
  \label{eq:potential_2}
\end{align}
When flow $\MBf$ has another routing path $\VBp$, we claim that
the Shapley value $\varphi_i(\CS,v_{g}^{f,\CP_f\cup\{\VBp\}}){\ge}\varphi_i(\CS,v_{g}^{f,\CP_f})$ for
$\forall i{\in}\VBp$ and $\forall \CS{\subseteq} \CI$.
This claim can be proved by the definition of Shapley value. In fact, we have
$v_{g}^{f,\CP_f\cup\{\VBp\}}(\CS){\ge}v_{g}^{f,\CP_f}(\CS)$. This is because
if the flow can use the new architecture when set of routing paths is
$\CP_f$, then the flow can use the new architecture when the set of routing paths is
$\CP_f{\cup}\{\VBp\}$. Also, when a critical ISP $i\in\CC(\VBp)$ does not
deploy the new architecture, the routing path $\VBp$ cannot use the new
architecture, so that
$v_{g}^{f,\CP_f\cup\{\VBp\}}(\CS)=v_{g}^{f,\CP_f}(\CS)$ when $i\not\in\CS$.
Together, we have
\[
  v_{g}^{f,\CP_f\cup\{\VBp\}}(\CT\cup
  \{i\})-v_{g}^{f,\CP_f\cup\{\VBp\}}(\CT)
  {\ge}
  v_{g}^{f,\CP_f}(\CT\cup\{i\}) - v_{g}^{f,\CP_f}(\CT)
\]
when $i{\not\in} \CT$.
According to Shapley value's formula below:
\begin{align*}
\varphi_i(\CS,v){=}
\sum_{\mathcal{T} \subseteq \CS \setminus \{i\}}\!\!\!\!
\frac{|\mathcal{T}|! (|\CS| {-} |\mathcal{T}|{-}1)!}{|\CS|!}
[v(\mathcal{T} {\cup} \{i\}) {-} v(\mathcal{T})],
\end{align*}
we have $\varphi_i(\CS,v_{g}^{f,\CP_f\cup\{\VBp\}}){\ge} \varphi_i(\CS,v_{g}^{f,\CP_f})$.
Combined with (\ref{eq:potential_2}), we have
$\Phi_{g}^{f,\CP_f\cup\{\VBp\}}(\tilde{\CC})\ge \Phi_{g}^{f,\CP_f}(\tilde{\CC})$.

Suppose $\CP_f{=}\{\vec{p}_f,
\VBp_1,{\ldots},\VBp_{|\CP_f|}\}$. If we add routing paths one-by-one, then we have a chain of inequalities
\[
  \Phi_{g}^{f,\{\VBp_f\}}(\tilde{\CC})\le
  \Phi_{g}^{f,\{\VBp_f, \VBp_1\}}(\tilde{\CC}) \le
  \ldots \le
  \Phi_{g}^{f,\{\VBp_f,\emptyset),\VBp_1,\ldots, \VBp_{|\CP_f|}\}}.
\]
Now, we have shown that
$\Phi_g(\bm{1})=\sum_{f\in\CF}\Phi_{g}^{f,\CP_f}(\tilde{\CC})$
increases as each flow $\MBf\in\CF$ has more routing paths $\CP_f$.

We then show $\Phi_l(\bm{1})$ will also increase as each flow $\MBf\in\CF$ has more routing paths. In fact, when there is no change of routing path, i.e. $\CP_f{\equiv}\{\vec{p}_f\}$, we have
$\delta_{i,f}(\CS){=}0$ for $\forall i{\not\in}\CS$. Thus the potential function $\Phi_l$ does not change as
more ISPs deploy the new architecture, i.e. $\Phi_l(\bm{a}){=}\Phi_l(\bm{0})$
for $\forall \bm{a}$ when $|\CP_f|{=}1$ for $\forall \MBf\in\CF$.
According to Lemma~\ref{lemma:potential_routing_path}, we have
$\Phi_l(\bm{1})\ge \Phi_l(\bm{0})$ when $|\CP_f|{\ge}1$.
Notice that $\Phi(\bm{1}){-}\Phi(\bm{0}){=}\Phi_g(\bm{1}){-}\Phi_g(\bm{0}){+}\Phi_l(\bm{1}){-}\Phi_l(\bm{0})$, and we have
\begin{align*}
  \begin{cases}
    \Phi(\bm{1}){-}\Phi(\bm{0}) {=} \sum_{\MBf{\in}\CF}\Phi_{g}^{f,\{\VBp_f\}}(\tilde{\CC}) &
    \text{when } \CP_f{=}\{\VBp_f\}, \forall f\in\CF\\
    \Phi(\bm{1}){-}\Phi(\bm{0}) {\ge}\sum_{\MBf{\in}\CF}\Phi_{g}^{f,\{\VBp_f\}}(\tilde{\CC}) &
    \text{when } \CP_f{\supseteq} \{\VBp_f\}, \forall f\in\CF.
  \end{cases}
\end{align*}
Therefore, suppose that a new architecture is ``deployable'' when each flow has
a fixed routing path, i.e. $\Phi(\bm{1}){-}\Phi(\bm{0}){\ge} 0$ when
$\CP_f{=}\{\VBp_f\},\forall \MBf{\in}\CF$. Then
$\Phi(\bm{1}){-}\Phi(\bm{0}){\ge} 0$ when
$\CP_f{\supseteq}\{\VBp_f\},\forall \MBf{\in}\CF$, which means the new
architecture is ``deployable'' when the flows have more routing paths.

\end{proof}
}

{\TON
\subsection{\bf Impact of Revenue Loss of Old Functionalities}
\begin{lem}
 When we consider the revenue loss in the form (\ref{eq:loss_function}), the
 Shapley value is
\begin{align*}
& \varphi_i(\CS,\tilde{v}) =
  \sum_{\MBf\in \CF^{new}(\CS)} \mathbbm{1}_{\{i\in\CC(\vec{p}_f)\}} \frac{\Delta_f(\CS)}{n_f(\CS)} \\
& - \sigma\sum_{\MBf\in\CF}\sum_{h\in\CF^{new}(\CS)}\sum_{\substack{j\in \vec{p}_f\\ j\in \CS}}
  \frac{w_fw_{h}}{|\CC(\vec{p}_{h})\cup \{j\}|} 
  \mathbbm{1}_{\{i,j\in\CC(\vec{p}_{h})\}}\\
& + \sum_{\MBf\in \CF^{new}(\CS)}\sum_{h \in \CF^{new}(\CS)} \sum_{\substack{j\in \vec{p}_f \\j\in \CS}}
  \frac{w_fw_{h}}{|\CC(\vec{p}_{h}){\cup} \CC(\vec{p}_f)|} \mathbbm{1}_{\{i,j\in\CC(\vec{p}_{h}){\cup} \CC(\vec{p}_f)\}}.
\end{align*}
\label{lemma:Shapley_revenue_loss}
\end{lem}

\begin{proof}[\bf Proof of Lemma~\ref{lemma:Shapley_revenue_loss}]
Similar to the proof of Lemma~\ref{lemma:shapley_routing_path}, we prove by
decomposing the valuation function $\tilde{v}(\CS)$, so that we can analyze the
game $G$ through a set of unanimity games.

To begin with, we revisit the value function $\tilde{v}$ where
\[
  \tilde{v}(\CS) = \sum_{f\in\CF}\Delta_f(\CS)
  -
  \sum_{i{\not\in}\CS}\sum_{f\in\CF}\delta_{i,f}(\CS)
  -
  \sum_{i\in\CS}\sum_{f\in\CF} \delta_{i,f}(\CS\backslash\{i\}).
\]
We define
$\delta_{i,f}^\prime(\CS)$ for all $i\in\CI$ as below: 
\begin{align*}
\delta_{i,f}^\prime(\CS)
  {=}
\begin{cases}
  -\sigma w_f \sum_{h{\in}\CF^{new}(\CS)}w_{h},  & i {\in} \vec{p}_f
  \text{ and }f{\in}\CF^{old}(\CS), \\
  0, & \text{otherwise}.
\end{cases}
\end{align*}
Note that $\delta_{i,f}^\prime(\CS)=\delta_{i,f}(\CS)$ when $i\not\in\CS$.
One can verify that for $f{\in}\CF^{old}(\CS)$, $\Delta_f(\CS){=}\sum_{i\in\CS}\delta_{i,f}^\prime(\CS){+}\sum_{i{\not\in}\CS}\delta_{i,f}^\prime(\CS)$. Then,
\begin{align*}
  \tilde{v}(\CS)
  {=} & \sum_{f{\in}\CF^{new}(\CS)}\Delta_f(\CS)
  {+} \left( \sum_{f{\in}\CF^{old}(\CS)}\Delta_f(\CS) {-}
    \sum_{i{\not\in}\CS}\sum_{f{\in}\CF}\delta_{i,f}(\CS) \right)\\
  &{-} \sum_{i\in\CS}\sum_{f\in\CF} \delta_{i,f}(\CS\backslash\{i\})\\
  {=} &\sum\limits_{f\in\CF^{new}(\CS)}\Delta_f(\CS)
    {+} \sum_{i\in\CS}\sum_{f\in\CF}\delta_{i,f}^\prime(\CS)
    {-} \sum_{i\in\CS}\sum_{f\in\CF}\delta_{i,f}^\prime(\CS\backslash\{i\}).
\end{align*}

After that, we rewrite the loss function as (for $\forall i\in\CI$)
\begin{align*}
 & \delta_{i,f}^\prime(\CS) {=} \sigma\mathbbm{1}_{\{i\in\vec{p}_f\}} \\
 & \left[ 
  \left(
  \sum_{h\in\CF^{new}(\CS)} w_{h} \right)w_f
  {-} \left(\sum_{h\in\CF^{new}(\CS)}w_{h} \right) \mathbbm{1}_{\{f{\in}\CF^{new}(\CS)\}}w_f
  \right].
\end{align*}
The first part is the revenue loss caused by other flows that can use the new
architecture. The second part is to compensate the loss of revenue after flow
$\MBf$ deploys the new architecture.
The condition for the first part to be non-zero is that all ISPs in the set
$\CC(\vec{p}_{h}){\cup} \{i\}$ deploy the new architecture.
The condition for the second part to be non-zero is that all ISPs in the set
$\CC(\vec{p}_f){\cup}\CC(\vec{p}_{h}){\cup}\{i\}$ all deploy the new architecture.
We can decompose the valuation function into these parts, and get the valuation
by checking whether each set of ISPs deploy the new architecture.

There are two parts: $\delta_{i,f}^\prime(\CS)$ and $\delta_{i,f}^\prime(\CS\backslash \{i\})$. The part that corresponds to $\sum_{i\in\CS}\sum_{\MBf\in\CF}\delta_{i,f}^\prime(\CS)$.
In particular, we decompose $\sum_{i\in\CS}\sum_{\MBf\in\CF}\delta_{i,f}^\prime(\CS)$ as
the following sum:
\begin{align*}
  & \sum_{i\in\CS}\sum_{\MBf\in\CF}\delta_{i,f}(\CS)\\
  =& \sigma \sum_{\MBf\in\CF}\sum_{h\in\CF^{new}(\CS)}\sum_{i\in\CS}
     \mathbbm{1}_{\{i\in\vec{p}_f\}}
     \left( w_{h}w_f - w_{h}w_f\MBBI_{\{f\in\CF^{new}(\CS)\}} \right)\\
  =& \sigma \sum_{\MBf\in\CF}\sum_{h\in\CF^{new}(\CS)}\sum_{i\in\vec{p}_f}
     \mathbbm{1}_{\{i\in\CS\}} \left( w_{h}w_f - w_{h}w_f\MBBI_{\{f\in\CF^{new}(\CS)\}} \right)\\
  =& \sigma \sum_{\MBf\in\CF}\sum_{h\in\CF} \sum_{i\in\vec{p}_f}
     (\mathbbm{1}_{\{\CC(\vec{p}_{h})\cup\{i\}\subseteq \CS\}}w_{h}w_f \\
   & ~~~~~ - \mathbbm{1}_{\{\CC(\vec{p}_{h})\cup\CC(\vec{p}_f)\cup\{i\}\subseteq \CS\}}w_{\vec{p}_{h}}w_f )
\end{align*}
Similarly, we rewrite $\delta_{i,f}(\CS\backslash\{i\})$ as the following sum:
\begin{align*}
  & \sum_{i\in\CS}\sum_{\MBf\in\CF}\delta_{i,f}(\CS\backslash\{i\})\\
  =& \sigma \sum_{\MBf\in\CF}\sum_{h\in\CF} \sum_{i\in\CS} \mathbbm{1}_{\{i\in\vec{p}_f\}} \times\\
    & \left( w_{h}w_\MBf\mathbbm{1}_{\{\CC(\vec{p}_{h})\subseteq \CS\backslash\{i\}\}}
     - w_\MBf w_{h} \mathbbm{1}_{\{\CC(\vec{p}_f)\cup\CC(\vec{p}_{h})\subseteq \CS\backslash\{i\}\}} \right)\\
  = & \sigma \sum_{\MBf\in\CF}\sum_{h\in\CF} \sum_{i\in\vec{p}_f}
     (
      \mathbbm{1}_{\{\CC(\vec{p}_{h})\cup\{i\}\subseteq \CS\}} \mathbbm{1}_{\{i\not\in\CC(\vec{p}_{h})\}} w_\MBf w_{h}\\
  & - \mathbbm{1}_{\{\CC(\vec{p}_f)\cup\CC(\vec{p}_{h})\cup\{i\}\subseteq \CS\}} \mathbbm{1}_{\{i\not\in\CC(\vec{p}_f)\cup\CC(\vec{p}_{h})\}} w_fw_{h}
    ).
\end{align*}

Now, the valuation function $\tilde{v}(\CS)$ is a weighted sum of
$\mathbbm{1}_{\{X\subseteq \CS\}}$ for different $X$'s.
The game for ISPs $\CS$ to distribute $\mathbbm{1}_{\{X\subseteq\CS\}}$ is a
unanimity game.
According to the paper\cite{ui2000shapley}, each ISP in $X$ will share $1/|X|$
of the value generated from these $X$ ISPs when all of them deploy the new architecture.
According to the additive property of Shapley value,
an ISP $i$'s Shapley
value counts all such sets $X$'s where $i\in X$. Then,
\begin{align*}
   & \varphi_{i}(\CS,\tilde{v})=\sum_{\MBf\in \CF^{new}(\CS)} \mathbbm{1}_{\{i\in\CC(\vec{p}_f)\}} \frac{\Delta_f(\CS)}{n_f(\CS)} \\
   & - \sigma \sum_{\MBf\in\CF}\sum_{h\in\CF} \sum_{j\in\vec{p}_f}
   ( \frac{\mathbbm{1}_{\{\CC(\vec{p}_{h})\cup\{j\}\subseteq \CS\}}}{|\CC(\vec{p}_{h})\cup\{j\}|} w_{h}w_f\mathbbm{1}_{\{i\in\CC(\vec{p}_{h})\cup\{j\}\}}\\
 &  ~~~~ - \frac{\mathbbm{1}_{\{\CC(\vec{p}_{h})\cup\CC(\vec{p}_f)\cup\{j\}\subseteq \CS\}}}{|\CC(\vec{p}_{h}){\cup}\CC(\vec{p}_f){\cup}\{j\}|}w_{h}w_f  \mathbbm{1}_{\{i\in\CC(\vec{p}_{h})\cup\CC(\vec{p}_f)\cup\{j\}\}} ) \\
   & + \sigma \sum_{\MBf\in\CF}\sum_{h\in\CF} \sum_{j\in\vec{p}_f}
     (
   \frac{\mathbbm{1}_{\{\CC(\vec{p}_{h})\cup\{j\}\subseteq \CS\}}}{|\CC(\vec{p}_{h})\cup\{j\}|}  w_hw_{f} \mathbbm{1}_{\{j\not\in\CC(\vec{p}_{h})\}}\times\\
  & ~~~~ \mathbbm{1}_{\{i\in\CC(\vec{p}_{h})\cup\{j\}\}}
    - \frac{\mathbbm{1}_{\{\CC(\vec{p}_f)\cup\CC(\vec{p}_{h})\cup\{j\}\subseteq \CS\}}}{|\CC(\vec{p}_f)\cup\CC(\vec{p}_{h})\cup\{j\}|}\times \\
  & ~~~~ \mathbbm{1}_{\{j\not\in\CC(\vec{p}_f)\cup\CC(\vec{p}_{h})\}} w_fw_{h}\mathbbm{1}_{\{i\in\CC(\vec{p}_f)\cup\CC(\vec{p}_{h})\cup\{j\}\}}
    )\\
     =& \sum_{\MBf\in \CF^{new}(\CS)} \mathbbm{1}_{\{i\in\CC(\vec{p}_f)\}} \frac{\Delta_f(\CS)}{n_f(\CS)} \\
 & - \sigma \sum_{\MBf\in\CF}\sum_{h\in\CF} \sum_{\substack{j\in\vec{p}_f\\j\in\CS}}
    \frac{\mathbbm{1}_{\{\CC(\vec{p}_{h})\subseteq \CS\}}}{|\CC(\vec{p}_{h})\cup\{j\}|} w_{h}w_f\mathbbm{1}_{\{i\in\CC(\vec{p}_{h})\cup\{j\}\}}\\ 
  & + \sigma \sum_{\MBf\in\CF}\sum_{h\in\CF} \sum_{\substack{j\in\vec{p}_f\\j\in\CS}}
  \frac{\mathbbm{1}_{\{\CC(\vec{p}_{h})\cup\CC(\vec{p}_f)\subseteq \CS\}}}{|\CC(\vec{p}_{h}){\cup}\CC(\vec{p}_f){\cup}\{j\}|} \times \\
  &~~~~w_{h}w_f \mathbbm{1}_{\{i\in\CC(\vec{p}_{h})\cup\CC(\vec{p}_f)\cup\{j\}\}} \\
  & + \sigma \sum_{\MBf\in\CF}\sum_{h\in\CF} \sum_{\substack{j\in\vec{p}_f\\j\in\CS}}
  \frac{\mathbbm{1}_{\{\CC(\vec{p}_{h})\subseteq \CS\}}}{|\CC(\vec{p}_{h})\cup\{j\}|} \times \\
  &~~~~ w_fw_{h}\mathbbm{1}_{\{j\not\in\CC(\vec{p}_{h})\}} \mathbbm{1}_{\{i\in\CC(\vec{p}_{h})\cup\{j\}\}}\\
 &
   -\sigma \sum_{\MBf\in\CF}\sum_{h\in\CF} \sum_{\substack{j\in\vec{p}_f\\j\in\CS}} \frac{\mathbbm{1}_{\{\CC(\vec{p}_f)\cup\CC(\vec{p}_{h})\subseteq \CS\}}}{|\CC(\vec{p}_f){\cup}\CC(\vec{p}_{h}){\cup}\{j\}|} \times \\
  &~~~~ \mathbbm{1}_{\{j\not\in\CC(\vec{p}_f)\cup\CC(\vec{p}_{h})\}} w_fw_{h}\mathbbm{1}_{\{i\in\CC(\vec{p}_f)\cup\CC(\vec{p}_{h})\cup\{j\}\}} \\
 =& \sum_{\MBf\in \CF^{new}(\CS)} \mathbbm{1}_{\{i\in\CC(\vec{p}_f)\}} \frac{\Delta_f(\CS)}{n_f(\CS)}\\
 &-\sigma \sum_{\MBf\in\CF}\sum_{h\in\CF} \sum_{\substack{j\in\vec{p}_f\\j\in\CS}}
  \frac{\mathbbm{1}_{\{\CC(\vec{p}_{h})\subseteq \CS\}}}{|\CC(\vec{p}_{h})\cup\{j\}|} \times\\
  &~~~~ w_fw_{h}\mathbbm{1}_{\{j\in\CC(\vec{p}_{h})\}}  \mathbbm{1}_{\{i\in\CC(\vec{p}_{h})\cup\{j\}\}}\\
 & + \sigma \sum_{\MBf\in\CF}\sum_{h\in\CF} \sum_{\substack{j\in\vec{p}_f\\j\in\CS}} \frac{\mathbbm{1}_{\{\CC(\vec{p}_f)\cup\CC(\vec{p}_{h})\subseteq \CS\}}}{|\CC(\vec{p}_f){\cup}\CC(\vec{p}_{h}){\cup}\{j\}|}\times\\
  &~~~~ w_fw_{h}\mathbbm{1}_{\{j\in\CC(\vec{p}_f)\cup\CC(\vec{p}_{h})\}}  \mathbbm{1}_{\{i\in\CC(\vec{p}_f)\cup\CC(\vec{p}_{h})\cup\{j\}\}}
\end{align*}

\vspace{0.1in}
All the above equations are simple algebra.
As a final step, because $\mathbbm{1}_{\{j\in\CC(\vec{p}_{h})\}}
\mathbbm{1}_{\{i\in\CC(\vec{p}_{h})\cup\{j\}\}} {=}
\mathbbm{1}_{\{i,j\in\CC(\vec{p}_{h})\}}$, we replace
$\mathbbm{1}_{\{j\in\CC(\vec{p}_{h})\}}
\mathbbm{1}_{\{i\in\CC(\vec{p}_{h})\cup\{j\}\}}$ by
$\mathbbm{1}_{\{i,j\in\CC(\vec{p}_{h})\}}$, and we replace 
$\mathbbm{1}_{\{j\in\CC(\vec{p}_f)\cup\CC(\vec{p}_{h})\}}
\mathbbm{1}_{\{i\in\CC(\vec{p}_f)\cup\CC(\vec{p}_{h})\cup\{j\}\}}$ by
$\mathbbm{1}_{\{i,j\in\CC(\vec{p}_{h})\cup\CC(\vec{p}_f)\}}$ in the final expression.
One can see that the derived $\varphi_i(\CS,\tilde{v})$ is equal to what is stated in our Lemma.
\end{proof}

\begin{lem}
  \label{lemma:potential_old_new}
Suppose conditions (\ref{eq:RevLossModel}) and (\ref{eq:CondNoIncreDeploy}) hold.  
If the revenue loss of flows $\Delta_f(\mathcal{S})$ satisfies (\ref{eq:CondRoutPath}) 
and the revenue loss ISPs satisfies (\ref{eq:loss_function}), 
the potential function of $G$ can be
\begin{align*}
  &\Phi(\bm{a}){=}
    \sum_{f{\in}\CF^{new}(\CS_{\bm{a}})} \frac{\Delta_f(\CS_{\bm{a}})}{|\CC(\vec{p}_f)|} \\
  & {-}\sigma \sum_{f\in\CF}\sum_{h\in\CF^{new}(\CS_{\bm{a}})}\sum_{i\in\vec{p}_f}
    \frac{w_fw_{h} \mathbbm{1}_{\{i\in \CC(\vec{p}_{h})\}}}{|\CC(\vec{p}_{h})\cup \{i\}|}  \\
& +\sigma \sum_{f\in\CF^{new}(\CS_{\bm{a}})}\sum_{h\in\CF^{new}(\CS_{\bm{a}})}\sum_{i\in\vec{p}_f}
  \frac{ \mathbbm{1}_{\{i{\in}\CC(\vec{p}_f){\cup}\CC(\vec{p}_{h)}\}} }{|\CC(\vec{p}_{h}){\cup} \CC(\vec{p}_f){\cup}\{i\}|}  w_{h}w_f 
.
\end{align*}
\end{lem}

\begin{proof}[\bf Proof of Lemma~\ref{lemma:potential_old_new}]
From the decomposition of the value function $\tilde{v}(\CS)$ as shown in the
proof of Lemma~\ref{lemma:Shapley_revenue_loss}, we directly derive
the potential function according to Lemma 3 (page 10) of the paper\cite{ui2000shapley}.
In particular, 
\begin{align*}
  &\Phi(\bm{a})=
    \sum_{\MBf\in\CF^{new}(\CS_{\bm{a}})}\mathbbm{1}_{\{\CC(\vec{p}_f)\subseteq \CS\}} \frac{\Delta_f(\CS_{\bm{a}})}{|\CC(\vec{p}_f)|} -\sum_{i\in\CI} a_ic_i\\
  & {-}\sigma \sum_{\MBf\in\CF}\sum_{h\in\CF}\sum_{i\in\vec{p}_f}
    w_{h}w_f
    \frac{\mathbbm{1}_{\{\CC(\vec{p}_{h}){\cup}\{i\}\subseteq\CS_{\bm{a}}\}}}{|\CC(\vec{p}_{h}){\cup} \{i\}|} \mathbbm{1}_{\{i\in \CC(\vec{p}_{h})\}} \\
& {+}\sigma \sum_{\MBf{\in}\CF}\sum_{h{\in}\CF}\sum_{i{\in}\vec{p}_f}
  w_{h}w_f
  \frac{\mathbbm{1}_{\{\CC(\vec{p}_{h}){\cup}\CC(\vec{p}_f){\cup}\{i\}{\subseteq}\CS_{\bm{a}}\}}}{|\CC(\vec{p}_{h}){\cup} \CC(\vec{p}_f){\cup}\{i\}|} \mathbbm{1}_{\{i{\in}\CC(\vec{p}_f){\cup}\CC(\vec{p}_{h})\}}
.
\end{align*}
One can see the terms in the potential function $\Phi(\bm{a})$ is similar to the terms
in the Shapley value $\varphi_i(\CS,\tilde{v})$.
One can check that
$\Phi(1,\bm{a}_{-i}){-}\Phi(0,\bm{a}_{-i}){=}u_i(1,\bm{a}_{-i}){-}u_i(0,\bm{a}_{-i}){=}\varphi_i(\CS,\tilde{v}){-}c_i$.
By definition, $\Phi$ is a potential function of the game $G$.
\end{proof}
Based on Lemma \ref{lemma:potential_old_new}, we study 
the impact of revenue loss on deployability of architectures via the
potential function.

\begin{proof}[\bf Proof of Theorem~\ref{theorem:deployability_old_new}]
  Under the condition
  \[
    \sum_{\MBf\in\CF}w_f\mathbbm{1}_{\{i\in\vec{p}_f\}}{\le}
    \sum_{\MBf\in\CF}\frac{w_f|\CC(\vec{p}_f)|}{|\CC(\vec{p}_f)|{+}\bar{N}_c{+}1},
  \]
  or equivalently after replacing the notation $\MBf$ to $h$,
\[
  \sum_{h\in\CF}w_{h}\mathbbm{1}_{\{i\in\vec{p}_h\}}{\le}
  \sum_{h\in\CF}\frac{w_{h}|\CC(\vec{p}_{h})|}{|\CC(\vec{p}_{h})|{+}\bar{N}_c{+}1},
\]
we have (multiply the same quantity on both LHS and RHS)
\begin{align*}
  &\sum_{h\in\CF}w_{h}\mathbbm{1}_{\{i\in\vec{p}_h\}}
  \frac{\sum_{h\in\CF} w_{h}\mathbbm{1}_{\{i\in\CC(\vec{p}_h)\}}}{\sum_{h\in\CF} w_h\mathbbm{1}_{\{i\in\vec{p}_h\}}}\\
  {\le} &
  \sum_{h\in\CF}\frac{w_{h}|\CC(\vec{p}_{h})|}{|\CC(\vec{p}_{h})|{+}\bar{N}_c{+}1}
  \left(\frac{\sum_{\MBf\in\CF} w_f\mathbbm{1}_{\{i\in\CC(\vec{p}_f)\}}}{\sum_{\MBf\in\CF} w_f\mathbbm{1}_{\{i\in\vec{p}_f\}}}\right).
\end{align*}
After re-organizing the terms, we get
\begin{align}
\sum_{h\in\CF}
     w_{h} 
  \left(
  \frac{\left(\frac{\sum_{\MBf\in\CF} w_f\mathbbm{1}_{\{i\in\CC(\vec{p}_f)\}}}{\sum_{\MBf\in\CF} w_f\mathbbm{1}_{\{i\in\vec{p}_f\}}}\right)}{|\CC(\vec{p}_{h})|{+}\bar{N}_c{+}1}
  {-}\frac{\mathbbm{1}_{\{i\in \CC(\vec{p}_{h})\}}}{|\CC(\vec{p}_{h})|} 
  \right){\ge} 0.
\end{align}
In addition, we have the following inequality that 
\begin{align}
  &   \sum_{\MBf\in\CF}\sum_{h\in\CF}\sum_{i\in\vec{p}_f}
     w_{h}w_f 
    \left(
\frac{\mathbbm{1}_{\{i\in\CC(\vec{p}_f){\cup}\CC(\vec{p}_{h})\}}}{|\CC(\vec{p}_{h}){\cup} \CC(\vec{p}_f){\cup}\{i\}|}
    {-}\frac{\mathbbm{1}_{\{i\in \CC(\vec{p}_{h})\}}}{|\CC(\vec{p}_{h}){\cup} \{i\}|} 
  \right)
   \nonumber\\
  &\ge
    \sum_{\MBf\in\CF}\sum_{h\in\CF}\sum_{i\in\vec{p}_f}
     w_{h}w_f \times \nonumber\\
  & ~~~~\left( -\frac{\mathbbm{1}_{\{i\in \CC(\vec{p}_{h})\}}}{|\CC(\vec{p}_{h})|} 
  + \frac{\mathbbm{1}_{\{i\in\CC(\vec{p}_f)\}}}{|\CC(\vec{p}_{h})|{+}|\CC(\vec{p}_f)|{+}1}  \right)
  \nonumber\\
  & \ge 
    \sum_{\MBf\in\CF}\sum_{h\in\CF}\sum_{i\in\vec{p}_f}
     w_{h}w_f 
 \left( -\frac{\mathbbm{1}_{\{i\in \CC(\vec{p}_{h})\}}}{|\CC(\vec{p}_{h})|} 
  {+} \frac{\mathbbm{1}_{\{i\in\CC(\vec{p}_f)\}}}{|\CC(\vec{p}_{h})|{+}\bar{N}_c{+}1}  \right)\nonumber\\
  & = \sum_{h\in\CF} \frac{-\mathbbm{1}_{\{i\in\CC(\vec{p}_{h})\}}}{|\CC(\vec{p}_{h})|} w_{h} \left(\sum_{\MBf\in\CF}\sum_{i\in\vec{p}_f}w_f\right)\nonumber\\
   & ~~~~ + \sum_{h\in\CF} \frac{1}{|\CC(\vec{p}_{h})|+\bar{N}_c+1}w_{h}
    \left(\sum_{\MBf\in\CF}\sum_{i\in\vec{p}_f}w_f\mathbbm{1}_{\{i\in\CC(\vec{p}_f)\}}\right)
  \nonumber\\
  & =
    \left(\sum_{\MBf\in\CF}\sum_{i\in\vec{p}_f}w_f\right) \times \nonumber\\
  & ~~~~ \left[
  \sum_{h\in\CF}w_{h} 
  \left(
  \frac{\left(\frac{\sum_{\MBf\in\CF} w_f\mathbbm{1}_{\{i\in\CC(\vec{p}_f)\}}}{\sum_{\MBf\in\CF} w_f\mathbbm{1}_{\{i\in\vec{p}_f\}}}\right)}{|\CC(\vec{p}_{h})|{+}\bar{N}_c{+}1}
  {-}\frac{\mathbbm{1}_{\{i\in \CC(\vec{p}_{h})\}}}{|\CC(\vec{p}_{h})|} 
    \right)
  \right]
    {\ge} 0.
  \label{eq:23}
\end{align}
Now, we explain how we get the above sequence of inequalities. The first inequality is because $|\CC(\vec{p}_{h}){\cup}\{i\}|{\ge}
|\CC(\vec{p}_{h})|$ and $|\CC(\vec{p}_{h}){\cup} \CC(\vec{p}_f) {\cup}\{i\}|{\le}
|\CC(\vec{p}_{h})|{+}\CC(\vec{p}_f)|{+}1$. The second inequality comes from the fact that
harmanic mean is less than or equal to arithmetic mean.
To elaborate, consider the value vector $(|\CC(\vec{p}_{h})|{+}|\CC(\vec{p}_f)|{+}1)_{\MBf\in\CF}$ and
the weight vector $(w_f|\CC(\vec{p}_f)|)_{\MBf\in\CF}$. Then the
weighted harmonic mean is less than the weighted arithmetic mean, i.e.
\begin{align*}
 & \frac{\sum_{\MBf{\in}\CF}|\CC(\vec{p}_f)|w_f}{\sum_{\MBf{\in}\CF}\frac{|\CC(\vec{p}_f)|w_f}{|\CC(\vec{p}_{h})|+|\CC(\vec{p}_f)|+1}}
  {\le}
  \frac{\sum_{\MBf{\in}\CF} (|\CC(\vec{p}_{h})|{+}|\CC(\vec{p}_f)|{+}1)|\CC(\vec{p}_f)|w_f}{\sum_{\MBf\in\CF} |\CC(\vec{p}_f)|w_f}\\
 & = |\CC(\vec{p}_{h})| +\bar{N}_c+1.
\end{align*}
Note that
$|\CC(\vec{p}_f)|{=}\sum_{i\in\vec{p}_f}\mathbbm{1}_{\{i\in\CC(\vec{p}_f)\}}$, then
we have the following inequality which leads to the second inequality in (\ref{eq:23}),
\begin{align*}
  \sum_{\MBf\in\CF} \sum_{i\in\vec{p}_f}\frac{\mathbbm{1}_{\{i\in\CC(\vec{p}_f)\}}w_f}{|\CC(\vec{p}_{h})|{+}|\CC(\vec{p}_f)|{+}1}
  \ge
  \sum_{\MBf\in\CF}\sum_{i\in\vec{p}_f} \frac{\mathbbm{1}_{\{i\in\CC(\vec{p}_f)\}}w_f}{|\CC(\vec{p}_{h})| +\bar{N}_c + 1}.
\end{align*}
  
We will compare the potential functions 
  when $\sigma{=}0$ and when $\sigma{>}0$ based on Lemma~\ref{lemma:potential_old_new}.
To distinguish the potential functions for different $\sigma$, we use
$\Phi_\sigma(\bm{a})$ to denote the potential function under
$\sigma$, when ISPs' action profile is $\bm{a}$.
Then, we have
\begin{align}
  \label{eq:potential_ineq}
  &\Phi_\sigma(\bm{1})-\Phi_0(\bm{1})=\sum_{\MBf\in\CF}\sum_{h\in\CF}\sum_{i\in\vec{p}_f}
     w_{h}w_f\times \nonumber\\
 &\left(
\frac{\mathbbm{1}_{\{i\in\CC(\vec{p}_f){\cup}\CC(\vec{p}_{h})\}}}{|\CC(\vec{p}_{h}){\cup} \CC(\vec{p}_f){\cup}\{i\}|}
    {-}\frac{\mathbbm{1}_{\{i\in \CC(\vec{p}_{h})\}}}{|\CC(\vec{p}_{h}){\cup} \{i\}|} 
  \right).
\end{align}
According to (\ref{eq:23}), we know $\Phi_\sigma(\bm{1})-\Phi_{0}(\bm{1})\ge 0$
for $\forall \sigma>0$.

Henceforth, if a new architecture is deployable when $\sigma=0$, i.e.
$\Phi_0(\bm{1}){\ge} \Phi_0(\bm{0}){=} 0$, then we have $\Phi_\sigma(\bm{1}){\ge}
\Phi_\sigma(\bm{0}){=}0$ when $\sigma {>} 0$, which means the architecture is
deployable when $\sigma{>}0$. 
\end{proof}
}

\subsection{\bf Partial Deployment in Sub-networks}
One can check that the above device-level 
deployment game is a potential game with the following potential
function:
\[
\Phi(\bm{\mathcal{A}})
=
\sum\nolimits_{{f} \in \CF}
\sum\nolimits_{m=1}^{n_f(\mathcal{S}_f (\bm{\mathcal{A}}))}
\frac{\tilde{\Delta}_f(m)}{m}
-
\sum\nolimits_{i \in \tilde{\mathcal{C}}} 
\sum\nolimits_{d\in \mathcal{A}_i} \tilde{c}_d.
\]
It holds that $\Phi(\bm{\CA}_{-i}\cup \CA_i)-\Phi(\bm{\CA}_{-i}\cup
\CA_i^\prime) = u_i(\bm{\CA}_{-i}\cup \CA_i) - u_i(\bm{\CA}_{-i}\cup
\CA_i^\prime)$ for $\forall \CA_i,\CA_i^\prime\subseteq \CD_i$ and $\bm{\CA}_{-i}\triangleq \bigcup_{j\in \tilde{\CC}\backslash\{i\}} \CA_j$.

Then, one can easily check we have the following lemma regarding to the
potential function of the game.
\begin{lem}
Under conditions (\ref{eq:CondRoutPath}) and (\ref{eq:RevLoss}), 
an architecture is deployable, 
i.e., $\Phi(\bigcup_{i \in \tilde{\mathcal{C}}}\mathcal{D}_i) \geq \Phi(\emptyset)$,
if and only if
\begin{align}
  \label{eq:condition_division}
\sum\nolimits_{{f} \in \CF}
\sum\nolimits_{m=1}^{|\mathcal{C} (\vec{p}_f)|}
\frac{\tilde{\Delta}_f(m)}{m}
\ge
\sum\nolimits_{i \in \tilde{\mathcal{C}}} 
\sum\nolimits_{d\in \mathcal{D}_i} \tilde{c}_d.
\end{align}
\end{lem}

With the above lemma, we can now prove the Theorem~\ref{thm:partial_deploy}.

\begin{proof}[\bf Proof of Theorem~\ref{thm:partial_deploy}]
We observe the condition (\ref{eq:condition_division}) is the same as the condition 
(\ref{eq:necessary_condition}). In particular, the left hand sides are the total
immediate benefit. Also, the right hand sides are 
the total launching cost to deploy the new protocols in all devices.
\end{proof}

\subsection{\bf Mechanism Design to Enhance Deployability}

\begin{proof}[{\bf Proof of Theorem \ref{thm:truth_telling}}]
Since we only consider one architecture and an ISP only has $\{0,1\}$ actions,
an ISP $i$ only needs to make one quote $q_i$ for all its networks.
  
In the proof, we will show that to quote higher or lower than an ISP's launching cost
will not yield a higher utility for the ISP. 
To begin with, we have some lemmas to show some properties of our mechanism.

We focus on the quoting decision of some ISP $i$. Suppose the ISP $i$ has two candidate
quotes, $q_i^H$ and $q_i^L$ where $q_i^L<q_i^H$. The quotes of any other ISP $j$
is $q_j$. Without loss of generality, the critical ISPs have indices
$\{1,\ldots, |\tilde{\CC}|\}$.
Then, we denote the quote profile $\bm{q}^L{=}(q_1,\ldots, q_{i-1}, q_i^L,
q_{i+1}, \ldots, q_{|\tilde{\CC}|})$, and $\bm{q}^H{=}(q_1,\ldots, q_{i-1},
q_i^H, q_{i+1}, \ldots, q_N)$. For presentation, we call some set $\CS_{\bm{a}}$
an ``equilibrium set'', if the corresponding action profile $\bm{a}$ is an equilibrium.

\begin{lem}
 \label{lemma:include} 
 $\CS^*(\bm{q}^H){\subseteq} \CS^*(\bm{q}^L)$.
\end{lem}
\begin{proof}
  Recall that $\CS^*(\bm{q}^H)$ and $\CS^*(\bm{q}^L)$ are the solutions of
  the following optimization problems:
  \begin{align}
    \label{opt:1}
  \CS^*(\bm{q}^H)=\arg\max_{\CS} && |\CS|,\nonumber\\
    \text{subject to} && \varphi_j(\CS, v)\ge q_j, \forall j\ne i \text{ and } j\in \CS, \nonumber\\
                      && \varphi_i(\CS, v)\ge q_i^H, \text{if }i\in \CS.
  \end{align}
    \begin{align}
    \label{opt:2}
  \CS^*(\bm{q}^L)=\arg\max_{\CS} && |\CS|,\nonumber\\
    \text{subject to} && \varphi_j(\CS, v)\ge c_j, \forall j\ne i \text{ and } j\in \CS, \nonumber\\
                      && \varphi_i(\CS, v)\ge q_i^L, \text{if }i\in \CS.
  \end{align}
The fact is that $\CS^*(\bm{q}^L)$ is the largest equilibrium when the cost
profile $\bm{c}{=}\bm{q}^L$. This will be proved in Theorem \ref{thm:max_objective}.
We claim that there will be some equilibrium set $\CS^*\supseteq
\CS^*(\bm{q}^H)$ when the launching cost profile $\bm{c}{=}\bm{q}^L$.

\noindent \underline{Case 1:} $i\in \CS^*(\bm{q}^H)$. The set
$\CS^*(\bm{q}^H)$ itself corresponds such an equilibrium.
Recall that the utility increment for ISP $i$ to deploy is $u_i(1,\bm{a}_{-i})-u_i(0,\bm{a}_{-i})=\varphi_i(\CS_{\bm{a}},v)-c_i$.
Note that $i\in
\CS^*(\bm{q}^H)$ will deploy and her benefit $\varphi_i(\CS^*(\bm{q}^H), v)\ge q_i^H>q_i^L=c_i$
is higher than her launching cost, so ISP $i$ does not want to deviate. For some
other ISP $j\ne i$ and $j\in \CS^*(\bm{q}^H)$, constraints in (\ref{opt:1})
guarantee a higher benefit than the launching cost, so such $j$ will not
deviate. For some other $j\ne i$ and $j\not\in \CS^*(\bm{q}^H)$, we have
$\varphi_i(\CS^*(\bm{q}^H), v)< q_i=c_i$, or else $\CS^*(\bm{q}^H)$ would not be
the largest set as an optimal solution of (\ref{opt:1}) since $\CS^*(\bm{q}^H)\cup\{j\}$ would be a
larger feasible solution. 

\noindent \underline{Case 2:} $i\not\in \CS^*(\bm{q}^H)$. Every ISP does not want
to deviate when a set $\CS^*(\bm{q}^H)$ of ISPs deploy, except ISP $i$. If
$\varphi_i(\CS^*(\bm{q}^H), v)<q_i^L$, then ISP $i$ also does not want to deviate.
If $\varphi_i(\CS^*(\bm{q}^H), v)\ge q_i^L$, then by {\it ``best-response''}, ISPs will
reach an equilibrium with a set of deployer $\CS^*\supseteq \CS^*(\bm{q}^H)$
because the game $G$ is a supermodular game\cite{topkis2011supermodularity}.

Since some $\CS^* \supseteq \CS^*(\bm{q}^H)$ is an
equilibrium set, as stated in Lemma \ref{lemma:super_eq}, it is contained in the
largest equilibrium set $\CS^*(\bm{q}^L)$, $\CS^*(\bm{q}^H){\subseteq} \CS^* {\subseteq} \CS^*(\bm{q}^L)$.
\end{proof}

\begin{lem}
  \label{lemma:selected}
  If $i\in \CS^*(\bm{q}^H)$, then
 $\CS^*(\bm{q}^H)=\CS^*(\bm{q}^L)$.
\end{lem}
This lemma states that if an ISP $i$ is selected with a higher quote $q_i^H$,
then the set of selected ISPs will be the
 same as that when ISP $i$ has a lower quote $q_i^L$.
\begin{proof}

Suppose on the contrary, $\CS^*(\bm{q}^H)\subsetneq \CS^*(\bm{q}^L)$. Then by
(\ref{eq:Shapley_monotone}) we have
\[\varphi_j(\CS^*(\bm{q}^L),
  v){\ge} \varphi_j(\CS^*(\bm{q}^H),v), \forall j.
\]
Also, because $i\in \CS^*(\bm{q}^H)$ we have $\varphi_i(\CS^*(\bm{q}^H),v){\ge}
q_i^H$, so
\[
\varphi_i(\CS^*(\bm{q}^L),
v){\ge} \varphi_i(\CS^*(\bm{q}^H),v){\ge} q_i^H.
\]
We could see that $\CS^*(\bm{q}^L)$ is a feasible solution to the problem
(\ref{opt:1}). Moreover, $|\CS^*(\bm{q}^L)|{>}|\CS^*(\bm{q}^H)|$, which is a
contradiction to the optimality of the set $\CS^*(\bm{q}^H)$ for the
problem (\ref{opt:1}).
\end{proof}

\begin{lem}
  \label{lemma:not-selected}
  Suppose $i{\not\in} \CS^*(\bm{q}^H)$. If $i{\in} \CS^*(\bm{q}^L)$, then $\varphi_i(\CS^*(\bm{q}^L),v) {<} q_i^H$.
   \end{lem}
This lemma states that if ISP $i$ is not selected with the higher quote $q_i^H$,
then its distributed benefit will be lower than $q_i^H$ even if she is selected
by the lower quote $q_i^L$.
\begin{proof}
 Suppose to the contrary that the ISP $i{\in} \CS^*(\bm{q}^L)$ and
 $\varphi_i(\CS^*(\bm{q}^L),v) {\ge} q_i^H$. Then we could see $\CS^*(\bm{q}^L)$
 is also a feasible set for the problem (\ref{opt:1}) because the constraints
 for the other ISPs are also satisfied. Furthermore, we know $\CS^*(\bm{q}^H)
 \subseteq \CS^*(\bm{q}^L)$ from Lemma \ref{lemma:include}. Moreover, $i\in
 \CS^*(\bm{q}^L)$ and $i\not\in \CS^*(\bm{q}^H)$. Therefore,
 $|\CS^*(\bm{q}^L)|{>} |\CS^*(\bm{q}^H)|$, which is a contradiction to the
 optimality of the set $\CS^*(\bm{q}^H)$ for problem (\ref{opt:1}).
\end{proof}
Now, we go back to our Theorem \ref{thm:truth_telling}. Denote ISP $i$'s utility
as $U_i(q_i, \bm{q}_{-i})$ when her quote is $q_i$ and other ISPs' quotes are
$\bm{q}_{-i}$. For notational convenience, we use $U_i(q)$ for $U_i(q, \bm{q}_{-i})$.

First, we show that for any ISP $i$, decreasing the quote down to $c_i$ will not hurt
her utility, i.e. $U_i(c_i)\ge U_i(q)$ for any $q\ge c_i$.  On
one side, suppose with the orginal quote $q_i\ge c_i$, the ISP is selected. Then
by Lemma~\ref{lemma:selected}, we could see that the selected set will not be changed
and ISP $i$'s utility will be the same. On the other side, if under the original quote
$q_i\ge c_i$, the ISP is not selected, then the original utility is $0$. But
with a lower quote, she is guaranteed to be distributed a benefit higher than $c_i$, and
therefore the utility $U_i(c_i)\ge U_i(q_i)=0$ when $q_i\ge c_i$. 

Second, we show that further decreasing the quote below $c_i$ will not improve her
utility, i.e. $U_i(c_i)\ge U_i(q)$ for any $q\le c_i$. On one hand, suppose the ISP is selected when quoting $c_i$, then by
Lemma~\ref{lemma:selected}, her
utility will be unchanged from quoting a lower value. On the other hand, if the
ISP is not selected by quoting $c_i$, according to
Lemma~\ref{lemma:not-selected}, the ISP will not be distributed more than $c_i$
when she quote lower
i.e. $U_i(q^L)=\varphi_i(\CS^*(\bm{q}^L),v) {-} c_i {<}0$.
Therefore, she will have negative utility if she is selected with a lower quote.

Now, we have shown that $U_i(c_i)\ge U_i(q)$ for any $q$. Therefore, to quote
$c_i$ is a weakly dominant strategy for ISP $i$, which is regardless of other
ISPs' quotes.
\end{proof}

\begin{proof}[{\bf Proof of Theorem \ref{thm:max_objective}}]
 We first show that optimization problem (\ref{eq:final_objective}) selects
 the largest equilibrium when $\bm{q}=\bm{c}$. To the contrary, suppose the
 largest equilibrium is $\bm{a}^\prime$ and the corresponding set of deployers is
 $\CS_{\bm{a}^\prime}$ which is not equal to $\CS^*(\bm{c})$. Then, $\CS^*(\bm{c})\subsetneq
 \CS_{\bm{a}^\prime}$ because $\CS_{\bm{a}^\prime}$ is the largest equilibrium
 set in a supermodular game like $G$. We further claim that $\CS_{\bm{a}^\prime}$ is a feasible
 solution to satisfy the constraints (\ref{eq:final_objective}). In fact, in the
 equilibrium, by \Cref{def:nash_eq}, the utility increment for an ISP $i\in
 \CS_{\bm{a}^\prime}$ to deploy is non-negative, i.e. $\varphi_i(\CS_{\bm{a}^\prime},v)-c_i\ge 0$.
 Now, we find a feasible solution $\CS_{\bm{a}^\prime}$ and
 $|\CS_{\bm{a}^\prime}|>|\CS^*(\bm{c})|$, which is a contradiction to the
 optimality of $\CS^*(\bm{c})$.

 Second, we show the largest equilibrium set $\CS^*(\bm{c})$ also maximizes the
 total revenue gain $v(\cdot)$. Formally, $\CS^*(\bm{c})$ is also the solution
 of the following optimization problem:
\begin{align}
  \label{problem:v}
& 
  \underset{\CS \subseteq \tilde{\CC}}{\text{maximize}} 
&& v(\CS),\nonumber\\
&  \text{subject to} && \varphi_i(\CS, v)\ge c_i, \forall i \in \CS.
\end{align}
In fact, we claim that some optimal solution to the problem (\ref{problem:v})
$\CS^\prime$ corresponds to an equilibrium $\bm{a}^\prime$ where
$\CS^\prime=\CS_{\bm{a}^\prime}$. Suppose on the contrary any optimal solution $\CS^\prime$ does
not corresponds to an equilibrium. Then, either of the following two cases will
happen 1) there exists some ISP $i\in
\CS^\prime$ such that $\varphi_i(\CS^\prime, v)<c_i$; 2) there is some ISP
$j\not\in \CS^\prime$ such that $\varphi_j(\CS^\prime\cup \{j\}, v)> c_j$. Case 1
will not happen because constraints in (\ref{problem:v}) are satisfied.
Therefore, we must be in Case 2 where we find some $j\not\in \CS^\prime$ such that $\varphi_j(\CS^\prime\cup
\{j\}, v)> c_j$. Because of (\ref{eq:Shapley_monotone}), we have another set
$\CS^\prime\cup\{j\}$ that satisfies the constraints in (\ref{problem:v}).
Furthermore, we have $v(\CS^\prime\cup\{j\}) > v(\CS^\prime)$ by the condition
of this Theorem. It means that we find another
feasible solution $\CS^\prime\cup\{j\}$ and $v(\CS^\prime\cup\{j\}) >
v(\CS^\prime)$, which is a contradiction to the optimality of $\CS^\prime$.
Since an maximizer of problem (\ref{problem:v}) comes some equilibrium. The
largest equilibrium $\overline{\bm{a}}^*$ will be the one that yields the
highest value $v(\CS_{\overline{\bm{a}}^*})$ because of the monotonicity of
$v(\cdot)$ stated by the condition of this Theorem.

\end{proof}

\begin{proof}[\bf Proof of Theorem~\ref{thm:tipping_algorithm}]
  We claim that in each iteration of the while loop in
  \Cref{alg:improved_selection}, the set $\CS$ will expand if it does not
  represent the largest equilibrium. If this claim holds, the
  while loop will terminate, because the number of ISPs is finite.

  For the above claim to hold, we only need to show that \Cref{alg:greedy_selection} returns a non-empty set $\CT$
  when the input $\CS$ corresponds to some equilibrium. Suppose on the contrary
  that $\CT=\emptyset$, then for the last ISP $j$ eliminated in Line 9, we have
  $\texttt{gap}(\CS\cup \{j\})=0$ because of the property governed by the while
  loop. Now, $\CS$ is no longer an equilibrium because $j$ is willing to deploy,
  which leads to a contradiction. When $\CT\ne \emptyset$, the equilibrium from
  $\CS\cup \CT$ in Line 4 of \Cref{alg:improved_selection} will be larger than
  $\CS$ because the ISPs in $\CS$ always want to deploy according to the supermodularity of the utility function.
\end{proof}
Our tipping set selection scheme keeps the property that the ISPs will
reach the largest equilibrium, so the desirable properties in
Theorem~\ref{thm:truth_telling} and \ref{thm:max_objective} still hold.
Meanwhile, the improved selection scheme reduces the number of assigned ISPs so that the mechanism is easier to operate. 
\bibliographystyle{IEEEtran}

{
  \bibliography{bib}
}